\documentclass[]{article}

\usepackage[T1]{fontenc} % Use modern font encodings

\usepackage{amsmath}
\usepackage{amssymb}
\usepackage{mathtools, cuted}
\usepackage{supertabular}
\usepackage[symbol]{footmisc}
\usepackage{multirow}
\usepackage{array}
\usepackage{stackengine}

\usepackage{graphicx}
\usepackage{caption}
\usepackage{subfigure} 
\usepackage{psfrag}
\usepackage{overpic}
\usepackage{float}
\usepackage{flafter}

\usepackage{enumerate}

\usepackage{color}
\usepackage[table]{xcolor}
\usepackage{tcolorbox}

\usepackage[]{algorithm2e}

\DeclareMathOperator*{\argmin}{\arg\,\min}
\def\yy{\mathbf{y}}
\newcommand{\norm}[1]{\left\lVert#1\right\rVert}

\graphicspath{{./FIGS/}}

\newtheorem{theorem}{Theorem}[section]
\newtheorem{definition}[theorem]{Definition}
\newenvironment{proof}{
    {\bf Proof:}}{\hbox{\ }\hfill $\blacksquare$\break %\hspace{-1\parindent}
    }

\newcommand{\figref}[1]{Figure~\ref{#1}}

\newcommand{\dst}{\delta}  % lennard jones dist
\newcommand{\dlo}{\underline{\dst}} 
\newcommand{\dhi}{\overline{\dst}} 

\newcommand{\acgW}{active constraint graph}
\newcommand{\acr}{active constraint region}
\newcommand{\rmc}{rigid molecular component}

\newcommand{\distall}{$C_1$}
\newcommand{\distexist}{$C_2$}
\newcommand{\ctwo}{\ref{eqn:preferredConstraints}}
\newcommand{\cone}{\ref{eqn:constraints}}

\newcommand{\toytwodone}{Toy1$\mathbb{R}^2$}
\newcommand{\toytwodtwo}{Toy2$\mathbb{R}^2$}
\newcommand{\toythreed}{Toy$\mathbb{R}^3$}

\newcommand{\bx}{\mathbf{x}}
\newcommand{\bp}{\mathbf{p}}

\newcommand{\psone}{A}
\newcommand{\pstwo}{B}
\newcommand{\pone}{a}
\newcommand{\ptwo}{b}

\newcommand{\ijx}{{\pone\ptwo}}
\newcommand{\lj}{f}  % lennard jones function
\newcommand{\ljlo}{\underline{\lj}} 
\newcommand{\ljhi}{\overline{\lj}} 
\newcommand{\Mn}{\mathcal{M}} % manifold

\definecolor{yelloworange}{RGB}{200,146,0}
\definecolor{grayoned}{RGB}{155,155,155}
\definecolor{green2d}{RGB}{0,150,0}

\newcommand{\rahul}[1]{\color{black}{#1}\color{black}}

\usepackage{fullpage}
\author{Rahul Prabhu$^\dag$ \and Meera Sitharam\footnote{These co-first authors contributed equally to this work} \and Aysegul Ozkan \and Ruijin Wu}
\title{Atlasing of Assembly Landscapes using Distance Geometry and Graph Rigidity}

\begin{document}
\maketitle
\begin{abstract}
\scriptsize{
This paper describes a novel geometric methodology for analyzing free-energy
and kinetics of assembly driven by short-range pair-potentials in an implicit 
solvent, and provides a proof-of-concept illustration of its unique capabilities. 

An \emph{atlas} is a labeled partition of the assembly landscape into a
roadmap of maximal, contiguous, nearly-equipotential-energy
conformational regions or macrostates, together with their neighborhood
relationships. The new methodology decouples the roadmap generation
from sampling and produces: (1) a query-able atlas of local potential
energy minima, their basin structure, energy barriers, and neighboring
basins; (2) paths between a specified pair of basins, each path being a
sequence of conformational regions or macrostates below a desired energy threshold;
and (3) approximations of relative path lengths, basin volumes
(configurational entropy), and path probabilities.
 
Results demonstrating the core algorithm's capabilities and high computational
efficiency have been generated by a resource-light, curated opensource
software implementation EASAL \cite{Ozkan:toms} (Efficient Atlasing and
Search of Assembly Landscapes, see software \cite{easalSoftware}, video \cite{easalVideo}
and user guide \cite{easalUserGuide}). Running
on a laptop with Intel(R) Core(TM) i7-7700 @ 3.60GHz CPU with 16GB of
RAM, EASAL atlases several hundred thousand
conformational regions or macrostates in minutes using a single compute core.
Subsequent path and basin computations each take seconds. A
parallelized EASAL version running on the same laptop with 4 cores
gives a 3X speedup for atlas generation.

The core algorithm's correctness, time complexity, and
efficiency-accuracy tradeoffs are formally guaranteed using modern
distance geometry, geometric constraint systems and combinatorial rigidity. 

The methodology further links the shape of the input 
assembling units to a type of intuitive and query-able bar-code of the
output atlas, which in turn determine stable assembled structures and 
kinetics. This succinct input-output relationship facilitates reverse 
analysis, and control towards design.

A novel feature that is crucial to both the high sampling efficiency and
decoupling of roadmap generation from sampling is a recently-developed
theory of convex Cayley (distance-based) custom parametrizations specific
to assembly, as opposed to folding.  Representing microstates with
macrostate-specific Cayley parameters, to generate microstate samples,
avoids gradient-descent search used by all prevailing methods. 
Further, these parametrizations convexify conformational regions or macrostates. 
This ratchets up sampling efficiency, significantly reducing number of
repeated and discarded samples. 

These features of the new stand-alone methodology can also be used to 
complement the strengths of prevailing methodologies including Molecular 
Dynamics, Monte Carlo, and Fast Fourier Transform based methods.
}
\end{abstract}

\section{Introduction} 
\label{sec:intro}
This paper describes a new geometric methodology for efficient atlasing and
search of assembly energy landscapes. Since many novel concepts are required to
even informally describe the new methodology, the reader is referred to the
\rahul{Table \ref{tab:terminology} } in Section \ref{sec:concepts}. The table
also includes a rough correspondence between new and established concepts.
Formal definitions can be found in Section \ref{sec:methods}. \emph{Assembly}
is defined broadly as starting from tethered or flexibly bonded rigid molecular
components, and driven by short-range potentials between atom-pairs in distinct
components in an implicit solvent (subsuming hard-sphere or sticky-sphere
models). 

A key concept central to the new methodology is the \emph{atlas} which is a
partition of the assembly landscape into macrostates, organized as a roadmap.
Each \emph{macrostate} is a maximal, contiguous, nearly-equipotential-energy
conformational region. {\bf Note:} We refer to molecular \emph{configurations}
or \emph{microstates} $X$ and $Y$ as being \emph{conformations} of each other
(in the same macrostate) if their potential energies are similar and any
continuous path between them does not have to cross significant energy
barriers, where `similar' and `significant' depend on the context, and the
level of discretization of the pair potentials (see Section
\ref{sec:geometrization}). Otherwise, they are in different macrostates. The
atlas further \emph{stratifies} macrostates by energy level.

The methodology provides resource-light, stand-alone algorithms that can also
be used to complement prevailing general ones such as Monte Carlo and Molecular
Dynamics, as well as specialized ones used in docking. The unique features are:

\begin{enumerate}
\item mitigating the curse of dimension in configurational entropy (free
energy) and kinetics computations by
    \begin{enumerate}
      \item decoupling exploration from sampling, i.e., generating 
	   an atlas of the landscape - including basins, barriers, paths 
		and their neighborhood relationships - with minimal sampling, using
		geometric constraints \cite{SJS:Handbook} and rigidity-based 
		roadmap;
      \item \emph{convexifying} macrostates using customized, 
      \emph{Cayley} parametrization\cite{SiGa:2010}, 
	  which is a distance-based internal coordinate representation of
	  assembly configurations that are constrained by inter-atomic distances (formally
	  defined in Section \ref{sec:convexification}); this
      parametrization achieves high sampling efficiency and accuracy, avoiding
      gradient-descent, and repeated or discarded samples; 
    \end{enumerate}
\item formally isolating succinct topological and geometric characteristics 
- intuitive \emph{bar-codes} - that differentiate assembly landscapes
from each other; 
\item exploiting characteristics that differentiate 
assembly landscapes from more complex folding landscapes;
\item establishing a connection between these succinct bar-codes
and the geometry of the input assembling structures, e.g., to facilitate 
labeling, querying, reverse analysis and design of landscapes;
\item mathematically proving efficiency and accuracy guarantees.
\end{enumerate}

The emphasis of this paper is not formal comparisons with prevailing
methodologies, nor demonstration of performance on benchmarks
\cite{Chill2014,cha2015accelerated} for macromolecular or spherical cluster
assembly systems (see Section \ref{sec:app:companionPapers} for a discussion on
companion papers). Rather, the emphasis here is on describing the new
methodology and core algorithms' rigorous geometric underpinnings, providing
both proofs of efficiency and accuracy, and proof-of-concept results
demonstrating its efficiency and unique features. The results have been
generated using an opensource software implementation EASAL (Efficient Atlasing
and Search of Assembly Landscapes) of the core algorithms (see software
\cite{easalSoftware}, video \cite{easalVideo} and user guide
\cite{easalUserGuide}).

The roadmap component of the atlas is \emph{refinable, and query-able} with
unique, intuitive ``street signs'' or macrostate labels, each of which is a
graph or network of geometric constraints \cite{SJS:Handbook}, called the
\emph{active constraint graph}. These constraints are imposed by the
short-range pair potentials that are \emph{active} in that macrostate, i.e.,
each constraint is between an atom-pair whose inter-atomic distance achieves
minimum energy, treated in the limit as a hard-sphere potential. Thus a
macrostate is an \emph{active constraint region} whose \emph{effective
dimension} can be determined using combinatorial rigidity \cite{SJS:Handbook}
of the active constraint graph. Consequently, the effective dimension becomes
a proxy for the energy level of the macrostate. The theory presented here
extends easily to longer range potentials with a slight modification to the way
the potentials are geometrized, as explained in detail in Section
\ref{sec:geometrization}. However, for ease of exposition, we use short-range
potentials throughout the paper.

Microstates or configurations, which are traditionally represented using
Cartesian parameters, are instead represented using \emph{Cayley} or
distance-based parameters that are customized to the active constraint graph of
the macrostates. As we show in Section \ref{sec:convexification}, generating
microstate samples within an active constraint region using Cayley parameters
avoids gradient-descent search used by all prevailing methods to sample
constrained regions. Further, under Cayley parametrizations, active constraint
regions or macrostates become convex with easily computable bounds. 

We show that the convexification technique applies readily to macrostates in
assembly, but not so readily to folding. In a convexifiable macrostate or
active constraint region, there is a collection of atom pairs (the Cayley
parameters $F$, which are not active constraints) satisfying the following
property: between any pair of configurations or microstates there is a path in
the macrostate along which the inter-atomic distances $F$ maintain a linear or
affine relationship. Thus, very roughly speaking, an assembly system can follow
straight paths, in a certain formal sense, and still avoid breaking energy
barriers, i.e., while remaining in the same macrostate (a contiguous, nearly
equipotential energy region). However, a folding system, lacking convexifiable
macrostates, may be forced to follow a convoluted path in order to remain in a
macrostate.

Convexification improves sampling efficiency for assembly landscapes,
significantly reducing the number of repeated and discarded samples. Overall,
the methodology directly addresses the curse of dimension and complexity of
landscapes while giving formal guarantees of efficiency, accuracy, robustness
and tradeoffs for the core algorithms (see Section
\ref{sec:results:complexity}). 

Furthermore, the methodology isolates \emph{input shape variables} of the
assembling units that directly influence \emph{landscape design variables},
including the number of macrostates and the average Cayley parameter value.
The values of the design variables yield a succinct \emph{bar-code} for the
output atlas.

Finally, a single-threaded software implementation of the core algorithms,
EASAL \cite{Ozkan:toms} (see software \cite{easalSoftware}, video
\cite{easalVideo} and user guide \cite{easalUserGuide}), atlases several
thousand (resp. hundred thousand) macrostates and computes entropy integrals of
reasonable accuracy in minutes (resp. a couple of hours) on a laptop with
Intel(R) Core(TM) i7-7700 @ 3.60GHz CPU with 16GB of RAM, running on a single
core, for bi-assemblies of up to 42-atom alpha-helices, and for clusters of up
to 24 identical sticky-sphere particles. Subsequent path computations each take
seconds. A recent proof-of-concept parallelized EASAL version running on the
same laptop, with 4 cores, gives a 3X speedup in preliminary tests. Efficiency
can improve significantly when the assembling entities are identical, by
exploiting symmetries in the landscape \cite{sym8010005}. Although the proof of
concepts results demonstrated in this paper only sample bi-assemblies of up to
42-atom alpha-helices, EASAL is more generally applicable to other
bi-assemblies of rigid molecular components of much larger size. EASAL has been
used to analyze assemblies of rigid molecular components with $n \approx 5000$
atoms \cite{Wu2012,virus2019}.

\subsection{Basic Concepts and Terminology}
\label{sec:concepts}
The following table gives the list of terms and their definitions
used in this paper.
\tablehead{\hline}
\tabletail{\hline}
\topcaption{\rahul{Basic concepts and terminologies.}} 
%to the sections in which they are defined.}}
\label{tab:terminology}
\begin{supertabular}{|p{.25\columnwidth}|p{.72\linewidth}|}
\emph{active constraint} & A pair of atoms from different \rmc s satisfying constraint \ctwo\ of the assembly problem (formally described in Section \ref{sec:geometrization}).\\\hline
\emph{assembly problem} & Problem (\cone, \ctwo) for general $k$ (formally described in Section \ref{sec:methods:furtherIO}). \\\hline
\emph{\rmc}, $n$, $k$ & The set of $n$ atom centers in a rigid molecule. $k$ \rmc s are inputs to the assembly problem (formally described in Section \ref{sec:methods:furtherIO}).\\\hline
\emph{assembly landscape} &  Complete set of solution configurations of the assembly problem (formally described in Section \ref{sec:methods:furtherIO}).  \\\hline
\emph{macrostate} & Maximal, contiguous, nearly-equipotential-energy conformational region.\\\hline
\emph{microstate} & A molecular configuration.\\\hline
\emph{dimension of assembly landscape} $d_\mathcal{C}$ & Dimension of the assembly landscape (formally described in Section \ref{sec:methods:furtherIO}).\\\hline
\emph{ambient dimension} $d_\mathcal{A}$ & Ambient dimension of assembly problem (formally described in Section \ref{sec:methods:furtherIO}).\\\hline
\emph{roadmap} & A directed acyclic graph whose nodes are active constraint regions. Edges go from parent to child active constraint regions (formally described in Section \ref{sec:stratification}).\\\hline
\emph{stratification} & Organization of the roadmap by dimension, i.e., energy level (formally described in Section \ref{sec:stratification}).\\\hline
\emph{active constraint region $R_G$} & A finite set of maximal, contiguous, nearly equi-potential energy regions (macrostates) that have the same set of active constraints (formally described in Section \ref{sec:stratification}).\\\hline
\emph{active constraint graph $G$} & The vertices are the atoms involved in active constraints, plus at least 3 atoms from each of participating \rmc. Two types of edges: (1) pairs $(a,b)$ that are active constraints 
and (2) all pairs $(a1, a2)$ such that both belong to the same \rmc (formally described in Section \ref{sec:stratification}). \\\hline
\emph{parent region in roadmap} & Higher dimensional, higher energy level, interior active constraint regions (formally described in Section \ref{sec:recursiveBoundarySearch}).\\\hline
\emph{child region in roadmap} & 1 lower dimensional (compared to parent), lower energy level, boundary active constraint regions (formally described in Section \ref{sec:recursiveBoundarySearch}).  \\\hline
\emph{atlas} & A topological roadmap of the assembly landscape (formally described in Section \ref{sec:convexity}).\\\hline
\emph{Cayley parameters} & Non-edges in the active constraint graph that make it a 3-tree (formally described in Section \ref{sec:convexity}).\\\hline
\emph{Cayley configuration} & A tuple of realizable length values of Cayley parameters for an active constraint graph (formally described in Section \ref{sec:convexity}).\\\hline
\emph{Cayley region} & The complete set of Cayley configurations in an active constraint region (formally described in Section \ref{sec:convexity}).\\\hline
\emph{chart} & The parameter map taking the (sampled) 
Cartesian configurations in an active constraint
region to the corresponding Cayley configurations (formally described in Section \ref{sec:convexity}).\\\hline
\emph{input shape variables} & The two variables \emph{width} and \emph{concavity} that describe the input \rmc s (formally described in Section \ref{sec:designVariable}).\\\hline
\emph{landscape design variables} & The variables used to facilitate control towards the design of the landscape (formally described in Section \ref{sec:designVariable}).\\\hline
\emph{witness} & The first configuration discovered in a region from the parent region, in which a new active constraint becomes active (formally described in Section \ref{sec:coreAlgorithm}).\\
\end{supertabular}
\color{black}

\subsection*{Organization}
The next Section \ref{sec:related} describes related work and delineates the
contributions of this paper. The Methods Section \ref{sec:methods} formally
describes the new methodology's distinct features and strategies - sketched in
the introduction - for generating (desired portions of) the atlas. The Results
Section \ref{sec:results} provides proof of concept illustrations of the
methodology's unique capabilities, using the EASAL software implementation
\cite{Ozkan:toms,easalVideo,easalSoftware}, including verification of the
theoretical efficiency and accuracy. These should be considered starting points
for the results and comparisons in the companion papers and future work (see
Section \ref{sec:app:companionPapers}, and Section \ref{sec:discussion}).

\section{Related Work and Contributions}
\label{sec:related}
Throughout this paper, we use two variables to differentiate between basic
types of assembly systems: $k$ and $n$ respectively denote the number of
assembling rigid molecular components (e.g. helices, clusters), and the maximum
number of points (e.g. atoms/particles) in any of them. The rigid molecular
components are assumed to be flexibly tethered or bonded to each other, with
the assembly being driven by weak interactions modeled as short-range,
Lennard-Jones pair potentials, in an implicit solvent.

We do not review the extensive literature on (ab-initio) simulation or
decomposition-based methods that are required to tractably deal with large
assemblies with $k \ge 25$, such as viral capsids. Direct computations of free
energy, binding affinity and kinetics of assembly are challenging even starting
from much fewer rigid molecular components. The challenges discussed below are
due to high dimensionality as well as topological and geometric complexity of
the potential energy landscape. 

\subsection{Exploiting Assembly Landscapes}
\label{sec:intro:ExploitingAssembly}
The simplest form of molecular assembly is site-specific docking, where a small
molecule initially in solution (the ligand) binds to a specific site in a much
larger molecule (the receptor). Diverse methods from computational geometry,
vision and image analysis have been used in site-specific docking algorithms
\cite{Bespamyatnikh, Choi2004, pmid1549581, Duhovny2002, pmid15980490,
Bolia2014, Bolia2016, Zhou2018}. Unlike the more general goals of the new
methodology itemized earlier, the goal of these algorithms is to find
site-specific docking configurations or microstates with optimal binding
affinity. While this depends on equilibrium free energy as well as kinetics
between stable states, docking methods arbitrarily restrict the region of
possible configurations or microstates to evaluate an approximate free energy
function. On the other hand, prevailing methods for more general free energy
and kinetics computation must incorporate both the depth and relative weighted
volumes (entropy) of basins and their topological relationships . These use
highly general approaches such as Monte Carlo (MC) and Molecular Dynamics (MD)
simulation.\cite{kaku, Head_Given_Gilson_1997, kurnikov1999harlem,
Andricioaei_Karplus_2001, Hnizdo_Darian_Fedorowicz_Demchuk_Li_Singh_2007,
Killian_Yundenfreund_Kravitz_Gilson_2007, Hnizdo_Tan_Killian_Gilson_2008,
Clark2009, Clark2009-Fragment, brooks2009charmm, Hensen_Lange_Grubmuller_2010,
GregoryS201199, King2012, Koppisetty2013, case2010amber, Jiang2019}

There exist a number of methods to explore the volume or configurational
entropy of the free energy basin landscape using the fact that effective
dimension of macrostates is a proxy for energy level. Small cluster assemblies
from spheres \cite{Arkus2009, Wales2010, Beltran-Villegas2011, Calvo2012,
Khan2012, Hoy2012, Hoy2014}, i.e., $n=1$ and $k$ arbitrary, ($k$ is typically
$\le 25$ without decomposition), are used in the study of mesoscale systems
such as $C_{60}$ molecule \cite{Doye1996, Hagen1993}, cDNA strands
\cite{Meng2010} and colloids interacting via depletion \cite{Gazzillo2006}.
These can be modeled as a short-ranged potentials particle system in which the
interaction length is much smaller than the system size. That is, their
Lennard-Jones wells are narrow and they can be treated as a Hard Sphere
assembly systems. For these systems, there exist a number of methods to
compute paths and kinetics as well as free energy and configurational entropy
of lower dimensional or lower energy macrostates \cite{Holmes-Cerfon2013,
Arkus2009, Wales2010, Beltran-Villegas2011, Calvo2012, Khan2012, Hoy2012,
Hoy2014}. Newer methods study the change in the energy landscape of spherical
particles when the interaction potentials change from a Lennard-Jones type
potential to sticky hard-sphere type potential \cite{Wales-Sticky2018,
trubiano2019canyons}. Taken together, these methods output trajectories of
sample configurations from free energy landscapes of a wide variety of systems.

However, these prevailing methods do not explicitly take advantage of the
relative simplicity of constraint systems that drive assembly compared to
folding. As we explain in detail in Section \ref{sec:convexification},
constraint graphs that arise in assembly yield convexifiable configurational
regions whereas a folding system has dense cycles of constraints that prevent
convexification. Although the energy and force models used by MC and MD differ
implicitly in assembly and folding, these methods miss out on critical
advantages by not explicitly exploiting special geometric properties of small
assembly configurational regions. In contrast, the new methodology specifically
exploits assembly constraint graph properties via \emph{Cayley
convexification}\cite{SiGa:2010}. 

\subsection{Decoupling Roadmap generation from Sampling}
\label{sec:intro:Decoupling} 
Larger assemblies are dimensionally intractable without recursively decomposing
into smaller ones \cite{Wu2012}. Recombining requires estimating not only the
equilibrium free energy but also the kinetics for each smaller intermediate
assembly. Understanding kinetics requires a \emph{topological roadmap} of
potential energy basins, i.e., the neighborhood and boundary relationships of
macrostates, especially when they feed into multiple basins. 

Yet, most prevailing methods either do not extract a comprehensive topological
roadmap even for small assemblies, or do so \emph{in conjunction with extensive
sampling} for configurational entropy or free energy computation. Decoupling
roadmap generation, i.e., topology extraction, from sampling is one of the key
achievements of the new methodology.

Ergodicity of methods such as MC and MD is unproven for configurational regions
of high geometric or topological complexity with low energy regions or
macrostates of low relative volume (low effective dimension) separated by high
energy barriers. Hence they require unpredictably long trajectories starting
from many different initial configurations or microstates for locating such
separated regions or macrostates. Starting from MC and MD trajectories, recent
heuristic methods infer a topological roadmap
\cite{Gfeller_DeLachapelle_DeLos_Rios_Caldarelli_Rao_2007,
Varadhan_Kim_Krishnan_Manocha_2006, Lai_Su_Chen_Wang_2009,
10.1371/journal.pcbi.1000415} and use topology to guide dimensionality
reduction \cite{Yao_Sun_Huang_Bowman_Singh_Lesnick_Guibas_Pande_Carlsson_2009}.
Methods for handling broken ergodicity include basin-hopping \cite{WALES20131}
(an optimization technique to find global minima), and the use of parallelism
\cite{earl2005parallel} or some combination of the two
\cite{Griffiths2019,Wales:Landscapes}. All of these methods are extremely
resource intensive and rely on heavy sampling.

A version of assembly arises in the robotics motion planning literature with
exponential time algorithms to compute a roadmap (a version of atlas) and paths
in general semi-algebraic sets \cite{bib:canny-roadmap, canny-alg,basu}, with
probabilistic versions to improve efficiency \cite{kavraki1, kavraki2}. For the
Cartesian configurational regions of non intersecting spheres, the works
\cite{Baryshnikov08022013, Kahle2011} characterize the complete homology,
viable only for relatively small point-sets or spheres, while more empirical
computational approaches for larger sets \cite{PhysRevE,
Bubenik10statisticaltopology} come without formal algorithmic guarantees. A
geometric rigidity approach was primarily used to characterize the graph of
contacts of arbitrarily large jammed sphere configurations in a bounded region
\cite{Kahle2012, Connelly:Jamming}.

Unlike the above approaches, the new methodology gives a comprehensive
\emph{atlas} of the assembly macrostates as fast as possible, by using
macrostate-specific, so-called, Cayley parametrization to decouple sampling
from exploration of landscape topology. In addition, unlike many former
approaches, our methodology is deterministic, its efficiency following from
exploiting special properties of those semi-algebraic sets that arise \rahul{in }
assembly.

While the work\cite{Jaillet2017} also uses the word `atlas' and refers to a
configurational region stitched together from custom-sampled subregions called
`charts', both the decomposition into subregions and the definition of charts
are entirely unrelated to our methodology. That work uses neither active
constraint regions, nor stratification by energy levels, nor Cayley parameters
for convexification and does not focus on assembly but systems such as
cyclo-octane, where loop-closure type constraints predominate. The paper
\cite{Holmes-Cerfon2013} uses a partition into active constraint regions that
they call `modes'. They compute minimum potential energy states and kinetics
using certain integrals over low dimensional macrostates of the interacting
particles.

The paper \cite{Holmes-Cerfon2013} formally showed that their (and our
methodology's) geometrization of such systems is physically realistic. However,
they directly search for hard-to-find minimum energy macrostates (effectively
zero dimensional) by walking one-dimensional macrostates. In contrast, our
methodology uses the dimensional stratification in the atlas: it starts from
Cayley convexified higher-dimensional interior (parent region) and recursively
locates easy-to-find boundary macrostates of exactly one less effective
dimension.

\subsection{Designing Landscapes} 
\label{sec:intro:designingAssembly} 
The prevailing methods described above are not particularly suited to reverse
analysis. By reverse analysis we mean: isolating and altering features of the
output landscape and free energy by tweaking input shape variables of the
assembly constituents and interactions. Our methodology links the shape of the
input assembling units to a type of intuitive and query-able bar-code of the
output atlas, such as the number of active constraint regions or macrostates
and their average Cayley parameter values. These landscape design variables
can be used to design stable assembled structures and kinetics. This succinct
input-output relationship facilitates reverse analysis, and control towards
design.

\subsection{Sampling and Computing Entropy Integrals} 
\label{sec:intro:VolumeIntegrals}
While the computation of potential energy of a given configuration is generally
straightforward, free energy computation requires accurate entropy computation,
i.e., relative (weighted) volume of potential energy basins, each of which is a
complicated mosaic of macrostates of varying effective dimension. Similarly,
kinetics requires the computation of path probabilities. As mentioned in
Section \ref{sec:intro:ExploitingAssembly} prevailing methods use exhaustive
sampling for path and volume computations. For $k=3$, there are bounds for
approximate configurational entropy using robotics-based methods without
relying on MC or MD sampling~\cite{GregoryS201199}. 

For small cluster assemblies from spheres, there exist a number of methods to
compute free energy and configurational entropy of lower dimensional or lower
energy macrostates \cite{Holmes-Cerfon2013, Arkus2009, Wales2010,
Beltran-Villegas2011, Calvo2012, Khan2012, Hoy2012, Hoy2014,
Holmes-Cerfon-2018}.

However, working with Cartesian configurations as microstates, they must deal
with macrostates that are comparable in topological complexity to the entire
Cartesian configurational region of small molecules such as cyclo-octane
\cite{Martin2010,Jaillet2017,Porta2007}.

In general, lower energy macrostates are probabilisitcally weighted higher, but
have lower effective dimension which - in principle - should make sampling
faster since time complexity of the volume integral computation (via sampling)
is exponential in the effective dimension of the macrostate. However, such
macrostates - which live within a much higher \emph{ambient dimension} - are
typically topologically complex. Ambient dimension is the maximum dimension of
any macrostate in the assembly configuration space, specifically a region with
no active constraints (discussed in Section \ref{sec:methods:furtherIO}).
Staying within such an effectively lower dimensional macrostate, a nearly
constant potential energy region, confounds most prevailing methods, forcing
them to rely on local linearization and energy gradient descent (to enforce the
energy constraints) leading to many discarded samples. For example the paper
\cite{Holmes-Cerfon2013} needs to solve a non-linear equation iteratively when
tracing a 1 or 2-dimensional constrained manifold by moving a Cartesian
configuration along its tangential direction and projecting it back to the
manifold. 

Stratified sampling \cite{dinner2017stratification}could be considered a
general way to address this problem by using stratified probability
distributions whose support is restricted to a constant potential energy
region. However, membership of a configuration in a region is characterized by
the tangent space (linearization using Eigen vectors) at that particular
configuration. Hence, the configuration needs to be sampled first, which does
not address the original difficulty of staying primarily within the bounds of
the desired region, i.e., computing the bounds of the support of the
probability distribution. In general, while the above-mentioned methods
address the accuracy issue in sampling effectively low dimensional regions,
they do so at the expense of many discarded samples and efficiency.

The new methodology addresses both the accuracy and efficiency issues via a
\emph{region-specific} Cayley convexification ensuring that sampling (for
volume computation) stays within macrostates without having to explicitly
enforce any constraints. For a given Cayley configuration sample, our
methodology computes the Cartesian configurations directly via evaluation,
minimizing discarded and repeated samples. Specifically, instead of sampling
in the Cartesian region which limits most methods, Cayley sampling permits our
methodology to sample macrostates of any dimension. 

On the other hand, higher energy regions or macrostates are effectively higher
dimensional; hence their volume computation by sampling is tractable only when
the number of rigid assembly constituents $k$ is small since the dimension of
the ambient space is exponential in $k$. For such higher dimensional regions or
macrostates, the new methodology does not have to rely entirely on sampling,
but utilizes the convexity arising from Cayley parametrization.

Finally, using fast, flexible sampling options \cite{Ozkan2014Jacobian}, our
methodology helps to rapidly find paths and compute entropy integrals through
multiple regions or macrostates of various energy levels or effective
dimensions, enabling kinetics computation.

\subsection{Delineating Contributions of Directly Related Work}
\label{sec:app:companionPapers}
\begin{figure*}
\includegraphics[width=0.9\textwidth, height=8cm]{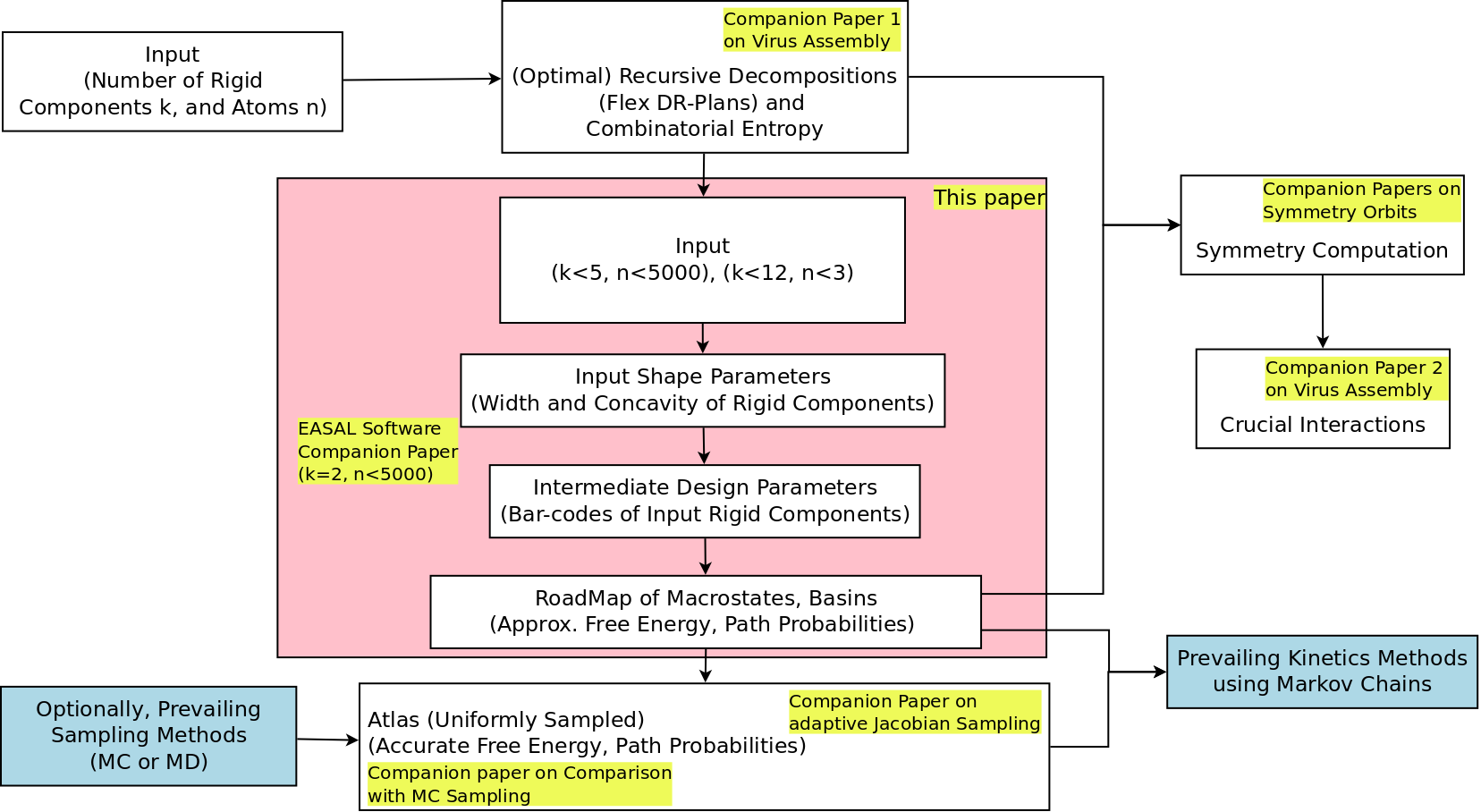}
\caption{\scriptsize Delineating contributions of directly related work.}
\label{fig:companionPaper}
\end{figure*}

Figure \ref{fig:companionPaper} shows the original contributions of this paper
and its connections with prevailing methods as well as companion papers whose
contents are sketched below. (1) A conference proceedings paper that provides
an early sketch of some aspects of the new methodology prior to the EASAL
software implementation. This paper does not contain results. (2) A
multi-perspective comparison of algorithmic variants of the new methodology
against metropolis Monte Carlo (MC) sampling for the assembly landscape of 2
transmembrane helices, in order to assess complementary strengths
\cite{Ozkan2014MC}. (3) Using an adaptive Jacobian strategy along with convex
Cayley parametrization for high-performance sampling towards entropy integral
computation \cite{Ozkan2014Jacobian}. (4) Prediction of assembly-crucial
inter-monomer-interface interactions with experimental mutagenesis validation
for 3 viral capsids ~\cite{Wu2012, virus2019} (Minute Virus of Mice (MVM),
Adeno-Associated Virus (AAV), and Bromo-Mosaic Virus (BMV)) - utilizing the
recursive decomposition (reverse assembly pathway) of the large viral capsid
assembly into smaller assembly intermediates \cite{sitharam:Assembly, mvs2006},
(5) Exploitation of symmetries \cite{sym8010005}. (6) Curated opensource
software and a user-developer manual that provides a summary of all of the
above, as well as a paper describing the software implementation, clearly
citing and summarizing the above papers as well as this manuscript
\cite{Ozkan:toms}.

\vspace{0.5cm}

\noindent {\bf Note:} For ease of exposition, we use 
``active constraint regions'' and ``configurations'' respectively to refer to
macrostates and microstates. Technically, these have different meanings when
expressed in Cartesian and Cayley parameters. Moreover, active constraint
regions may not be contiguous, while macrostates are. However, as shown in Sections
\ref{sec:convexity} and \ref{sec:realization}, our use of Cayley
parametrization ensures that this ambiguity is not an issue when the context
is clear.

\section{Methods}
\label{sec:methods}
The new methodology uses several key strategies to generate the assembly
configurational regions. Section \ref{sec:geometrization} discusses
geometrization of short-range Lennard-Jones potentials as active geometric
constraints, by which effective dimensions of active constraint regions become
proxies for energy levels. Section \ref{sec:stratification} introduces a
dimensional classification of active constraint regions, called Thom-Whitney
stratification, using active constraint graphs.

In Section \ref{sec:recursiveBoundarySearch} we describe a recursive method
that along with stratification decouples the generation of the roadmap from
sampling.  In Section \ref{sec:convexification} we describe how to find  a
lower dimensional boundary of a given region, using a a distance based
region-specific Cayley parametrization to convexify the region.

In Section \ref{sec:designVariable} we establish a connection between the input
molecular geometry and the output features (a type of bar-code of the output)
via landscape design variables.

Section \ref{sec:coreAlgorithm} sketches the core algorithm for 2 input rigid
molecular components ($k =2$, $n$ arbitrary). This is followed by algorithmic
variants for multimers ($2 < k < 12$, $n$ arbitrary) in Section
\ref{sec:AlgorithmVariantkg2} and for cluster assembly from spherical particles
($k \le 24$, $n=1$), with improvements when the assembling components are
identical in Section \ref{sec:AlgorithmVariantn1}; and we briefly describe
under what conditions and how the approach can be extended to the general case
($k$ arbitrary, $n$ arbitrary) in Section \ref{sec:AlgorithmVariantnk}.

In Section \ref{sec:methods:immediateResults} we discuss algorithmic extensions
to compute basins, barriers, paths, and volumes.\\

\noindent{\bf Note:} most of the illustrative figures in this section are
generated by the EASAL implementation of the new methodology
\cite{easalSoftware,Ozkan:toms,easalVideo}.

\subsection{Geometrization and Input Setup}
\label{sec:geometrization}
\begin{figure*}[htpb]
\centering
\includegraphics[scale = 0.4]{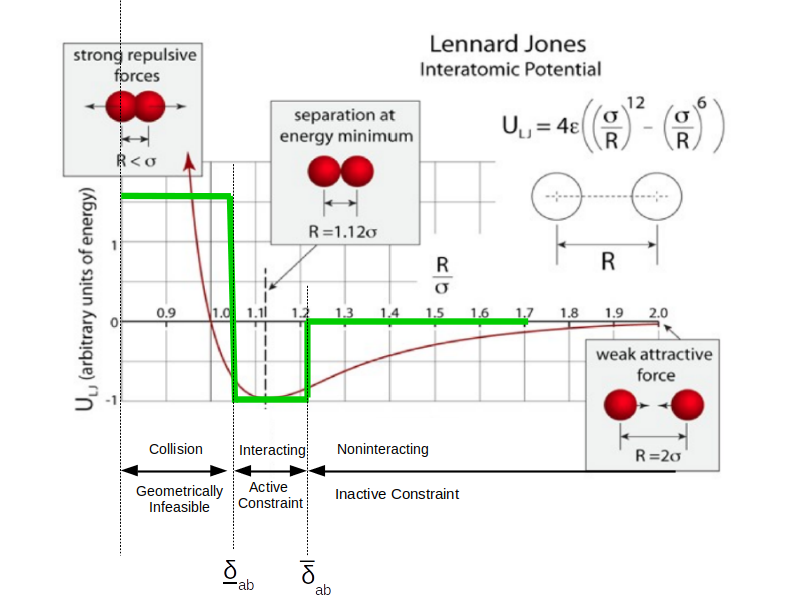}
\caption{\scriptsize \textbf{Geometrization}: The short-range Lennard-Jones potential
function is discretized into 3 main regions. (1) Large pairwise inter-atomic 
distances at which no force is exerted between the atoms. Such atom pairs are 
called inactive constraints. (2) Small inter-atomic pairwise distances which 
are prohibited by the interatomic repulsion, which is called a collision.
(3) The interval between these, known as the well, in which stable interactions
or active constraints are formed.
$\varepsilon$ is the depth of the Lennard-Jones potential well and $\sigma$ is
the distance at which the interatomic force is zero.  
See Section \ref{sec:geometrization}.}
\label{fig:geometrization}
\end{figure*}

With $\dst_\ijx = ||a - b||$, the distance between the centers of atoms $a$ and
$b$, we \emph{geometrize} the inter-atomic short-range Lennard-Jones potential energy
terms into 3 main regions (see \figref{fig:geometrization}):
\begin{enumerate}
\item [i] Large distances at which nearly zero force is exerted between the atoms, such
atom pairs, called \emph{inactive constraints}, correspond to atom pairs
$(a,b)$ such that $\dst_\ijx > \dhi_\ijx $, $\dhi_\ijx \in \mathbb{R}_+$.

\item [ii] Very close distances that are prohibited by inter-atomic repulsion
or inter-atomic collisions, corresponding to pairs $(a,b)$ such that $\dst_\ijx
< \dlo_\ijx$, $\dlo_\ijx \in \mathbb{R}_+$.

\item [iii] The interval between these, known as the short-range Lennard-Jones well, in
which bonds are formed. Atom pairs in the well are called \emph{active
constraints} and corresponding to the preferred distance $\dlo_\ijx \le \dst_\ijx
\le \dhi_\ijx$.

\end{enumerate}
\begin{equation}
\label{eqn:LJ}
   L(\dst_\ijx) := \begin{cases}
      \infty,&  0 \leq \dst_\ijx < \dlo_{\ijx}\\
      \lj_\ijx(\dst_\ijx)  & \dlo_\ijx \leq \dst_\ijx \leq \dhi_\ijx \\
      \ljhi_\ijx \approx 0, & \text{otherwise}
      \end{cases}, \; \lj_\ijx < \ljhi_\ijx
\end{equation}

The {\em steric} or {\em collision} distance $\dlo_{\ijx}$ keeps atom-centers
apart; $\dhi_{\ijx}$ is the distance beyond which the Van der Waals forces are
no longer relevant. The interval $[\dlo_{\ijx}.. \dhi_{\ijx}]$ is called the
\emph{well}. When $\dlo_{\ijx} = \dhi_{\ijx}$, $L$ is a \emph{Hard-Sphere}
potential. The function $L$ is discretized (see Equation \ref{eqn:LJ}) so that
$\lj_\ijx(\dst_\ijx) \equiv \ljlo_\ijx$ is constant. 

{\bf Note:} For longer range potentials, the LJ \emph{well} can be discretized
into several intervals according to the energy level. Assuming that the well
has $m$ levels, the active constraint region with one constraint will have $m$
sub-regions. Each sub-region will have constant energy. A region with $i$
active constraints will have $m\times i$ such sub-regions. For ease of
exposition, we use $m=1$ in the rest of the paper.\\

The assembly problem can now be viewed as exploring the feasible relative
positions of a collection of $k$ point sets, each of size $n$, in $R^3$
that are mutually constrained by distance intervals. Here, each point set
corresponds to a \rmc\ such as helices, each point in the
point set corresponds to the center of an atom and the distance
constraints correspond to the discretized short-range Lennard-Jones potentials.

\subsubsection{Geometrization Example \toythreed}
\label{sec:geometrizationExample}
\begin{figure}[htpb]
\centering
\subfigure[]{\label{fig:sliceC} \includegraphics[width=0.45\columnwidth]{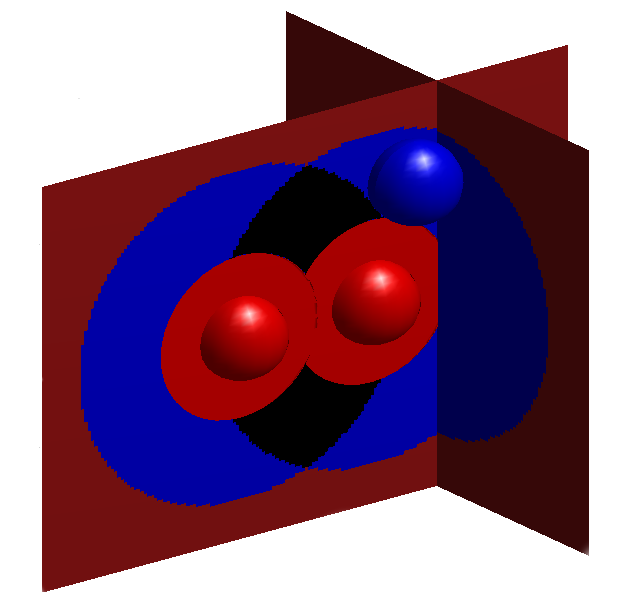}}
\subfigure[]{\label{fig:sliceEnergy} \includegraphics[width=0.45\columnwidth]{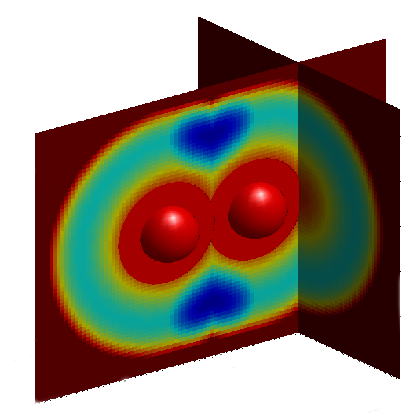}}
\caption{\scriptsize \textbf{Geometrization example \toythreed}: 
(a) shows slices through the energy landscape of the discretized ($m=1$) short-range Lennard-Jones
(LJ) potential for an assembly system consisting of two \rmc s, $A$ (red with 2
atoms) and $B$ (blue with 1 atom).  
(b) shows slices through the energy landscape of a more finely 
discretized (large $m$) short-range Lennard-Jones
(LJ) potential for the same assembly system ($B$ not shown).
See text in Section \ref{sec:geometrizationExample}.
}
\label{fig:geometrizationExample}
\end{figure}

To concretely illustrate the notion of geometrization, consider an input system
consisting of two rigid molecular components: $A$ consists of two atoms $a_1$ and $a_2$
with centers at a fixed distance $\dst_{12} := \|\bp_1-\bp_2\| = 2$; $B$ is a
single atom $b$ with center $q$ (see \figref{fig:geometrizationExample}). 
The atom steric radii are 1/2, i.e, the
center of $b$ must maintain a distance of at least 1 from the centers of
$a_1$ and $a_2$. The potential energy $E$ of this system is the sum of the
short-range Lennard Jones potentials: $E = L(\dst_{a_1b}) + L(\dst_{a_2b})$. 

Since the energy is invariant under Euclidean rigid transformations,
we may fix $\bp1 := (0, 0, 0)$ and $\bp2 := (-2, 0, 0)$. \figref{fig:sliceC}
illustrates the discrete ($m=1$) geometrization of the energy
landscape. The lower short-range Lennard-Jones bound prevents $q$ from falling into the red region
surrounding the red atoms $a_1$ and $a_2$. Similarly $b$ does not interact
with $a_1$ and $a_2$ when its center is in the outer burgundy region
corresponding to inactive configurations. The lowest energy is obtained when
$q$ lies in the well of both $a_1$ and $a_2$ and two constraints are active.
The lowest state corresponds to a
black torus whose two transversal 4-sided sections are shown on the
front-facing slice. The torus is formed as the intersection of the effectively
2-dimensional blue spherical shells of the two energy wells. 
Figure \ref{fig:sliceEnergy} uses a finer discrete geometrization with large $m$.

\subsubsection{Assembly Problem, Input and Output}
\label{sec:methods:furtherIO}
\noindent With this setup, the core method takes as input an assembly system which
consists of up to four items.
\begin{enumerate}
\item A collection of $k$ \rmc s with $n$ atoms each. In the
context of molecular assembly, each $A_j$, $1 \le j \le k$, is specified as the
positions of at most $n$ atom-centers.

\item Distance interval constraints on point pairs $(a, b)$ given by $\dlo_\ijx, \dhi_\ijx
\in \mathbb{R}_+$.

\item Optional non-pairwise constraints imposed globally on the configuration. For
instance, implicit solvent (water or lipid bilayer membrane) effects
\cite{Lazaridis_Karplus_1999, Lazaridis_2003, Im_Feig_Brooks_2003}, are
specified as a constraint on the assembly configuration.

\item Optional: set of active constraint regions of interest, specified 
by their active constraint graphs.

\end{enumerate}

Consider two \rmc s $\psone$ and $\pstwo$, of size $n$, in $R^3$, that are
mutually constrained by distance intervals. For atoms $a\in A$ and $b\in B$, define
$\dst_\ijx$ as $\norm{a-T(b)}$, where 
$T \in SE(3)$ is a Euclidean orientation preserving isometry, and $SE(3)$ is
the special Euclidean group in $\mathbb{R}^3$.
$T$ is feasible if the following hold:

\begin{align}
\forall (\pone \in \psone, \ptwo \in \pstwo),\qquad& \dst_\ijx \ge \dlo_\ijx& \dlo_\ijx \in \mathbb{R}_+ \tag{\distall}\label{eqn:constraints}\\
\exists (\pone \in \psone, \ptwo \in \pstwo),\qquad& \dst_\ijx  \le \dhi_\ijx, & \dhi_\ijx \in \mathbb{R}_+.\tag{\distexist}\label{eqn:preferredConstraints}
\end{align}

\noindent Constraint \cone\ implies that $T$ is infeasible when there exists a
pair $(\pone, T(\ptwo))$ that is too close. Constraint \ctwo\ implies that at
least one pair $(\pone, T(\ptwo))$ is within a \emph{preferred} distance
interval.  When $T$ is feasible, the \emph{Cartesian configuration} $T(\pstwo)$
is called a \emph{realization} of the constraint system (\cone, \ctwo).
Assembly problem (\cone, \ctwo) asks for a description of the \emph{realization
space} or the Cartesian configurational region of feasible $T$'s.
This is the \emph{assembly landscape} $\mathcal{C}$.

When $k=2$, the ambient dimension of Assembly Problem (\cone, \ctwo) $d_\mathcal{A}$
is 6, namely, the dimension of $SE(3)$. In general $d_\mathcal{A}$ is $6(k-1)$, where $k$ is
the number of \rmc s. When $\dhi_\ijx -\dlo_\ijx \approx 0$ the effective
dimension $d_\mathcal{C}$ of the assembly landscape is $d_\mathcal{A} - (k-1) = 5(k-1)$.\\

\noindent\textbf{Note:} Generally, when we refer to the assembly problem, we mean 
assembly problems with \emph{small assembly systems} as input ($n$ large and $k$
small, or $n$ small and $k$ large). When the input system has large $n$ and 
large $k$, we get a folding system. Thus, in this paper we look at assembly and 
folding as all encompassed within the various ranges of $n$ and $k$. 

We make this distinction to highlight the fact that when both $n$ and $k$ are 
large, the active constraint graphs could be complicated enough to not permit
Cayley convexification. The core algorithm in Section
\ref{sec:coreAlgorithm} and its variants in Section \ref{sec:AlgorithmVariantkg2} 
and Section \ref{sec:AlgorithmVariantn1}, directly atlas small assembly systems, 
leveraging Cayley convexification. The algorithm variant in Section 
\ref{sec:AlgorithmVariantnk}  decomposes the larger system in a folding 
problem using the so-called DR-plans to take advantage of Cayley convexification.

The main output of the methodology is the topological, dimensional, and
geometric structure of the assembly landscape $\mathcal{C}$ i.e., the set of
all feasible $T \in SE(3)$ satisfying (\cone, \ctwo). This is specified
as a roadmap which includes a stratification of active constraint regions by
dimension or energy levels, and their neighborhood relationships.
Each active constraint region is labeled by its active constraint graph and
comes with its Cayley and Cartesian configurations.
The set of Cartesian configurations is visualized as the \emph{sweep} (see
\figref{fig:pctreeSweep}). The sweep shows one of the rigid molecular components 
$\psone$ together with all feasible Cartesian configurations $T(\pstwo)$ of $\pstwo$ 
traced out.\\

\noindent{\bf Note:}
although Assembly Problem (\cone, \ctwo) is defined for $k=2$ \rmc s, Sections
\ref{sec:AlgorithmVariantkg2}, \ref{sec:AlgorithmVariantn1}, and
\ref{sec:AlgorithmVariantnk} show how it generalizes to arbitrary $k$.
Furthermore, the exposition of this paper assumes the case where radii
$\rho_a$, $\rho_b$ are assigned to atoms $a$ and $b$ respectively, and the
constants $\dhi_\ijx$ and $\dlo_\ijx$ in (\cone, \ctwo) (generalized to arbitrary $k$) 
are functions of $\rho_a + \rho_b$. 

\subsection{Stratified Roadmap}
\label{sec:stratification}
We partition the assembly landscape into active constraint regions
(nearly-equipotential-energy regions or macrostates).
\begin{definition}
\label{def:ACG}
For a configuration $T$ in the assembly landscape $\mathcal{C}$, i.e., a feasible
configuration for Assembly Problem (\cone, \ctwo), the vertices of an \emph{active 
		constraint graph} are the atoms involved in active constraints (atom pairs
$(a,b)$ satisfying \ctwo), but additionally include at least 3 atoms from each of 
$A$ and $B$. The two types of edges are (1) pairs $(a,b)$ that are active constraints 
for $T$ and (2) all pairs $(a1, a2)$ and $(b1, b2)$ where $a_i \in A$ and $b_i \in B$ 
(see \figref{fig:pctreeACG}. An 
\emph{active constraint region} $R_G$ is the set of all configurations $T$ with the 
same active constraint graph $G$.
\end{definition}

The active constraint regions are organized as a partial order (directed
acyclic graph or DAG) that captures their boundary relationships (see
\figref{fig:exampleStratification} and \figref{fig:pctreeACG}). A node of the
DAG is an active constraint region, and a directed edge captures a boundary
relationship. In particular, the active constraint graph of a region is a
subgraph of the active constraint graph of its children boundary regions. 

More active constraints correspond to lower potential energy and the lowest
potential energy is attained at the bottoms of the potential energy basins
(discussed in detail, later in this section). The DAG is thus a refinable
topological \emph{roadmap} with unique ``street signs'' or region labels, which
are the active constraint graphs. 

In addition to indicating a boundary relationship, the DAG edges between two
nodes in the roadmap indicate a dimensional relationship (see
\figref{fig:exampleStratification}): a lower dimensional child region is the
boundary of a parent region one dimension higher (one fewer active constraint).
Next we describe how the roadmap is organized into dimensional strata.

\subsubsection{Generic Rigidity-based Stratification of Active Constraint Regions}
\label{sec:rigidity}
Active constraint graphs are analyzed using combinatorial rigidity theory
(\rahul{discussed in Section \ref{sec:app:rigidity} of the Appendix})
\cite{SJS:Handbook,CombinatorialRigidity}. 

In particular, generically, the sum of the dimension of the active constraint
region and the number of edges in its active constraint graph is the ambient
dimension $d_\mathcal{A}$, which is $6(k-1)$, where $k$ is the number of \rmc s
(see Section \ref{sec:genericity}).  Since the active constraint regions
satisfy distance constraints which are quadratic polynomial equations and
inequalities, the assembly landscape is a \emph{semi-algebraic set} (a union of
sets defined by polynomial inequalities). This permits a so-called
\emph{Thom-Whitney stratification} \cite{Kuo}.

\begin{figure*}[htpb]
\centering
%Fig1
\subfigure[]
{\label{fig:exampleStratification} 
\begin{overpic}[scale=.24]{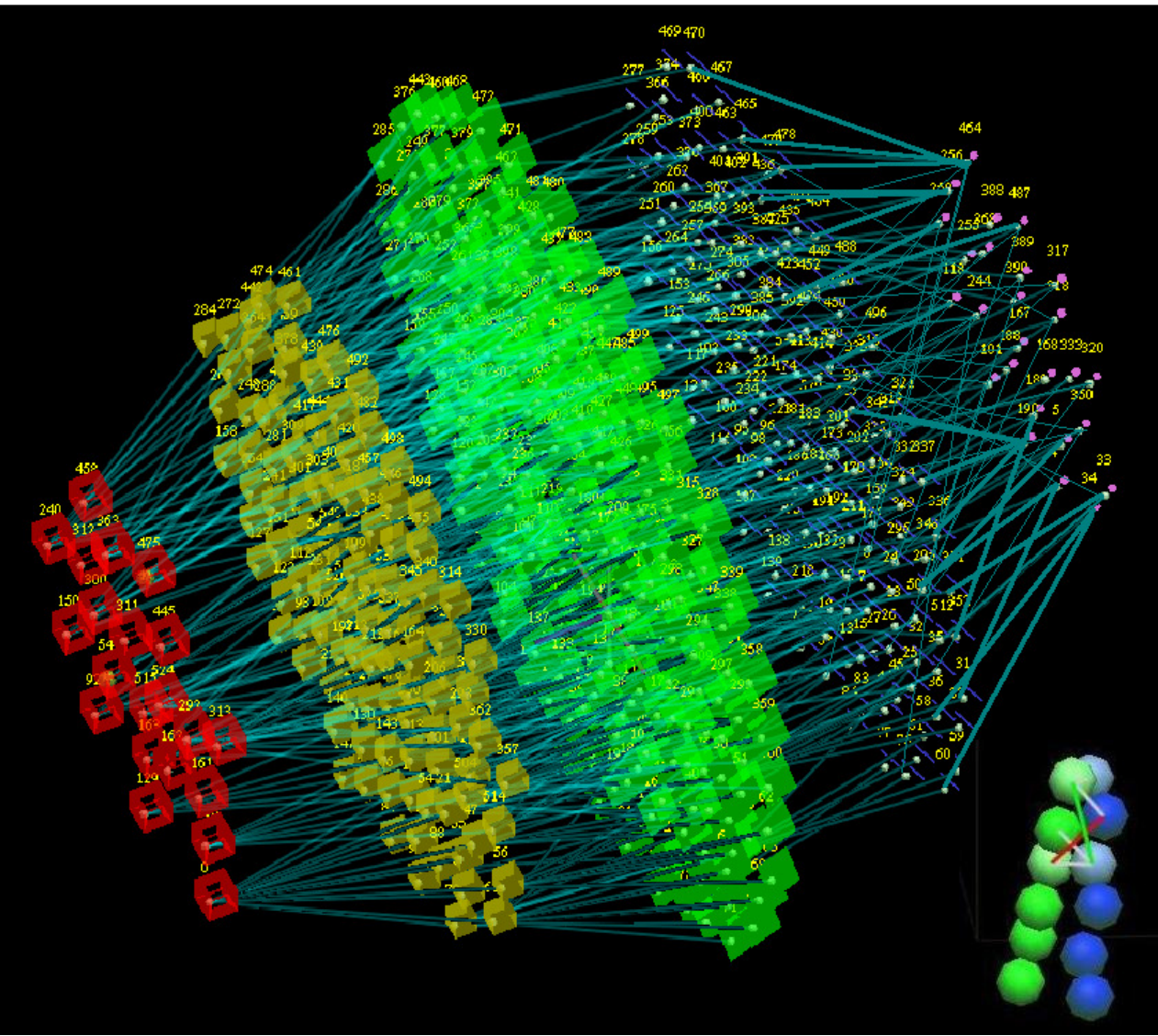}
\put (5,93) {\color{black}{--- Decreasing energy --->}}
\end{overpic}
}
%Fig2
\subfigure[]{
\begin{overpic}[scale=.225,tics=10]{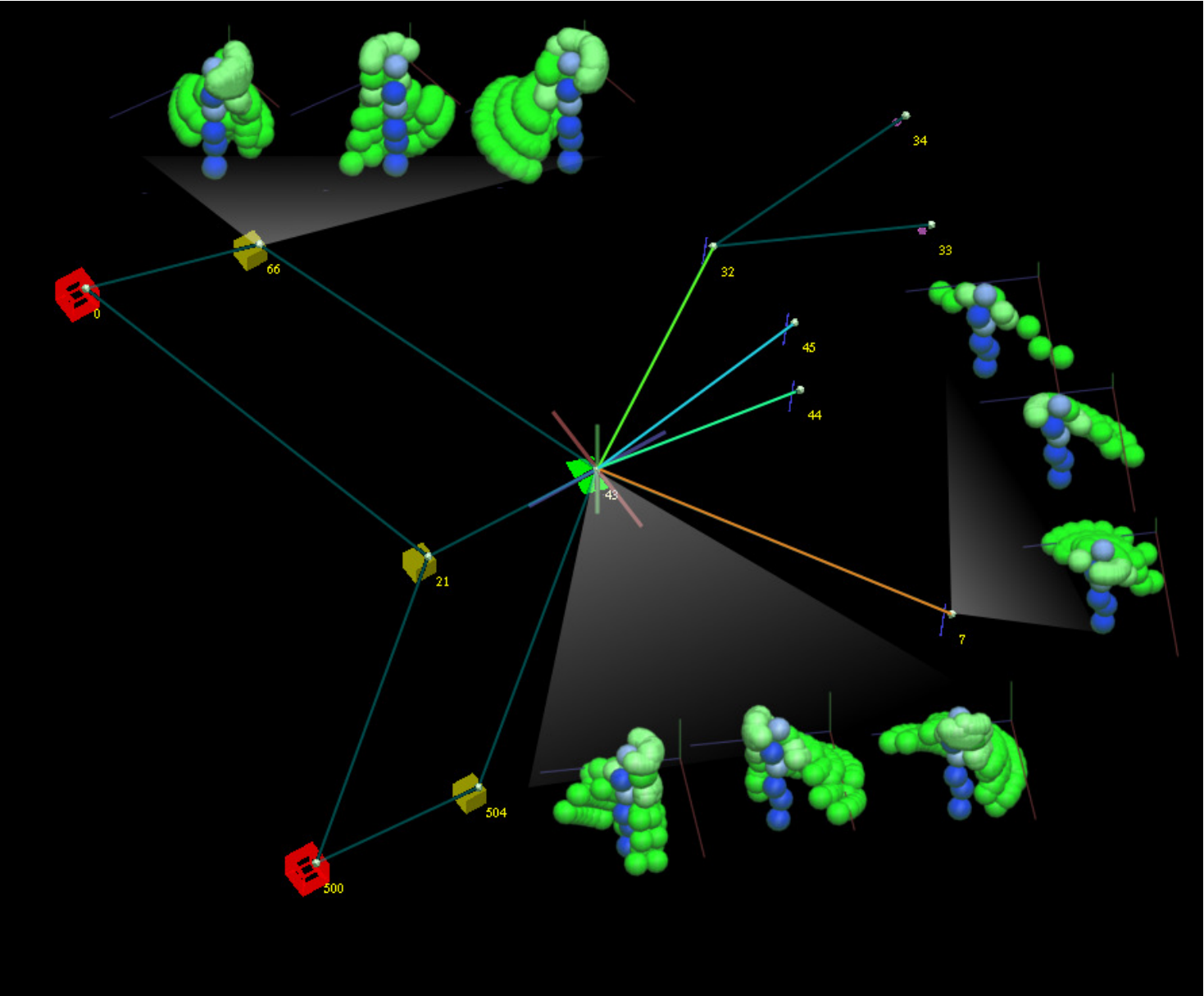}
     \put (15,85) {\color{black}{Sweep of different flips}}
     \put (72,25) {\color{grayoned}{1D node}}
     \put (40,30) {\color{green2d}{2D node}}
     \put (23,20) {\color{yelloworange}{3D node}}
     \put (3,2) {\color{red}{4D node}}
\end{overpic}
\label{fig:pctreeSweep}
}
%Fig3
\subfigure[]{
\begin{overpic}[scale=.35,tics=10]{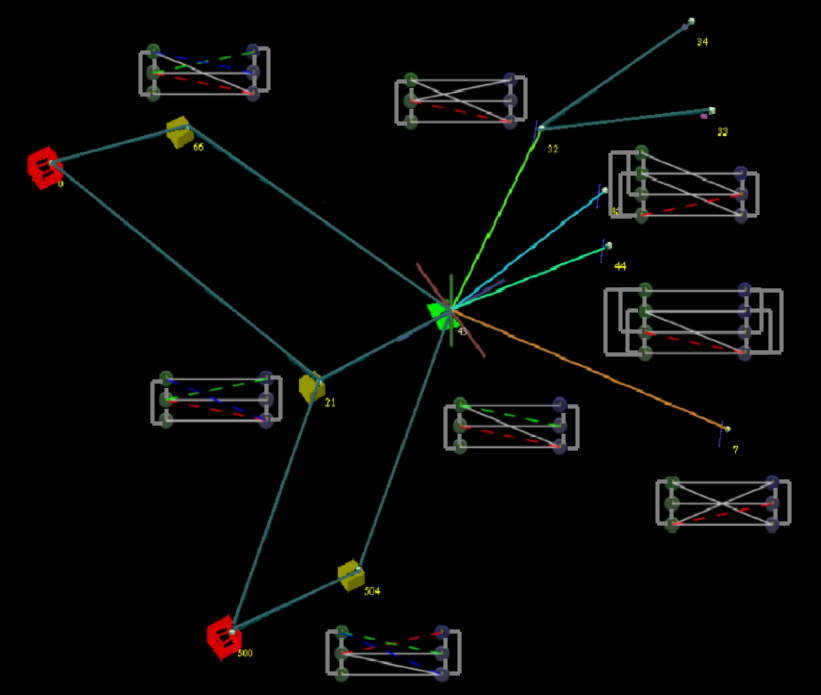}
     \put (69,17) {\color{grayoned}{5 active}}
     \put (65,12) {\color{grayoned}{constraints}}
     \put (75,28) {\color{grayoned}{1D node}}
     \put (50,25) {\color{green2d}{4 active}}
     \put (48,21) {\color{green2d}{constraints}}
     \put (46,37) {\color{green2d}{2D node}}
     \put (60,2) {\color{yelloworange}{3 active constraints}}
     \put (45,15) {\color{yelloworange}{3D node}}
     \put (12,10) {\color{red}{4D node}}
\end{overpic}
\label{fig:pctreeACG}
}
\caption{\scriptsize \textbf{Stratification of the Roadmap}:
(a) shows a portion of the roadmap (up to 4D regions) for the inset pair of input \rmc s. 
The nodes represent active constraint regions and are colored by their dimension. 
The red nodes represent 4D active constraint regions (each with 2 active 
constraints). Each successive stratum (from left to right) contains boundary regions 
one dimension lower (with one additional active constraint), until we reach 
0D regions (pink nodes), each of which contain finitely many rigid 
configurations.
As the dimension of the active constraint region decreases
(as we add more active constraints) its potential energy decreases (see 
Section \ref{sec:genericity}).
(b) Active constraint regions in the roadmap shown with their Cartesian 
configuration sweeps. Each sweep is the union of Cartesian configurations 
in the corresponding active constraint region. Each sweep within a node shows a different 
flip (defined in Section \ref{sec:realization}).
(c) Ancestors and descendants of a 2D active constraint region, shown 
with their active constraint graphs.
}
\label{fig:NestedRegions}
\end{figure*}

A Thom-Whitney stratification of an assembly landscape 
is a partition into active constraint regions $R_G$
that are grouped into strata $X_d$ by their dimension $d$ (see
\figref{fig:exampleStratification}). $X_d = \bigcup\limits_G R_G$ where the
number of active constraint edges in $G$ is $d_\mathcal{A} - d$.  For all $d$,
$R_{G_d} \subset X_d$ is \emph{effectively $d$-dimensional}, i.e., it has
dimension $d$ when all its $d_\mathcal{A}-d$ active constraints are treated as
Hard-Sphere wells of zero width.

\subsubsection{Genericity Assumption}
\label{sec:genericity}
In the above discussion, we assumed that 
generically the effective dimension of an active constraint region plus the
number of active constraints is the ambient dimension $d_\mathcal{A}$, i.e., the number of active
constraints is generically the co-dimension of the region. This is
justified because, in Assembly Problem (\cone, \ctwo), generically,
the active constraint edges are not implied by
(dependent on) the rest of the active constraint graph. 
In other words, implied distances are not active constraints, and
inactive constraints (implied or not) do not restrict the dimension of active constraint regions.
For the special case of Assembly Problem (\cone, \ctwo), in which sets $A$ and $B$ are
centers of non intersecting spheres of generic distinct radii, these
assumptions are an unproven conjecture, for which counterexamples haven't been
encountered. For $\mathbb{R}^2$ this conjecture has been proven \cite{connelly:2019:StickyDisks}.
When the radii are all the same, several simple counterexamples (e.g.,
crystalline structures with high coordination number) exist
where active constraints are implied by other active constraints. 

Non-generically, there could be active constraints regions of effective
dimension $d$ whose number of active constraints exceeds $d_\mathcal{A} - d $,
i.e., the active constraint system is over constrained, or some of whose active
constraints are dependent. For entropy calculations, such regions should be
explicitly tracked since although dependent constraints do not alter the
effective dimension, they can diminish the set of configurations. For
simplicity, our exposition ignores such over-constrained regions in the
stratification and assumes that all regions of effective dimension $d$ are
obtained by choosing exactly $d_\mathcal{A} - d $ active constraints, i.e.,
they are assumed to be independent.

\subsubsection{Stratification Example \toytwodone}
\label{sec:toytwodone}
\begin{figure}[htpb]
\centering
\includegraphics[scale=0.35]{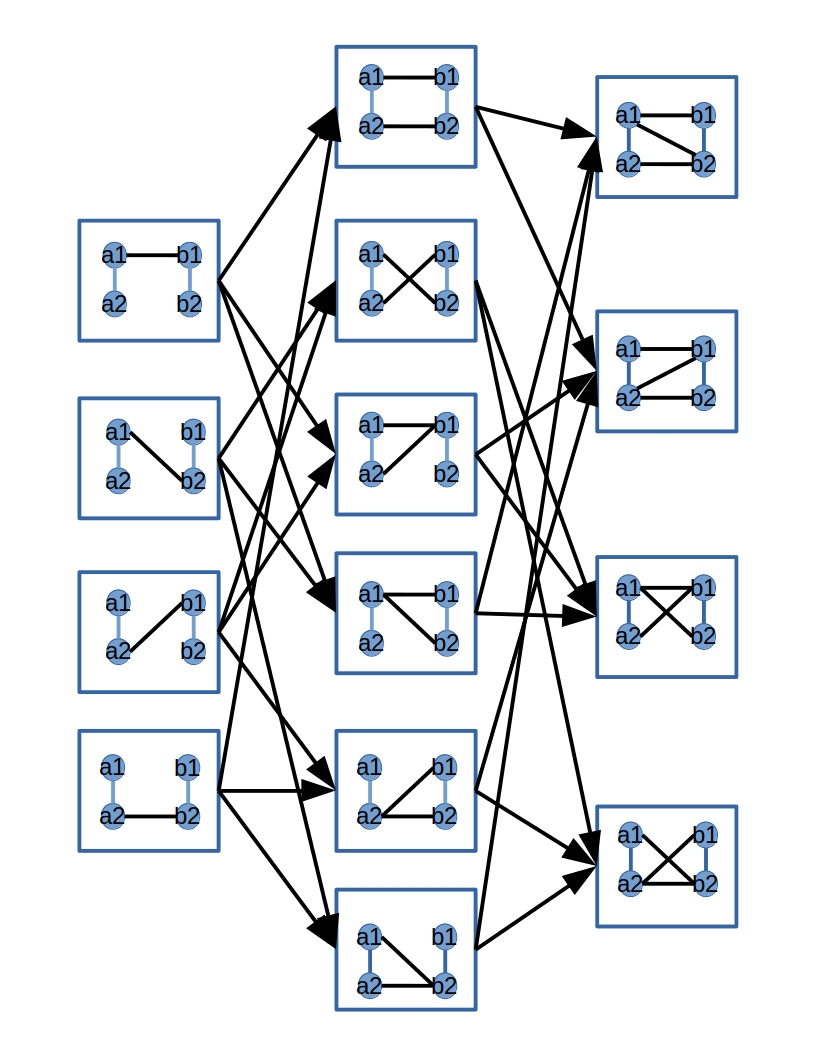}
\caption{\scriptsize \textbf{Stratification Example}: Roadmap of the (toy-sized) configurational
region of Example \toytwodone\ of Section \ref{sec:toytwodone}. 
The nodes of the DAG represent active constraint regions and DAG edges 
connect a region to a boundary region, one dimension lower. Each node 
box displays the active constraint graph of its corresponding region. 
The stratum $X_2$ contains regions in the left most column, which are
2D active constraint regions, containing configurations with 2 degrees
of freedom. The stratum $X_1$ contains the regions in the middle column, 
which are 1D active constraint regions, containing configurations with
one degree of freedom. 
The stratum $X_0$ contains the regions in the right column, which
are 0D regions, containing finitely many rigid configurations
(see Section \ref{sec:toytwodone}). }
\label{fig:2DToy}
\end{figure}

To illustrate the concept of stratification further, consider Assembly Problem
(\cone, \ctwo) in $\mathbb{R}^2$ with two \rmc s $\psone$ and $\pstwo$;
$\psone$ contains two atoms - $\pone_1$ and $\pone_2$ - and
$\pstwo$ contains two atoms - $\ptwo_1$ and $\ptwo_2$. The ambient space is
$SE(2)$ of dimension 3. A complete stratification of the Cartesian region is
shown in \figref{fig:2DToy}. The three strata are organized as a DAG, with
nodes representing active constraint regions and labeled by their corresponding
active constraint graphs. In $\mathbb{R}^2$, the minimum number
of atoms required from each \rmc\ in Definition \ref{def:ACG} is 2 instead of
3. As in Definition \ref{def:ACG}, the vertices in the active constraint graph
are atoms participating in the active constraints that define $R$. The edges
are of two types, (i) between atoms in the same \rmc\ and (ii) the active
constraints, between atoms in different \rmc s.

The stratum $X_2$ consists of the 4 regions in the left column, which are 2D
regions, each containing configurations with 2 degrees of freedom.  Adding an
extra active constraint to any of these regions yields 1D regions in the middle
column. The stratum $X_1$ consists of the 6 regions in the middle column, each
containing configurations with one degree of freedom. Adding an extra active
constraint to any of these regions yields 0D regions in the right column, each
containing finitely many rigid configurations.  The stratum $X_0$ consists of
the 4 0D regions in the right column.  A DAG edge represents a boundary
relationship of the parent interior region to a child boundary region of one
lower dimension. This is also illustrated in \figref{fig:exampleStratification}
where moving from left to right, we add more constraints and decrease the
dimension of the configurational regions.

\subsubsection{Advantages of Stratified Roadmap}
\label{sec:AdvantagesStratification}
There are many advantages of partitioning and dimensionally stratifying the
assembly landscape into active constraint regions and labeling them with unique
``street signs'' which are the active constraint graphs. 

(1) The labels avoid repeated sampling of assembly landscape regions
(shown in Theorem \ref{thm:completeness}). 

(2) The labels facilitate querying, sampling, and refining individual active
constraint regions.

(3) The roadmap gives neighborhood and boundary relationships between active
constraint regions, which allows (in conjunction with Cayley convexification
discussed in Section \ref{sec:convexification}) for recursive search for
boundaries one dimension lower. Stratification also facilitates the decoupling of
roadmap finding (including byproducts such as basins, barriers, paths, etc.)
from sampling density (results in Section \ref{sec:results:decoupling}).

(4) From the roadmap several byproducts can be directly computed. These
include locating and mapping the structure of different types of potential
energy basins, and their neighborhood relationships (algorithm in Section
\ref{sec:methods:BasinStructure} and results in Section
\ref{sec:results:potentialEnergy}). In addition, the roadmap allows us to find
paths between active constraint regions (algorithm in Section
\ref{sec:methods:ConfPaths} and results in \ref{sec:results:ConfPaths}).

(5) The roadmap connects the input shape variables and landscape design variables
(discussed below), facilitating designing of assembly landscapes. Specifically,
it allows us to design active constraint regions via active constraint graphs
(results in Section \ref{sec:results:design}).

%%%%%%%%%%%NewSection%%%%%%%%%%%%%%%%
\subsection{Recursive Exploration Decoupled from Sampling}
\label{sec:recursiveBoundarySearch}
This section discusses the exploration from
higher dimension/energy (interior/parent) to lower dimension/energy
(boundary/child) recursively. The exploration is enabled by
(1) neighborhood relationships between active constraint regions as given by the
roadmap, (2) stratification by dimension/energy level, and (3) Cayley convexification
(discussed in Section \ref{sec:convexification}).
\subsubsection{Advantages of Recursive Exploration}
\label{sec:advantagesRS}
Recursive exploration has 3 main advantages.

(1) Searching for boundaries one dimension less at every
stage (boundary detection is explained in detail in Section
\ref{sec:algorithms}), has a higher chance of success than looking
for the lowest dimensional active constraint regions directly, as
illustrated in Section \ref{sec:results:interiorPoint}). 

(2) When a new \emph{child} region of one dimension less is found, all its
higher dimensional \emph{ancestor} regions are immediately discovered since
they correspond to a subset of the active constraints.  So, even if a ``small''
(hard-to-find) region is missed at some stage, if any of its descendants are
found at a later stage, say via a larger (easy-to-find) sibling, the originally
missed region is discovered.

Assume, for example, that regions with active constraint graphs $G$, $G \cup
\{a\}$ and $G \cup \{a\} \cup \{b\}$ are found, by successive boundaries of one
lower dimension, but the boundary corresponding to active constraint region
with graph $G \cup \{b\}$ is not detected during exploration of the region for $G$.
It will however be discovered when the ancestors of the region $G \cup \{a\}
\cup \{b\}$ are added.

(3)  Recursive search allows us to find \emph{hypostatic basins} (algorithm in
\ref{sec:methods:BasinStructure}), that have non-generically high energy and
high configurational volume. These basins have no lower dimensional boundaries,
i.e., they are flat local minima. These are difficult to locate or recognize
both by standard gradient descent methods relying on a high energy gradient or by
exploration starting from a generic, low energy basin and breaking small energy
barriers.

These observations show how our methodology decouples roadmap generation from
the sampling of the configurational region, since finding even low effective
dimensional regions is no longer heavily dependent on the sampling accuracy or
completeness (see results in Section \ref{sec:results:interiorPoint}).
\subsubsection{Recursive Search Example \toythreed}
\label{sec:CartesianDisconnectedness}
\begin{figure}[htpb]
\centering
\includegraphics[scale=0.2]{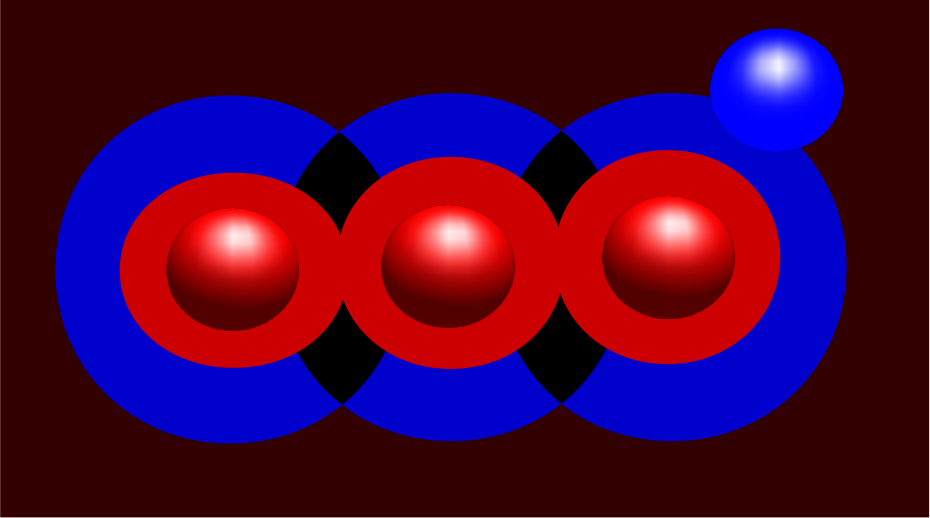}
\caption{\scriptsize \textbf{Advantages of Recursive Search}:
The geometrized energy landscape of an input assembly system with two \rmc s 
$A$ (red with 3 atoms) and $B$ (blue with a single atom). In this example, 
the two black lowest energy regions are disconnected. However, they are both 
boundaries of the middle blue, higher energy region, and hence reachable from 
it by a parametrized search (Section \ref{sec:convexification}), and also 
reachable from either the left or the right blue regions through
ancestor addition (Section \ref{sec:advantagesRS}).
}
\label{fig:slice} 
\end{figure}
Consider an input system similar to the one described in Example \toythreed.
with \rmc\ $A$ augmented by a third atom with center $(-4,0,0)$. Then the
minimum energy regions form two unconnected, parallel tori (black cross section shown in 
\figref{fig:slice}). Since the regions are not connected, MC would need at
least \emph{two} random jumps/initial states to reach both regions. Our
methodology treats both regions as lower dimensional boundaries that are
guaranteed to be encountered in a parametrized exploration of any \emph{one} of
the three starting higher dimensional connected regions.
%%%%%%%%%%%NewSection%%%%%%%%%%%%%%%%
\subsection{Cayley Parametrization for Efficiency}
\label{sec:convexification} 
Locating boundary regions with minimal sampling during recursive search, as
well as refinable sampling for volume computation are both challenging due to
the disconnectedness and complexity of Cartesian active constraint regions (see
\rahul{Section \ref{sec:correspondanceCayleyCart}}). To address this challenge,
we use a theoretical framework developed in the paper\cite{SiGa:2010} (see
\rahul{Section \ref{sec:app:convexity} of the Appendix}). Specifically, we map (many to one) a
$d$-dimensional Cartesian active constraint region $R$, to a convex region of
$\mathbb{R}^d$ called the \emph{Cayley region} of $R$. Exploration of this
convex region is direct and does not require enforcing the active constraints.
In addition, the inverse map, taking an element of the Cayley region, a
\emph{Cayley configuration} to its finitely many corresponding Cartesian
pre-image configurations is efficient to compute. For $k=2$, each Cayley region
corresponds to up to 8 contiguous Cartesian regions or macrostates called
\emph{flips} (defined in Section \ref{sec:realization}) and each Cayley
configuration corresponds to up to 8 easily computable Cartesian configurations
or microstates.

\subsubsection{Convex Cayley Parametrization Theory}
\label{sec:convexity}
\begin{figure}[htpb]
\centering
\includegraphics[scale=0.2] {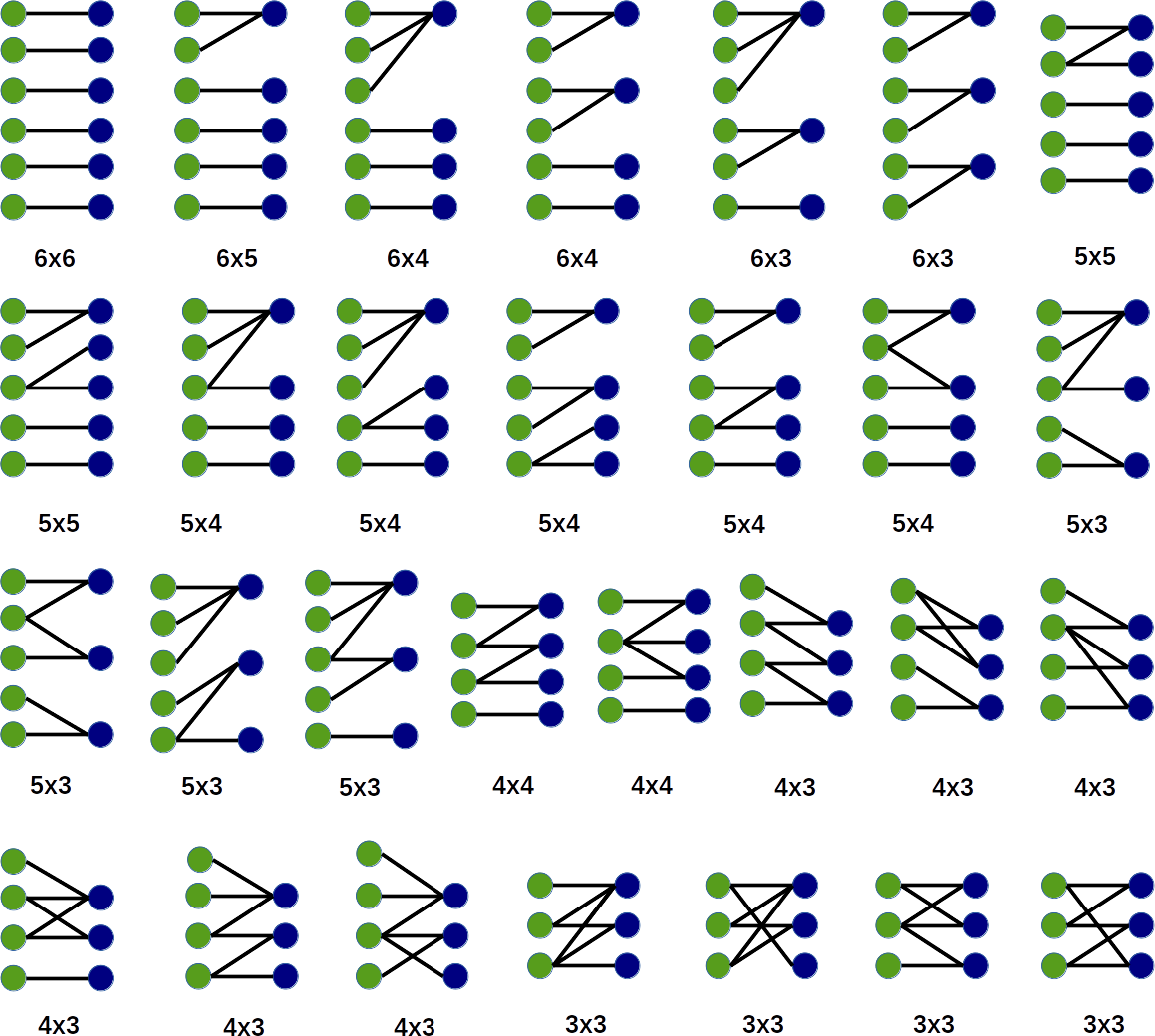}
\caption{\scriptsize \textbf{Exhaustive List of all Active Constraint Graphs 
for generic 0D regions in Assembly Problem (\cone, \ctwo)}: 
In each graph above, 
the vertices of the same color represent atoms
in the same \rmc\ and form a complete graph (whose edges are not shown).
Edges between vertices of different colors, indicate atom pairs 
whose distances are in the short-range Lennard-Jones'
well, i.e., the constraint is active. For $k=2$, all active constraint
graphs are isomorphic to subgraphs of the ones shown. The graphs above are
generically, minimally rigid and correspond to 0D active constraint
regions. The label $m_1 \times m_2$ below each active constraint graph
indicates that $m_1$ atoms in the first \rmc\ and $m_2$ atoms in the
second \rmc\ participate in active constraints (see Section \ref{sec:convexity}).}
\label{fig:3-trees}
\end{figure}

Define a \emph{non-edge} of a graph $G$ as a vertex pair not connected by an
edge in $G$. One way to represent a higher dimensional active constraint
region, i.e., with flexible configurations, is to use the realizable lengths of
non-edges of its active constraint graph $G$. By choosing non-edges whose
addition make $G$ minimally rigid (\rahul{see Section \ref{sec:app:rigidity} of
the Appendix}), a tuple of realizable non-edge length values, a \emph{Cayley
configuration}, corresponds to finitely many Cartesian configurations.  The
non-edge length coordinates of a Cayley configuration are its \emph{Cayley
parameters} (see \rahul{Section \ref{sec:app:convexity} of the Appendix}). Finding
Cayley parameters for an active constraint graph $G$ (for $k=2$) reduces to
picking a (isomorphic)  graph from \figref{fig:3-trees}, with minimal number of
vertices, (a minimally rigid graph, \rahul{see Section \ref{sec:app:rigidity} of 
the Appendix}) for which $G$ is a subgraph. Thus, each active constraint region
has its own region-specific set of Cayley parameters.

Consider the map from the Cartesian active constraint region of $G$ to the
corresponding tuples of Cayley parameters (for those readers familiar with
algebraic topology, we note that this map is a branched covering map
\cite{wiki:coveringSpace, wiki:branchedCovering} that preserves dimension).
The image of this map is the Cayley region of the active constraint graph $G$.
Sampling a Cayley region reduces to incrementing the Cayley parameters while
staying within the region: the active constraints need not be explicitly
enforced.

\begin{definition}[Chart]
A \emph{chart} for an active constraint region $R$ is the map taking a
Cartesian configuration in $R$ to its corresponding Cayley configuration.
\end{definition}

Certain classes of graphs, such as \emph{partial 3-trees} (\rahul{defined in
Section \ref{sec:app:rigidity} of the Appendix}), yield convex Cayley regions with
easily computable bounds \cite{SiGa:2010} (\rahul{see Theorem \ref{thm:SiGa} in Section
\ref{sec:app:convexity} of the Appendix}).  Sampling in such active constraint regions
with partial 3-tree graphs reduces to incrementing the Cayley parameters while
ensuring that we stay within these easily computable bounds, and then computing
the corresponding Cartesian configurations. Again, no explicit enforcing of
active constraints is necessary. Section \ref{sec:coreAlgorithm} describes how
to sample regions and find Cartesian configurations in regions whose active
constraint graphs are not partial-3-trees.

For small $k$, almost all active constraint graphs arising from Assembly
Problem (\cone, \ctwo) are partial 3-trees and thus their regions have convex
Cayley parametrizations. For $k=2$, all the active constraint graphs with 1, 2
and 3 active constraints (5D, 4D and 3D atlas regions) are partial 3-trees.
These are subgraphs of the graphs shown in \figref{fig:3-trees}. $86\%$ of
active constraint graphs with 4 active constraints (2D atlas regions) and
$70\%$ of active constraint graphs with 5 active constraints (1D atlas regions)
are partial 3-trees. Since, regions with 6 active constraints (0D atlas
regions) have generically only finitely many (at most 8) Cartesian
configurations in them, Cayley parametrization is irrelevant.  \rahul{See Section 
\ref{sec:app:convexity} of the Appendix for a more detailed discussion, including an
example, on Cayley convexification}.

\begin{figure}[htpb]
\centering
\begin{overpic}[scale=.3,tics=10]{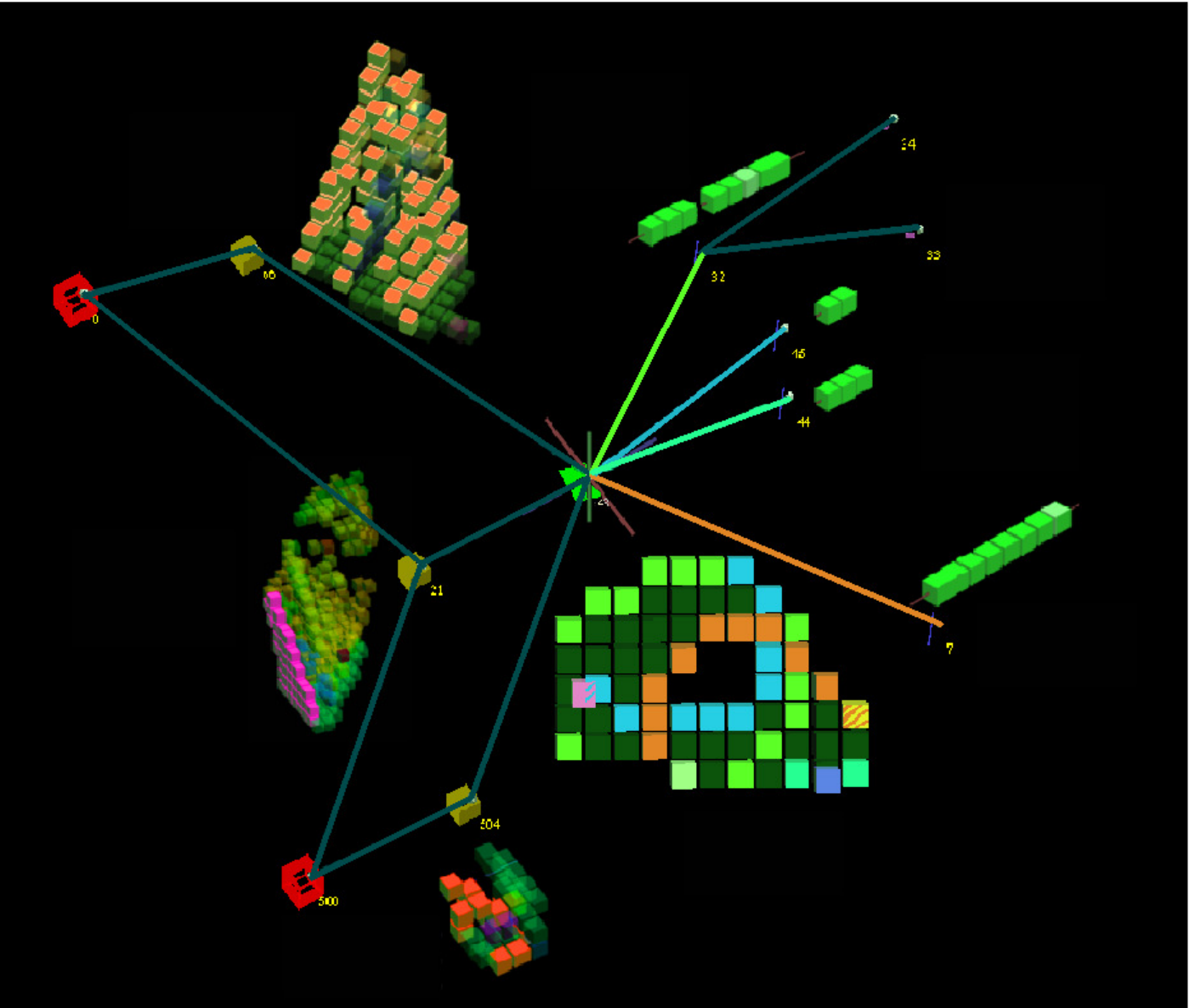}
     \put (50,75) {\color{grayoned}{1D Cayley}}
     \put (75,35) {\color{grayoned}{1D Cayley}}
     \put (75,28) {\color{grayoned}{1D node}}
     \put (76,20) {\color{green2d}{2D Cayley}}
     \put (76,15) {\color{green2d}{with hole}}
     \put (55,42) {\color{green2d}{2D node}}
     \put (04,80) {\color{yelloworange}{3D Cayley}}
     \put (48,05) {\color{yelloworange}{3D Cayley}}
     \put (42,15) {\color{yelloworange}{3D node}}
     \put (5,5) {\color{red}{4D node}}
\end{overpic}
\caption{\scriptsize \textbf{Portion of the atlas with Cayley Regions}: 
Active constraint regions in the atlas represented as nodes
colored by their dimension, shown with their Cayley regions.
The grid of little cubes next to each node
delineates the Cayley region of that node. Each little cube is
a Cayley configuration. Consider the 2D active constraint
region in the center. This region has has no Cayley configurations in the middle (a
hole) since every Cartesian configuration, corresponding to these Cayley configurations,
violates Constraint \cone. These
violations are caused by atom pairs that are neither Cayley parameters nor
edges of the active constraint graph. Such holes typically also have a
convex Cayley parametrization. The Cayley configurations highlighted with different
colors are points adjacent to their children (boundary) regions albeit using
different Cayley parameters (see Section \ref{sec:convexity}).
}
\label{fig:pctreeSpace}
\end{figure}

\begin{figure*}[htpb]
\centering
\subfigure[]{\includegraphics[scale=0.07]{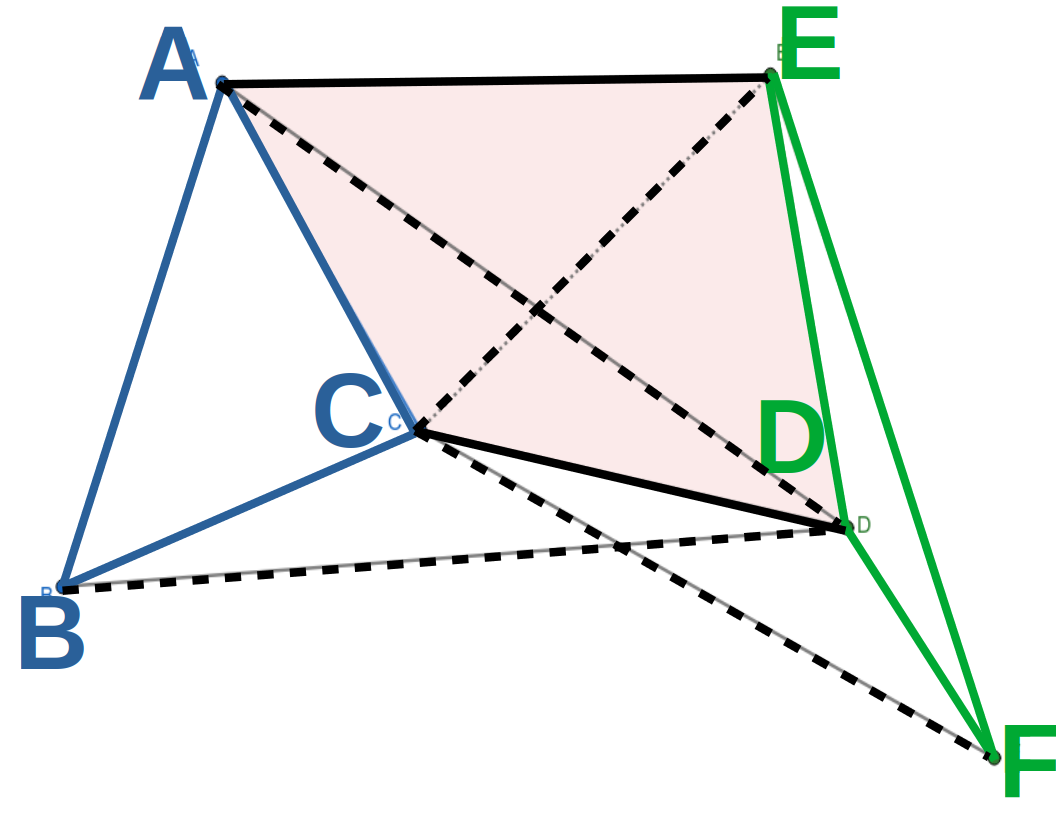}}
\subfigure[]{\includegraphics[scale=0.07]{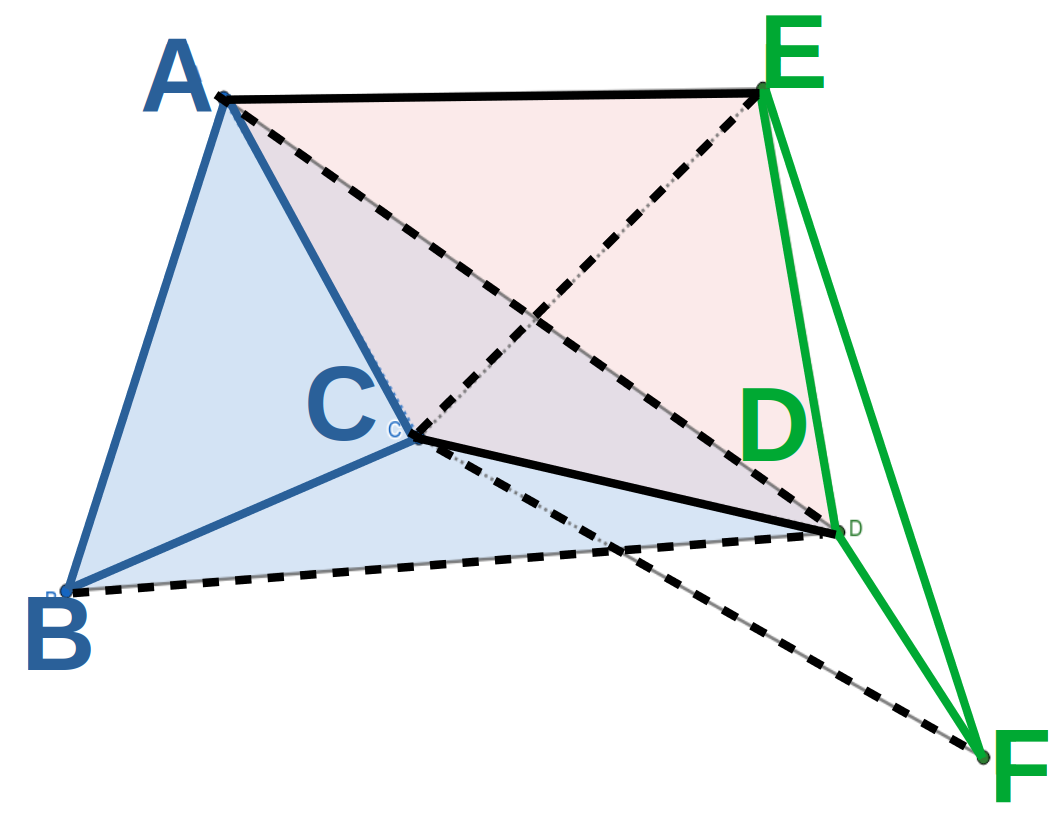}}
\subfigure[]{\includegraphics[scale=0.07]{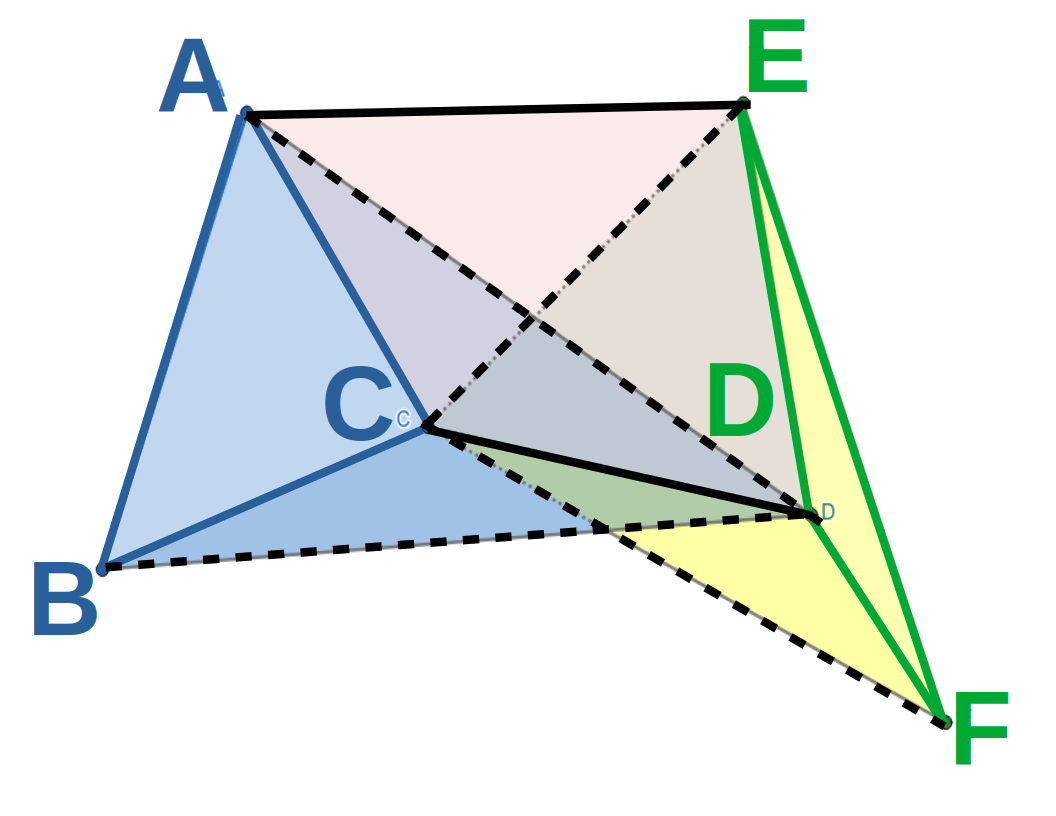}}
\subfigure[]{\includegraphics[scale=0.07]{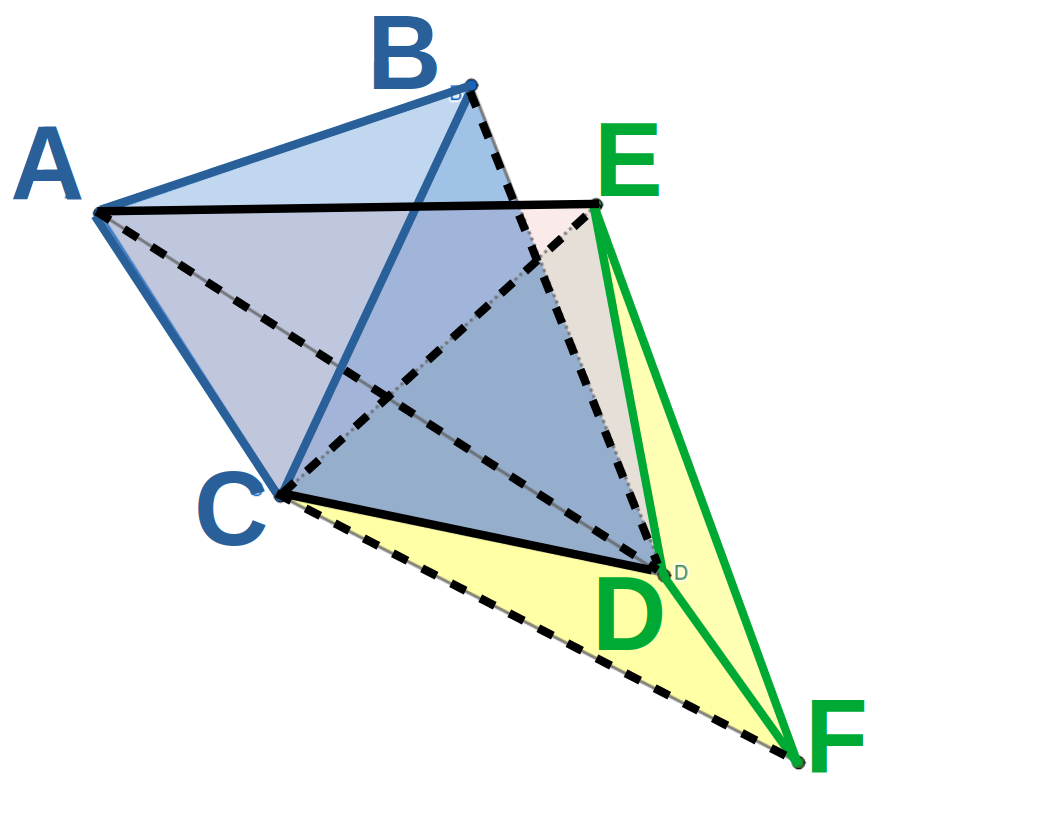}}
\subfigure[]{\includegraphics[scale=0.07]{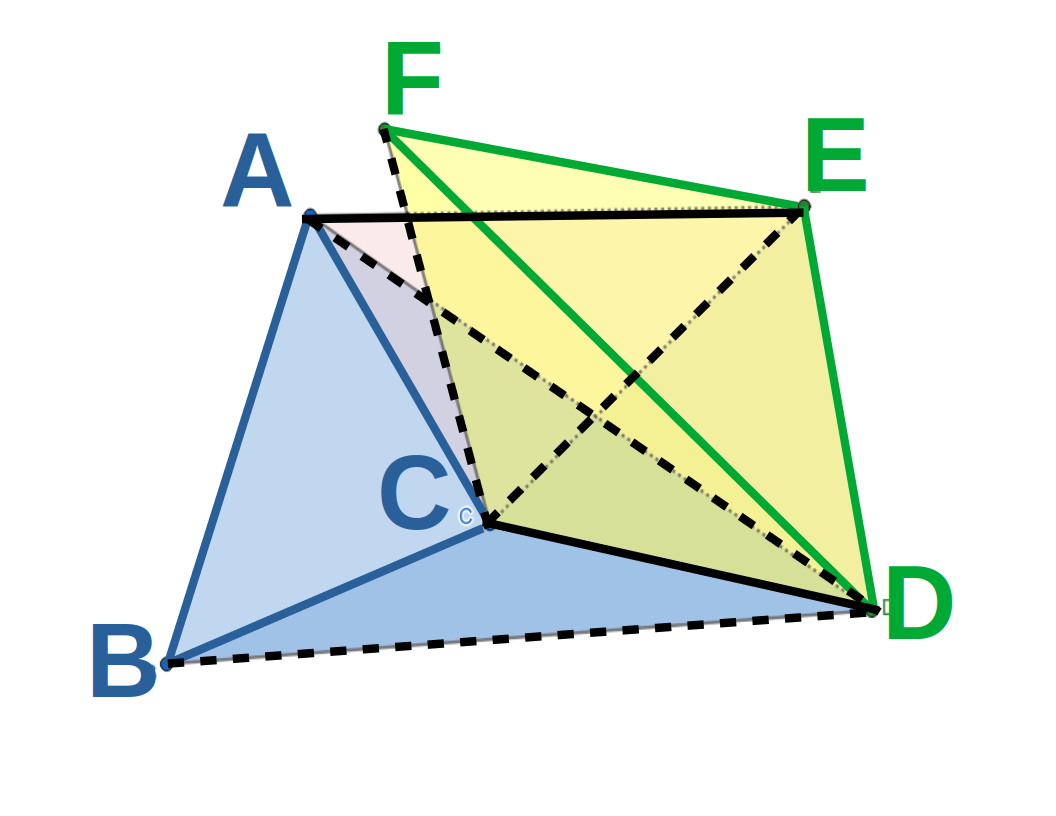}}
\subfigure[]{\includegraphics[scale=0.07]{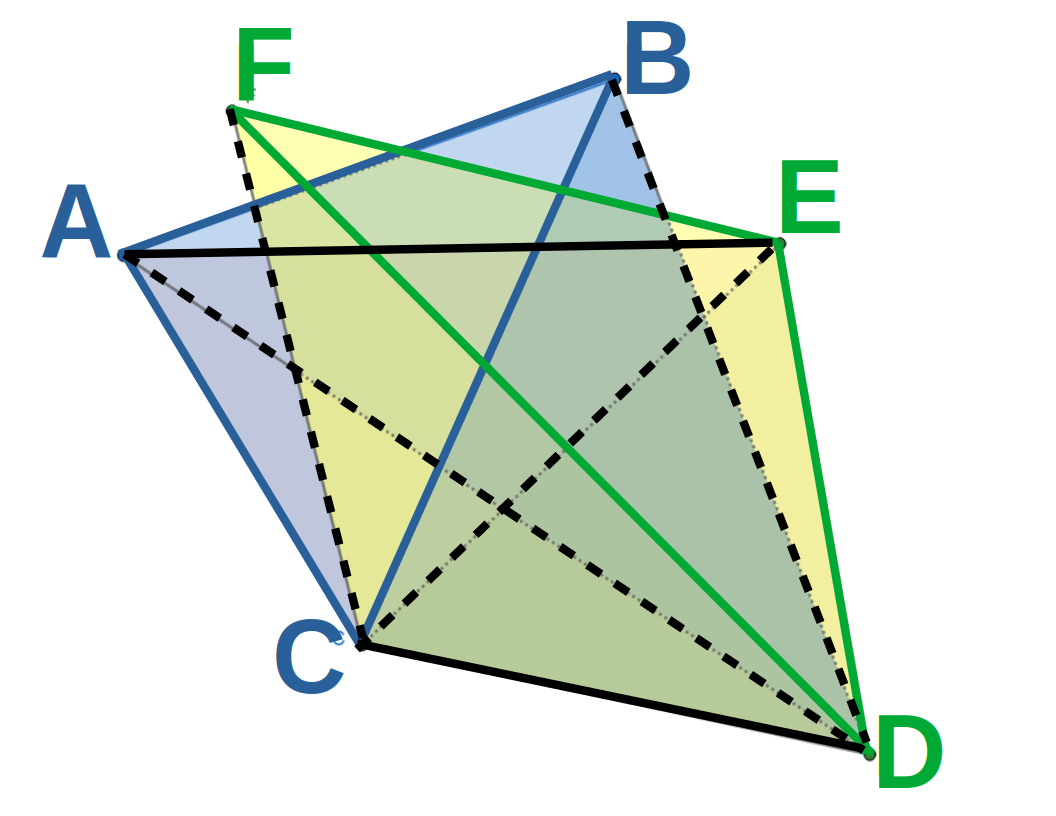}}\\
\subfigure[]{\includegraphics[scale=0.1]{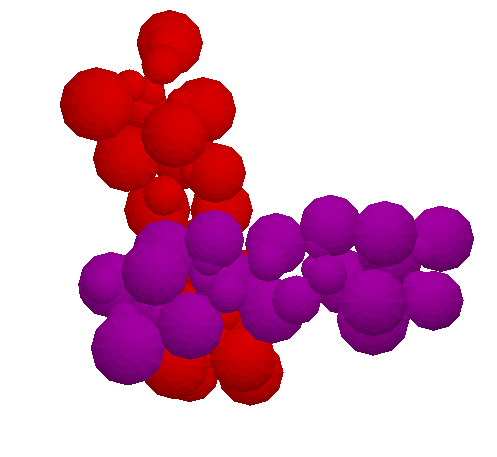}}
\subfigure[]{\includegraphics[scale=0.1]{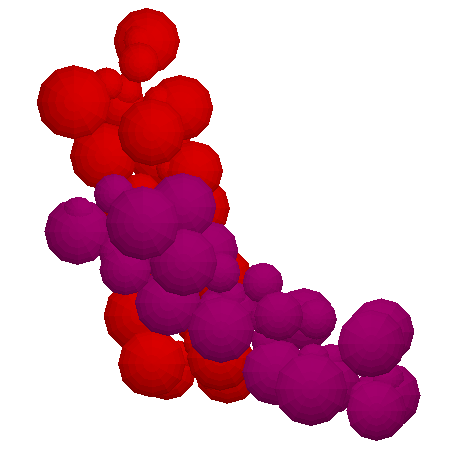}}
\subfigure[]{\includegraphics[scale=0.1]{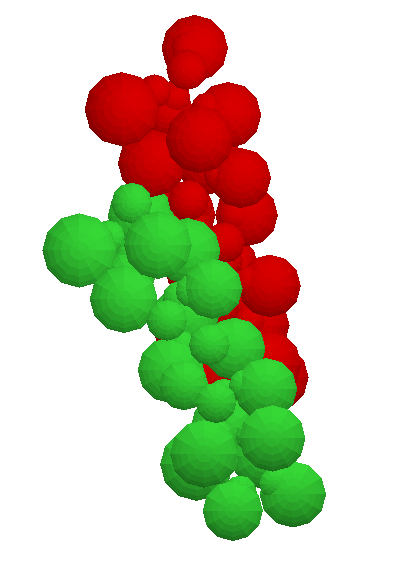}}
\subfigure[]{\includegraphics[scale=0.1]{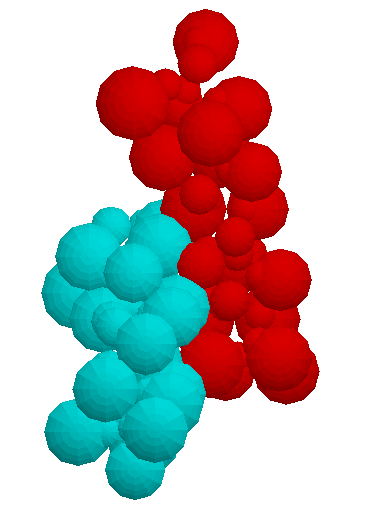}}

\caption{\scriptsize \textbf{Illustration of Flips}:
(a) through (f) show the same active constraint graph, where
vertices A, B, and C belong to one \rmc\ and vertices D, E, and F
belong to the other. Blue and green edges denote edges within the
first and second \rmc s respectively. Solid black edges denote
active constraints, while dotted black edges denote Cayley parameters. 
This graph has two active constraints and thus depicts a 4D active
constraint region.
The edges of the active constraint graph form 3 tetrahedra.
(a) through (c) highlight the different tetrahedra and 
(c) through (f) show four different realizations of the 3 tetrahedra.
(g) through (j) show the corresponding Cartesian configurations of the 
realizations shown in (c) through (f).
(a) Tetrahedron 1 highlighted in red.
(b) Tetrahedra 1 and 2 are highlighted in red and blue respectively.
(c) Tetrahedra 1, 2, and 3 are highlighted in red, blue, and yellow respectively.
(d) Tetrahedron 2 is reflected across the plane ACD, but Tetrahedra 1 and 3 are in the same positions as in (c).
(e) Tetrahedron 3 is reflected across the plane CED, but Tetrahedra 1 and 2 are in the same positions as in (c).
(f) Tetrahedra 2 and 3 are both reflected, but Tetrahedron 1 is in the same position as in (c).
(g) through (j) show 4 different flips in an example 4D active constraint region. 
See text in Section \ref{sec:realization}.}
\label{fig:flips}
\end{figure*}

\begin{definition}[Atlas]
An \emph{atlas} of an assembly landscape consists of the roadmap, and, for each active constraint region,
(a) the Cayley parameters 
		of the chart, and (b) a set of Cayley configurations and
		their corresponding Cartesian configurations. 
\end{definition}

Although most active constraint graphs arising in Assembly problem (\cone,
\ctwo) have convex Cayley regions, the feasible region is often (when
$\dlo_\ijx >0$ in Constraint \cone) a non-convex subset created by cutting out
regions defined by other constraints of type \cone. Each such constraint
resulting in a such a cut-out region is neither an active constraint nor a
Cayley parameter in the active constraint graph.  However, the cut-out regions
typically have a (potentially different) convex Cayley parametrization. This
can be seen in \figref{fig:pctreeSpace} where the configurational region of the
node in the center has a cut-out hole because of constraint violations by atom
pairs that are not a part of the active constraint graph. 

\subsubsection{Computing Cartesian Configurations from a Cayley Configuration}
\label{sec:realization}
The inverse map from Cayley configurations to Cartesian configurations can be
computed easily since 3-trees (obtained by the addition of Cayley parameter
non-edges to the active constraint graph) are realized by solving three
quadratics at a time.  This in conjunction with convexity of the Cayley region
allows for efficient sampling of the typically disconnected and topologically
complex Cartesian region (\rahul{see Section \ref{sec:correspondanceCayleyCart}}).

Realizing a complete 3-tree with $i$ tetrahedra means finding the coordinates
of $i$ new points, one at a time, using 3 distance constraints between a new
point and 3 already placed points.  Each new point is the solution of the
quadratic system for intersecting 3 spheres resulting in two possible
placements of the new point. This yields $2^i$ possible Cartesian
configurations of the Cayley configuration. A \emph{flip} associated with an
active constraint region consists of all Cartesian configurations in that
region restricted to one of these $2^i$ placements (see \figref{fig:flips}).
For $k=2$, there are always 3 tetrahedra and as a result, every Cayley
configuration has up to 8 Cartesian configurations (\rahul{see Theorem
\ref{thm:realization} in Section \ref{sec:app:realization} of the Appendix}).

Flips could be disconnected and form distinct (Cartesian) macrostates.  In
addition, multiple flips could be connected to form a single macrostate when
the tetrahedra flatten, which happens at the convex Cayley bounds (for those
readers familiar with algebraic topology, we note that the Cayley
parametrization map is a branched cover. \rahul{See Section
\ref{sec:app:convexity} in the Appendix}).

\subsubsection{Correspondence between Cayley and Cartesian Regions}
\label{sec:correspondanceCayleyCart}
\begin{figure*}[htpb]
\centering
\subfigure[]{\label{fig:KBSpace}\includegraphics[scale=0.23]{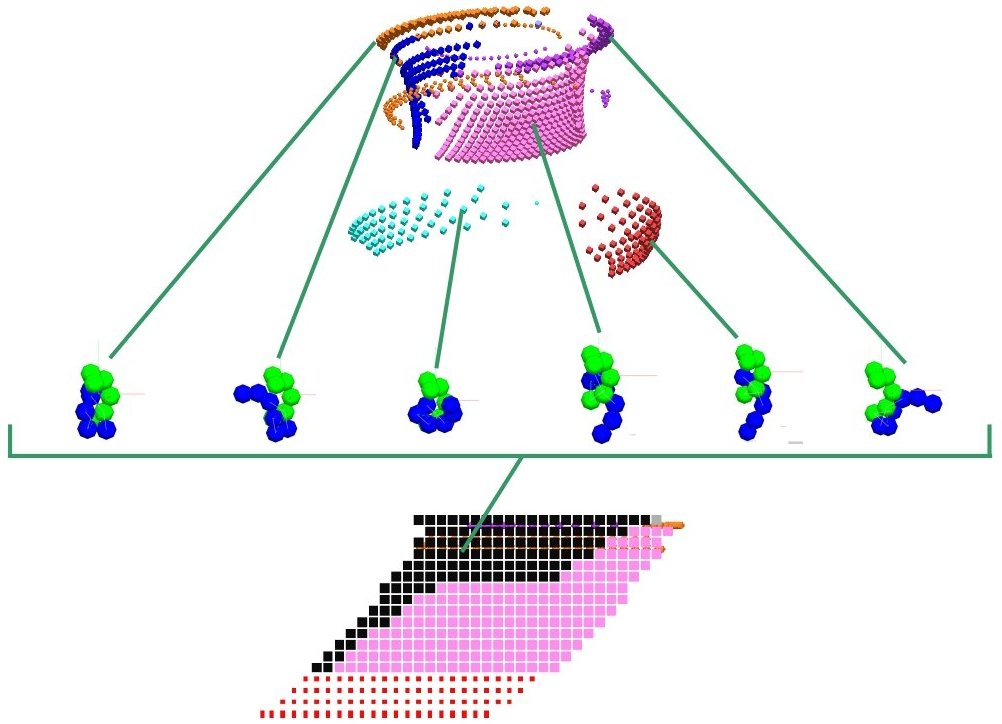}}
\subfigure[\label{fig:KBboundary}]{\includegraphics[scale=0.28]{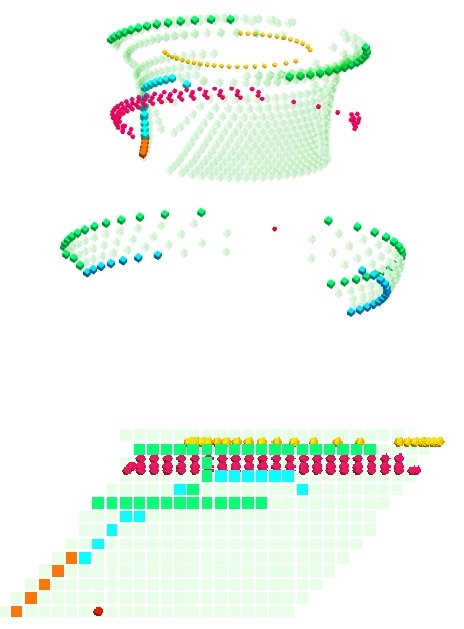}}

\subfigure[\label{fig:FunnelSpace}]{\includegraphics[scale=0.23]{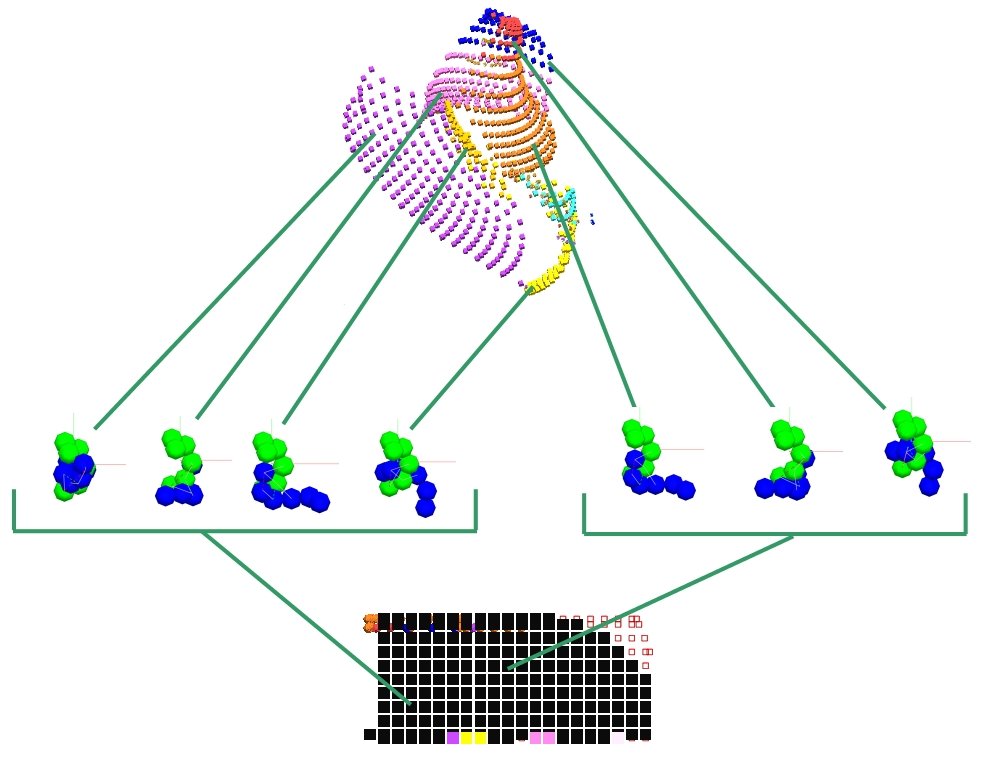}}
\subfigure[\label{fig:FunnelBoundaries}]{\includegraphics[scale=0.28]{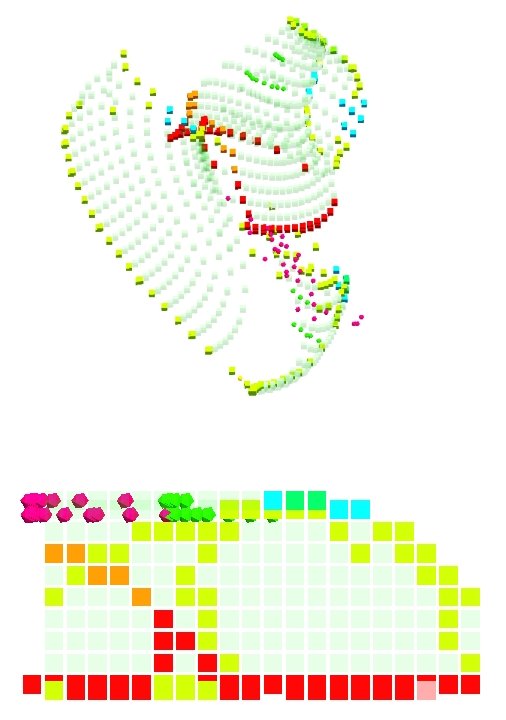}}
\caption{\scriptsize \textbf{Cartesian and Cayley Regions}: 
(a) and (c) show the Cartesian (top) and Cayley (bottom)
regions of two 2D active constraint regions in the 6 pocketed input
described in Section \ref{sec:expSetup}.
Each point in the bottom is a Cayley configuration.
Each point in the top represents a Cartesian configuration, some of which are shown in the middle (see text in 
Section \ref{sec:correspondanceCayleyCart}). Notice that the same active constraint
region is convex in the Cayley parametrization while it is highly non-convex
and disconnected in the Cartesian parametrization. In the Cartesian region, the
set of configurations of a single color represent a flip. Since the
Cayley parametrization is a many to one map, one Cayley configuration may
potentially correspond to several configurations in the Cartesian
parametrization. The black points in the Cayley region indicate configurations
that correspond to multiple Cartesian configurations.
Figures (b) and (d) show the boundaries of the active constraint region shown in (a) and (c)
respectively. The transparent
points show the interior of the region and the colored points show the boundary
regions where new constraints become active. Different colored points show different
boundary regions where a different constraint becomes active.
(see Section \ref{sec:correspondanceCayleyCart}).
}
\label{fig:boundaries}
\end{figure*}

\figref{fig:KBSpace} and \figref{fig:FunnelSpace} show the stark difference in
the topological complexity between the Cayley and Cartesian parametrization of
two typical 2D active constraint regions. Each point in the bottom is a Cayley
configuration. Each point in the top is a Cartesian configuration in the
quaternion representation, i.e., as a point six dimensional space (3
translational and 3 rotational dimensions), projected onto the 3 translational
axes. 

\figref{fig:KBboundary} and \figref{fig:FunnelBoundaries} show the boundary or
child regions of the active constraint regions of \figref{fig:KBSpace} and
\figref{fig:FunnelSpace} respectively, where an extra constraint becomes
active. 

A Cayley configuration at the boundary of a parent region can map to multiple
Cayley configurations in either the same or multiple children regions (via
different flips and different region-specific Cayley parametrizations for the
children). In the same way, multiple Cayley configurations at the same boundary
in the parent region may map to the same Cayley configuration in a child region
(via the same or different flips) due to different Cayley parametrization for
parent and child.

\subsection{Bar-Code: Input Shape Variables and Landscape Design Variables}
\label{sec:designVariable}
We establish a formal and intuitive connection between the topology of assembly
landscapes (output) and the geometry of the assembling \rmc s(input). The
geometry of the input  \rmc s is described by two \emph{input shape
variables}, the \emph{width} of the \rmc\ and its \emph{concavity}. A
\rmc\ is considered \emph{narrow} if the width of the entire \rmc,
defined as the spread of the \rmc\ across its principal axis, is comparable
to the average diameter of the `hard-sphere' (described in Section
\ref{sec:stratification}) corresponding to constituent atoms. It is considered
\emph{wide} otherwise.

To measure the concavity of an input \rmc, take a closed smooth manifold
(\emph{skin}), enclosing the minimum volume containing all the hard spheres
corresponding to each of the constituent atoms in the \rmc. For any two
points on the skin of the \rmc, consider the difference between their geodesic
and Euclidean distance. \emph{Concavity} is the average of all such differences
across all pairs of points. \emph{Convex} (resp. \emph{concave}) \rmc s have
low (resp. high) concavity. We can discretize concavity computation by
reducing it to a discrete set of points for each atom that has at least a part
of its surface on the skin.

The input shape variables can be tuned to ensure that specific active
constraint regions and basins are present by influencing the landscape design
variables (atlas bar-code). These are the number of active constraint regions
and the average Cayley parameter range, where the average
is taken over a set of active constraint regions of interest.  The average
Cayley parameter range for an active constraint region determines its volume.
Furthermore, concavity of the input \rmc, an input shape parameter, is directly
correlated with the average Cayley parameter range and the volume of the active
constraint regions. 

Results in Section \ref{sec:results:design}) show how the bar-code influences
other dependent properties of the output landscape such as volume of regions
and basins.

\subsubsection{Advantages of Region-Specific Cayley Parametrization}
\label{sec:boundaryDetection}
There are 4 main advantages of region-specific Cayley parametrization.

1. Avoiding gradient descent: Since the Cayley region of an active constraint
region is convex, and exact bounds are easily computed, sampling the region
is as simple as taking a step within the convex Cayley bounds and computing the
Cartesian configurations of the new Cayley configuration. This computation
time is constant for any given $k$ as described in Section
\ref{sec:realization}. This avoids gradient-descent search used by all
prevailing methods to enforce the active constraints. With Cayley
parametrization no constraints need to be enforced, drastically improving
sampling efficiency and minimizing discarded samples.

2. Boundary detection:
Sampling using region-specific parametrization reveals more configurations (see
results in Section \ref{sec:results:RegionSpecific}) and hence finds more lower
dimensional boundary regions, decoupling the generation of the roadmap from
sampling. 

3. Volume computation: A child or boundary region may have negligible volume
compared to the parent or interior region (as the child region has one lower
effective dimension, for example a narrow channel in the assembly landscape).
Yet, it deserves careful sampling when computing free energy and entropy since
it has a lower potential energy level.  Region-specific parametrization ensures
better accuracy for sampling such regions as shown in Section
\ref{sec:results:RegionSpecific}.

Cayley convexification aids efficient approximate and exact volume computation.
Approximate volume computation is used in the comparison of relative volumes of
potential energy basins and computation of path probabilities (see algorithm in
Section \ref{sec:methods:approximateVolume} and results in Section
\ref{sec:results:approximateVolume}). Exact volume is used for entropy
computation (see algorithm in Section \ref{sec:methods:exactVolume} and results
in Section \ref{sec:results:ExactVolume}).

4. Path Probabilities:
To assign probabilities to paths between active constraint regions (
discussed in Sections \ref{sec:methods:ConfPaths} and
\ref{sec:methods:PathProb}) we can use 
two distinct partitions of an active
constraint region into subregions, one based on the relative volumes of the
parent regions and the second based on the relative volumes of children
regions. These partition volumes along with paths between active constraint
regions yields approximate path probabilities .

%%%%%%%%%%%NewSection%%%%%%%%%%%%%%%%
\subsection{Algorithms}
\label{sec:algorithms}
This section discusses the key algorithms implemented in the EASAL software
\cite{Ozkan:toms, easalVideo, easalSoftware}. We first present the core
algorithm and analyze its correctness and time complexity. This is followed by
3 variants of the core algorithm for different assembly regimes: (1) arbitrary
$n$ and $2<k<12$; (2) $n=1$ and $k<24$; and the most general version (3) with
arbitrary $n$ and $k$.

\subsubsection{Core Algorithm}
\label{sec:coreAlgorithm}
We describe the core algorithm for arbitrary $n$ and $k=2$ \rahul{(Algorithm
\ref{alg:sampleAtlasNode} in Section \ref{sec:app:algorithm} of the Appendix}),
leveraging the capabilities described in the prior sections.  The algorithm
starts by generating all possible active constraint graphs with 1 or 2
(depending on user input) active constraints yielding 5D or 4D regions
(represented as root nodes) in the roadmap.

Starting with an initially generated root node, the algorithm then (i)
recursively (by depth first search) generates the roadmap by discovering active
constraint regions of decreasing dimensions; (ii) uses Cayley convexification
of the region to efficiently compute bounds for Cayley parameters a priori
(before computing the Cartesian configurations), and coarsely samples Cayley
configurations in this convex region; (iii) efficiently finds the (finitely
many) Cartesian configurations of the sampled Cayley configurations; (iv)
creates a boundary regions of 1 lower dimension when a new constraint becomes
active detected as a collision a posteriori (after computing the Cartesian
configurations). Configurations where a new constraint becomes active are
\emph{witness} configurations for the corresponding boundary or child region;
(v) all ancestors of 0D regions are introduced into the roadmap, i.e., those
missed by coarse sampling are later discovered by ancestor addition through 
their descendants, and sampled.

Thus, the algorithm, through coarse sampling, generates the roadmap, decoupling
roadmap generation from sampling (as discussed in Section
\ref{sec:AdvantagesStratification}).  Subsequently, a desired active
constraint region can be sampled using different variants of Cayley sampling
with any desired level of refinement.  Optionally, as detailed in the
pseudocode in \rahul{Section \ref{sec:app:algorithm} of the Appendix}, 
the sampling can proceed along
with the roadmap generation, resulting in a complete atlas. 

\subsubsection*{Handling Non-Convexifiable Regions}
When the active constraint graph of a region is not a partial 3-tree, 
we first drop
constraints one at a time, until the active constraint graph becomes a partial
3-tree. In doing so, we end up in an ancestor region, with a partial 3-tree
active constraint graph and a convex Cayley parametrization. Note that since
non-partial 3-trees potentially arise only when exploring active
constraint regions with 4 or 5 active constraints (2D and 1D atlas nodes
respectively), it is always possible to drop one or two constraints to reach an
ancestor region which has a partial 3-tree active constraint graph. 

Once in the ancestor region, we trace along rays to populate the lower
dimensional region by searching in the ancestor region. For example, to find a
2D boundary region which does not have a partial 3-tree active constraint graph
or a convex parametrization, we drop one constraint resulting in a 3D parent
region that is guaranteed to have a convex Cayley parametrization. We then
perform multiple one dimensional searches in the 3D region (ray tracing) in
order to locate the target 2D boundary (child) region along each of these rays,
using binary search. This generalizes to any dimension and region in the sense
that ray tracing is robust when searching for and populating a region one
dimension lower. By recursing on the thus populated region, we find further
lower dimensional regions.

\subsubsection{Core Algorithm Accuracy}
\label{sec:results:CompletenessAnalysis}
The discussion in the preceding sections proves the following theorem showing
that the core algorithm finds the complete atlas of the assembly landscape (up
to a tolerance specified as part of the input); overwhelmingly generates only
feasible configurations, i.e., minimizes discarded samples; and only explores or
samples any active constraint region once.

\begin{theorem}{(Completeness, Tightness and Efficiency)}
\label{thm:completeness}
The core algorithm 
\begin{enumerate}
	\item creates the complete atlas for the assembly landscape of Problem (\cone, \ctwo) up to a tolerance 
specified as part of the input;
	\item only creates non-empty active constraint regions (through a witness configuration);
	\item only samples feasible configurations in active constraint regions with desired refinement;
    \item explores active constraint regions only once.
\end{enumerate}
\end{theorem}

We emphasize a few notable features of the core algorithm.  Lower-dimensional
(boundary) regions are detected by detecting collisions.  Binary search then
determines, up to a tolerance, a witness configuration for creating this
boundary region, ensuring that any created active constraint region is
non-empty.  Since children regions inherit the constraints of their parent
regions, no region can be missed by not exploring the descendants of an empty
region.

Once the boundary region is found, the core algorithm explores it using its
region-specific Cayley parametrization. Therefore, even regions that have
negligible volume in the parent's parametrization, through their own
parametrization, are as exhaustively sampled as the parent region.

In rare cases when there is no convex Cayley parametrization, the region is
sampled as part of the parent region to ensure good sampling coverage. 

In the core algorithm, we use a hash table to keep track of previously found
and sampled active constraint regions. This ensures that active constraint
regions, found through multiple parents, are not explored or sampled more than
once.

Convexity guarantees that all sampled configurations (other than binary search
due to collisions) are valid and none are discarded. 

During binary search to locate a boundary region, after a collision, some
samples may be discarded. Since this happens only once per newly discovered
region from each ancestor region, the number of discarded samples is bounded by
the number of edges in the roadmap DAG.

\subsubsection{Complexity Analysis}
\label{sec:CA}
\noindent \textbf{Roadmap Generation\\}
By Theorem \ref{thm:completeness} (1), (2), and (4)
the time to generate the roadmap is linear in the number of edges of the
roadmap DAG, since roadmap generation is decoupled from sampling.
In the worst case, the number of regions or nodes, $r$, of the roadmap can be as large as $O(k^2 \cdot n^{12k})$.
Since the
complexity cannot be less than the output size, this achieves the best possible
complexity. Usually, $r$ is a much smaller number of regions specified as part
of the input by designating a set of active constraints of interest.

\noindent \textbf{Sampling an Active Constraint Region\\}
By Theorem \ref{thm:completeness} (1) and (3)
the sampling time grows linearly in the number of feasible configurations, which 
in turn grows polynomially with sampling density and exponentially
in the dimension.  The dimension of an active constraint region is bounded by
$6(k-1)$, where  $k$ is the number of \rmc s.  However, for any given $k$ the
dimension is a constant; hence the sampling time complexity is polynomial
in the sampling density.  

Sampling time grows quadratically in $n$ the number of atoms in each \rmc.
This is because an a posteriori collision check involves, at worst, checking
every atom pair (one from each \rmc) for violation of the Constraint \cone
of the assembly problem. 
Note that this can be improved by using more sophisticated collision detection algorithms.

Thus, given a step size $t$ as a fraction of the range for each Cayley parameter, the
complexity of sampling a region is $O((\frac{1}{t})^{6(k-1)} \times n^2)$, showing the
expected tradeoff between complexity and accuracy \cite{Ozkan2011}.
The constant of proportionality
hidden in the big $O$ depends on one of the two independent landscape design variables, i.e., the
average Cayley parameter range as discussed in Section
\ref{sec:designVariable} (see Results in Section
\ref{sec:results:complexity}).

\noindent \textbf{Sampling the Landscape\\}
Putting these factors together, if $r$ is the number of regions to sample,
given as part of the input by specifying a set of active constraints of
interest, the complexity of sampling all these regions is $O(r \times
(\frac{1}{t})^{6(k-1)} \times n^2)$. For a given number of \rmc s $k$, the
complexity is linear in the number of regions $r$, polynomial in the sampling
density $\frac{1}{t}$ and quadratic in the maximum number of atoms in a \rmc\
$n$.  Note that number of regions $r$ is one of the landscape design variables.
The number of atoms $n$ is a constituent of the input shape variables.\\

A recent parallelized version of the EASAL software (using a modified version
of the core algorithm) takes advantage of the partition of the landscape to
sample active constraint regions in parallel without repeat sampling. The
algorithm inherently does not present any obstacles to achieving a linear
speedup with the number of compute cores. Early testing of this method has
shown a 3X speedup on a 4 core machine.

\subsubsection{Algorithm Variant for arbitrary $n$ and $2<k<12$}
\label{sec:AlgorithmVariantkg2}
%%%%%%%%%%%NewSection%%%%%%%%%%%%%%%%
Assembly Problem (\cone, \ctwo) generalizes as follows. The constraint \cone\
remains the same, the constraint \ctwo\ now enforces a tree as the minimal
active constraint graph, with one vertex (representing an atom) in each of the
$k$ \rmc s. There are three distinct methods for tackling this variant of the
problem. 

In the first method we start by generating the assembly landscape of two \rmc
s, say $A_1$, $A_2$. Then, treating each configuration $c_{A_1A_2}$ in the
atlas $A_1A_2$ as a \rmc\ we generate the atlas of $c_{A_1A_2}A_3$. This way
we add one \rmc\ at a time till we have added all the \rmc s. With this method,
at each stage, the number of atlases needed to be generated grows by a factor
of the number of rigid configurations discovered at each stage. However,
most of these configurations may not be valid further down the line.
Thus this method does not scale well when the number of rigid configurations
generated is large.

\begin{figure*}[htpb]
		\includegraphics[width=\textwidth]{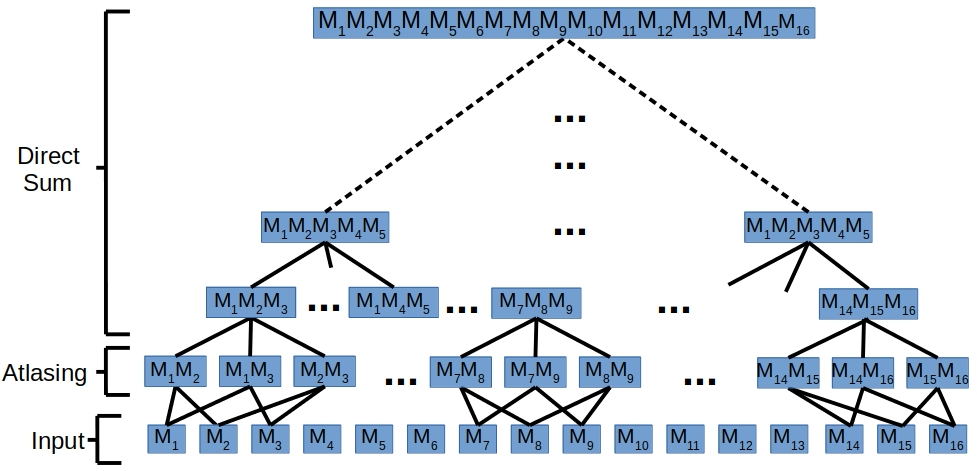}
		\caption{Tournament tree showing the assembly of 16 \rmc s, $A_1$ through
		$A_{16}$. The input are the leaves of the tree, the internal nodes adjacent
		to the leaves are atlases generated using the core algorithm. All
		other internal nodes are obtained through a direct sum of its children.
		The root node is the final atlas of all the input \rmc s. 
		See text in Section \ref{sec:AlgorithmVariantkg2}.}
		\label{fig:tournamentTree}
\end{figure*}

The second method is to generate, through atlasing, all the $\binom{k}{2}$
atlases of pairs of \rmc s, namely $A_i$ and $A_j$ for $1 \le i, j \le k$ and
$i\ne j$. We then recursively merge by taking direct sums of these atlases,
aligning the common \rmc s between them, and building larger atlases at every
stage. This sets up a \emph{tournament tree}, whose leaves are the input \rmc
s, its internal nodes adjacent to the leaves are atlases obtained through
atlasing, using the core algorithm, and all other internal nodes are obtained
through a merging or direct sum of its children (see Figure
\ref{fig:tournamentTree}).  The root node is the final atlas of all the input
\rmc s. At every internal node, the complexity of search is at most the product
of the search complexity at all of its children.

For a given set of input \rmc s, there are  potentially several different 
tournament trees. The optimal among these is the one which has the least
search complexity at the root of the tree. Thus, picking of the best
tournament tree can be setup as an optimization problem which minimizes
the search complexity at the root of the tree.

Let us assume we are merging the atlases of $A_1A_2$, $A_2A_3$ and $A_1A_3$ to
obtain the atlas $A_1A_2A_3$.  For a configuration $(c_{A_1A_2}, c_{A_1A_2})$
in the direct sum, let configuration $c_{A_1A_2}$ belong to the active
constraint region of the atlas $A_1A_2$ with graph $G_{12}$ and let
configuration $c_{A_1A_3}$ belong to the active constraint region of the atlas
$A_1A_3$ with graph $G_{13}$. Now the configuration $(c_{A_1A_2}, c_{A_1A_2})$
is a potential witness for the active constraint region with graph $G_{12}\cup
G_{13}$ in the atlas $A_1A_2A_3$ for the 3 \rmc s. While we are guaranteed that
all the configurations in the atlas for $A_1A_2A_3$ are contained in this
direct sum, some configurations need to be removed.  Specifically, we know that
collision constraint \cone\ holds between $A_1$ and either of the other two,
but the configurations where \cone\ does not not hold between $A_2$ and $A_3$
should be removed. These configurations can be eliminated using a simple
collision check.

\begin{figure}[htbp]
\centering
\subfigure[]{\label{fig:2molecule}\includegraphics[width=0.18\textwidth]{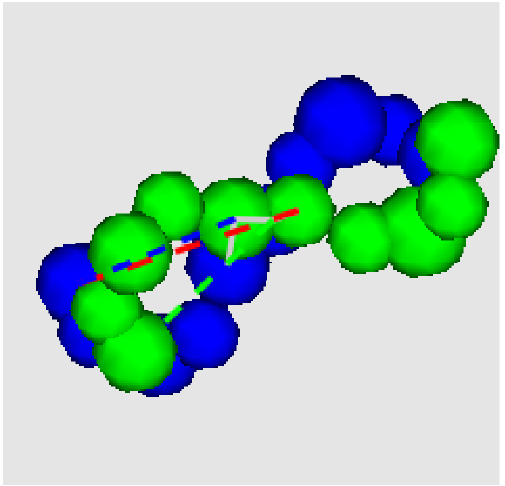}}
\subfigure[]{\label{fig:2molSweep}\includegraphics[width=0.18\textwidth]{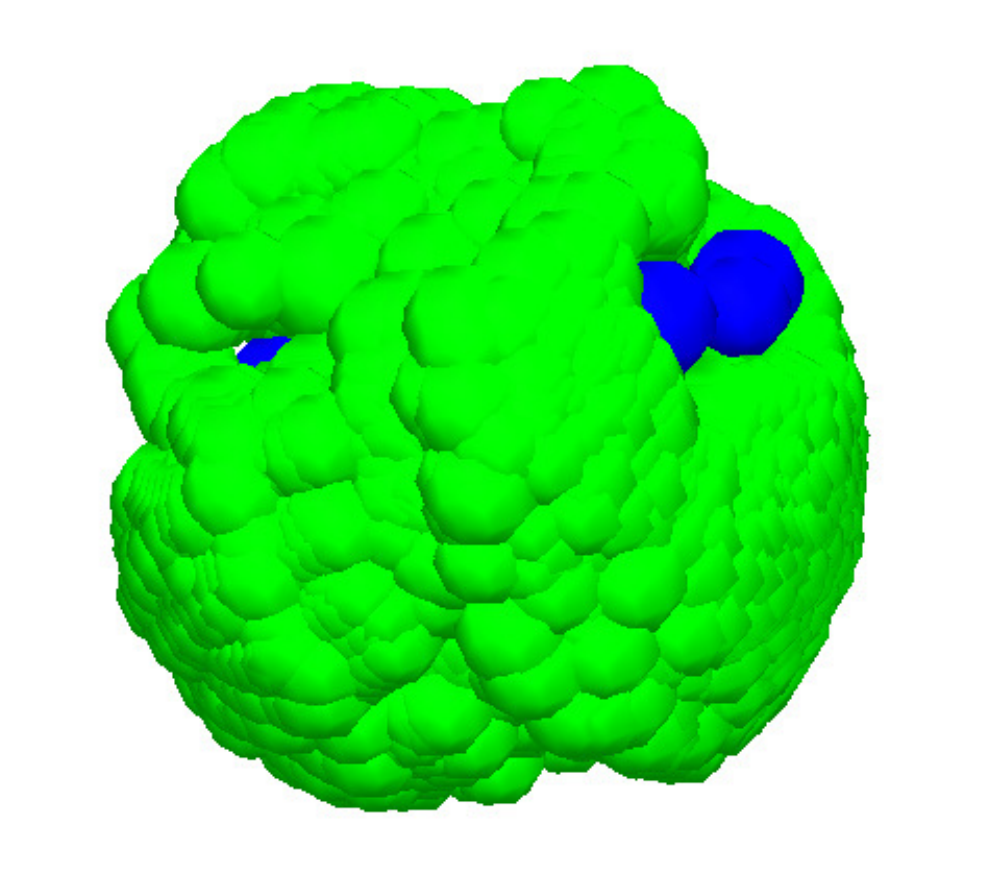}}
\subfigure[]{\includegraphics[width=0.18\textwidth]{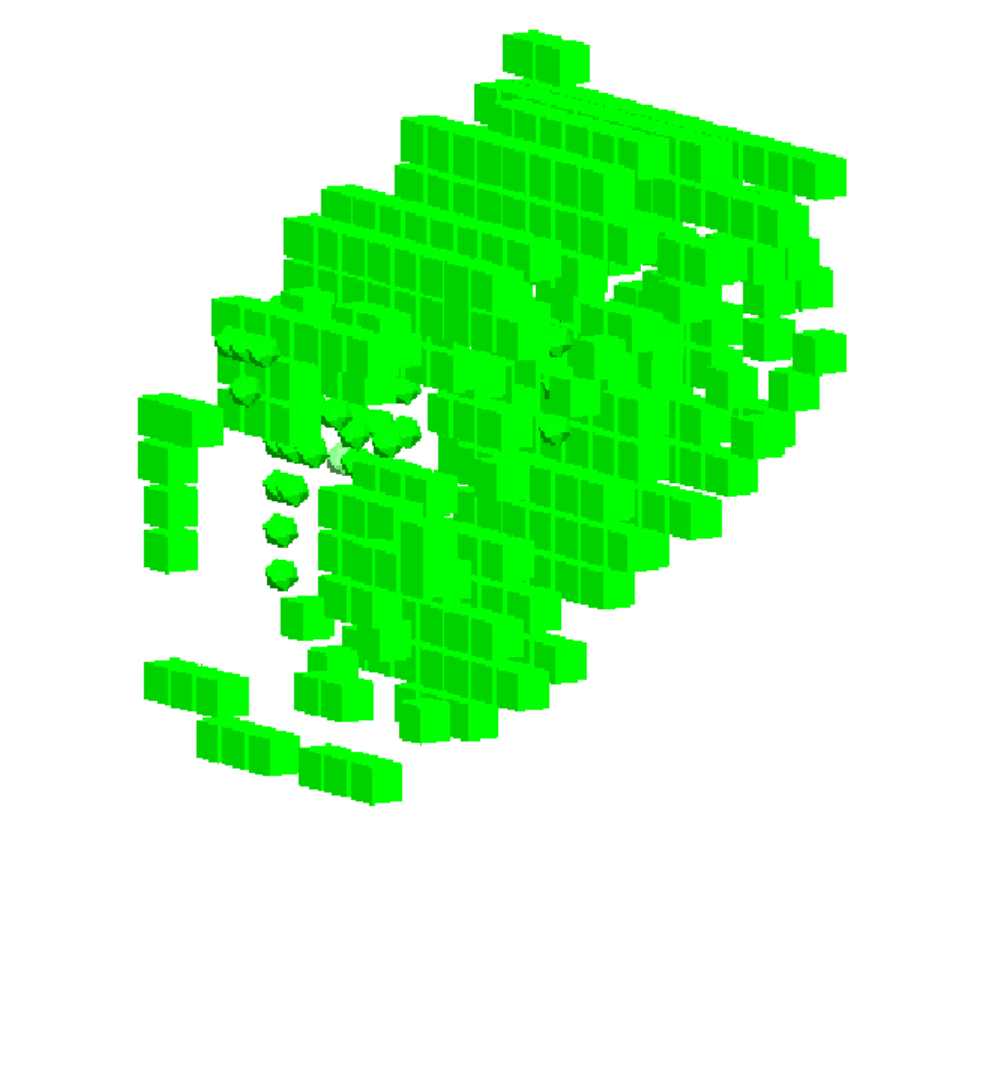}}
\subfigure[]{\includegraphics[width=0.18\textwidth]{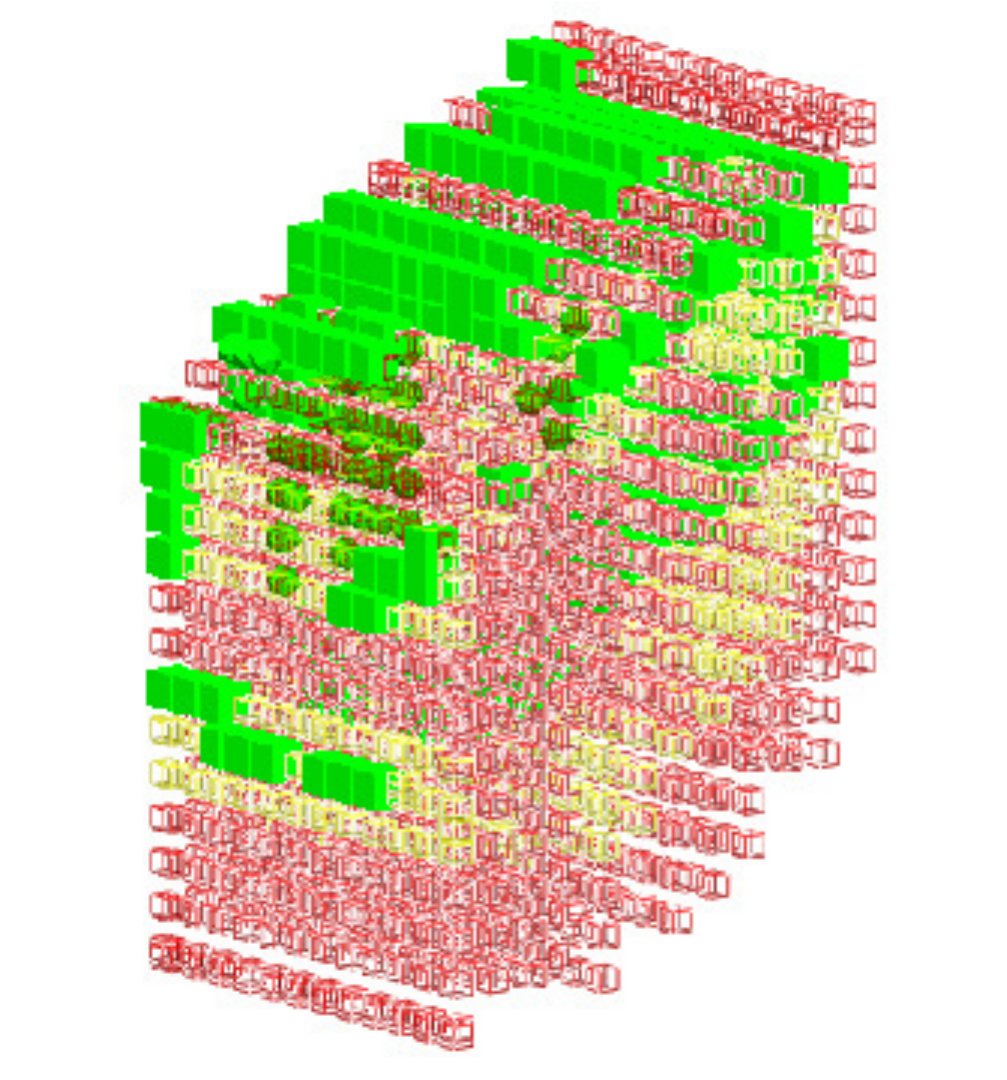}}

\subfigure[]{\includegraphics[width=0.18\textwidth]{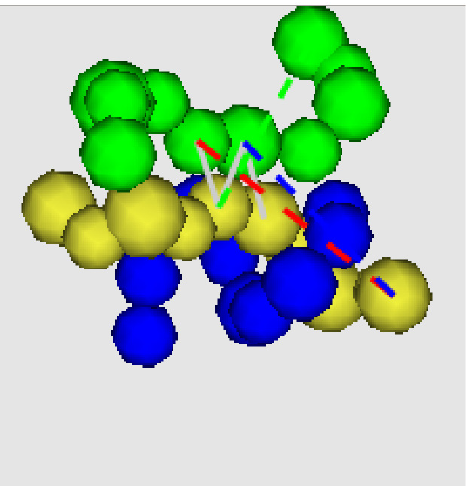} }
\subfigure[]{\includegraphics[width=0.18\textwidth]{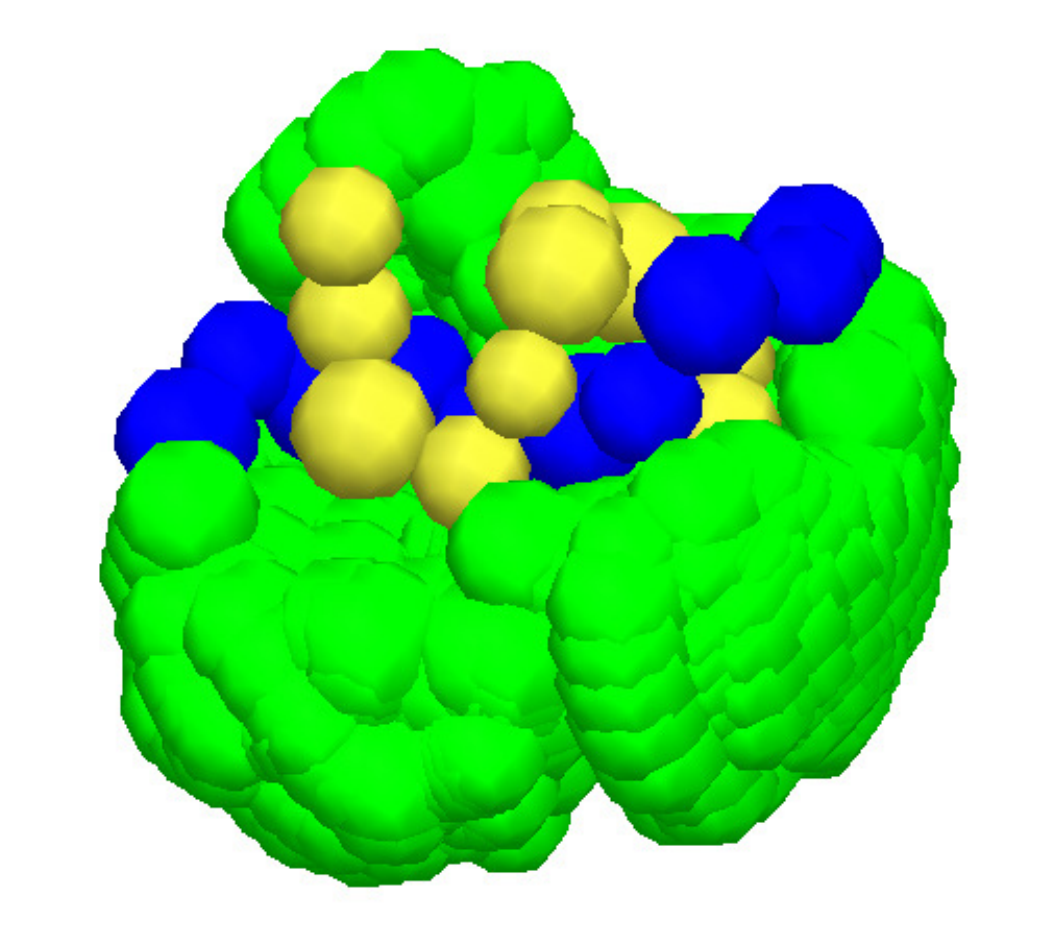} }
\subfigure[]{\includegraphics[width=0.18\textwidth]{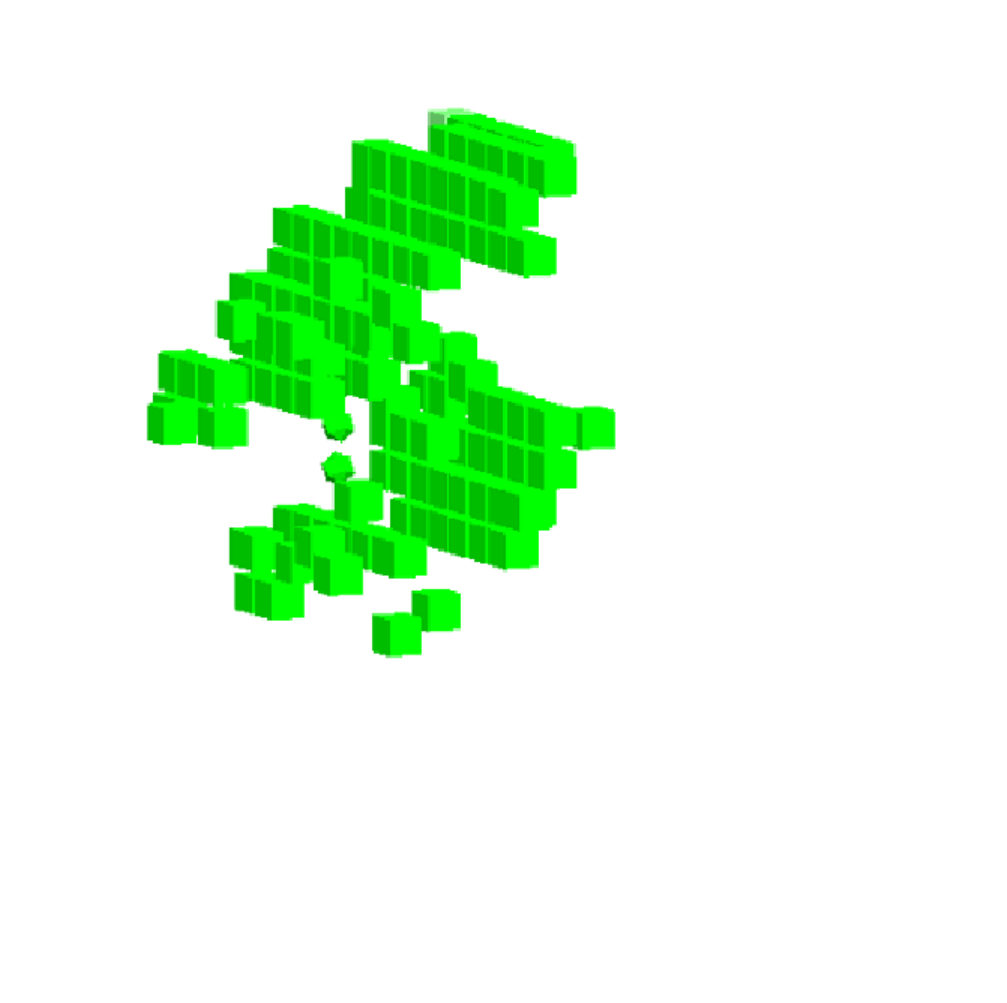}}
\subfigure[]{\includegraphics[width=0.18\textwidth]{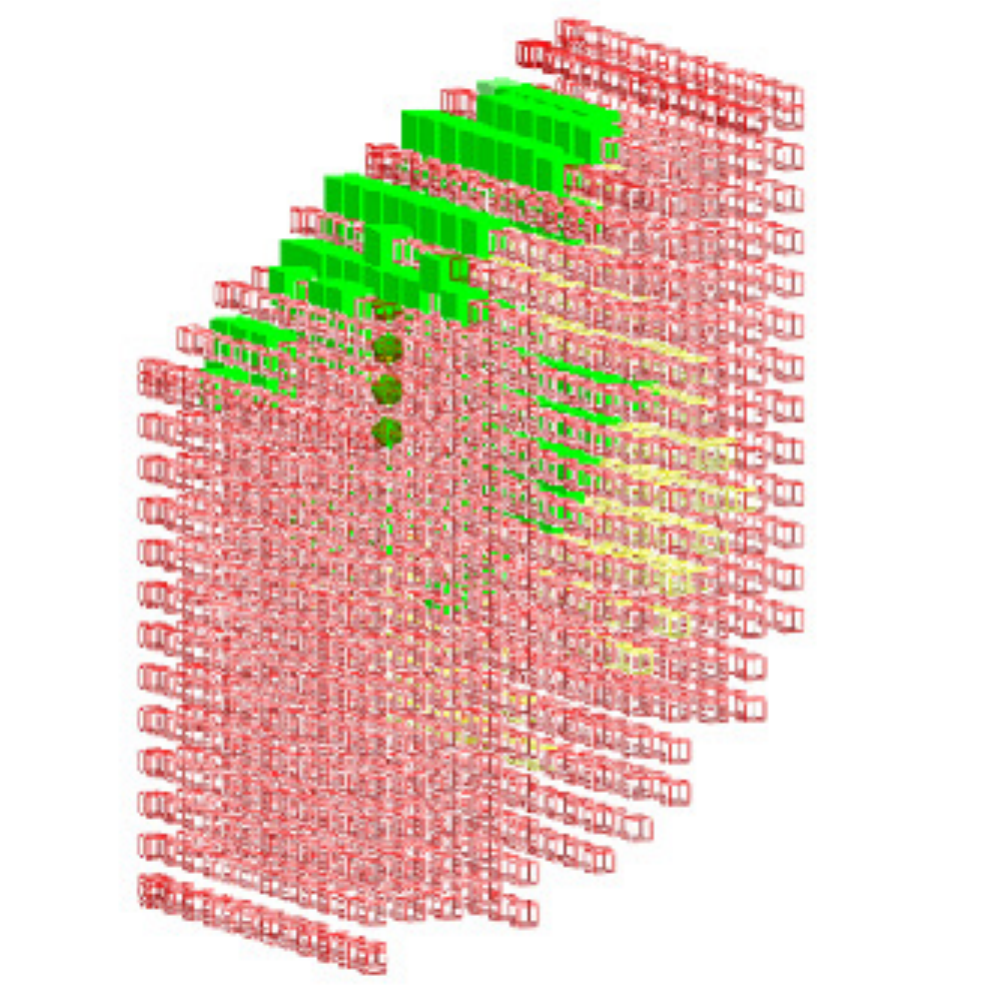} }
\caption{ \scriptsize \textbf{Algorithm Variant for arbitrary $n$ and $k=3$}:
(a) shows a particular configuration in an active constraint region
(with graph $G_{12}$ having 3 active constraints)
in the assembly region of two out of the three \rmc s.
(b) The blue \rmc s is held fixed while the green one swept to show
all feasible Cartesian configurations in the same region.
(c) shows, for the same region, the feasible portion of the Cayley region 
(green blocks), while (d) in addition includes colliding (red) configurations. 
(e) shows a particular configuration in an active constraint region
in the assembly landscape of the three \rmc s with graph $G_{12} \cup G_{13}$.
This configuration contains 
the same 3 active constraints between the blue and the green \rmc\ $G_{12}$
(but a different configuration is used for ease of visualization).
(f) A third (yellow) \rmc\ is fixed in orientation to the blue while the green is
swept to show all the feasible Cartesian configurations. 
(g) Shows all the valid Cayley configurations and (h) in addition shows the
collision configurations (red).
Notice in (g) that some portion of the green blocks of (c) are missing. These represent
configurations that are infeasible due to collision with the third \rmc.
Some of these configurations actually belong to the active constraint region with
graph $G_{12} \cup G_{13} \cup G_{23}$, and need to be moved there
(see Section \ref{sec:AlgorithmVariantkg2}).
All figures are screenshots from the EASAL software 
\cite{easalVideo,easalSoftware,Ozkan:toms}.
}
\label{fig:ss23mol}
\end{figure}

Second, configurations in an active constraint region with graph $G_{12} \cup
G_{13}$ may additionally have a graph of active constraints $G_{23}$ between
$A_2$ and $A_3$. These configurations are witnesses for a new active constraint
region with graph $G_{12} \cup G_{13} \cup G_{23}$. However, this active constraint
region need not be re-atlased, we can simply use the atlas of $A_2A_3$ to look up
the active constraint region with the constraint graph $G_{23}$ and merge it.
This method is illustrated in \figref{fig:ss23mol} for $k=3$.

This method can be further optimized by indexing or hashing the active constraint regions
(according to their active constraint graphs) and sample points (according to
some Cartesian configuration metric). Indexing the active constraint regions
allows us to look at a tiny portion of the assembly landscape instead of the
whole thing at the same time giving constant time lookup when the active
constraint graph is known.  Indexing and hashing the sample points allows us to do range
operations on them to either select or eliminate entire sets of configurations
in an active constraint region, instead of having to look up individual ones
every time we do a direct sum.

In this method, the number of atlases generated is always $\binom{k}{2}$ and all
further atlases generated are done using simply by indexing and searching the
previously generated atlases. 
If all $A_i$'s are identical, this method gains from the consequent
symmetry in the landscape, since we just need to sample the
configurational region of only one pair of molecular units and essentially
reuse them. 

In the third method, we start by creating \emph{root nodes}, whose active
constraint graphs are the complete set of trees. There are $k^{(k-2)}$ trees
with one vertex (representing an atom) in each \rmc\ $A_i$. If there are $n$
atoms in each $A_i$, there are $n^k$ different active constraint graphs that
are isomorphic to the same tree, and hence $k^{(k-2)} \times n^k$ root nodes.
Starting from the root nodes, we use the core algorithm, to recursively, by
depth first search, generate the roadmap of the assembly landscape.  In this
method, we directly sample the $6(k-1)$ dimensional space through Cayley
convexification. Thus, for very large values of $k$ this method quickly becomes
intractable. However, when all the $A_i$'s are identical, the number of root
nodes is vastly reduced because the number of non-isomorphic (unlabeled) trees
is much less than $k^{(k-2)}$.

In fact, even when the \rmc s are not identical, we can exploit the
existence of symmetries in the structure of the \rmc s to efficiently
sample. The paper \cite{sym8010005} studies all such symmetries in assembly
landscapes. When each \rmc\ is an identical singleton sphere,
symmetries can be further exploited, leading to a much simpler algorithm, as
explained in the next section.

%%%%%%%%%%%NewSection%%%%%%%%%%%%%%%%
\subsubsection{Algorithm Variant for $n=1$ and $k\le 24$}
\label{sec:AlgorithmVariantn1}
This variant of Assembly Problem (\cone, \ctwo) is used in the study of
particle cluster assembly (see discussion in \ref{sec:intro:Decoupling}). It is
much simpler than the general input to our methodology since the presence of
many isomorphic active constraint graphs lead to a large symmetry group for the
assembly landscape.  Utilizing these symmetries for this variant of the problem
is discussed in the paper \cite{sym8010005} and summarized in \rahul{ Section 
\ref{sec:app:symmetries} of the Appendix.}

In particular, the symmetry (isomorphic active constraint graphs) allows the
potential stratified roadmap to be created \emph{apriori}, with only one copy
of each unlabeled active constraint graph, i.e., the \emph{fundamental domain}
of the full roadmap. The core algorithm proceeds as usual but its function is
to remove all descendants of a region that has no witness configurations
(\rahul{see Algorithm \ref{alg:Sticky} in Section
\ref{sec:app:stickySphereAlgorithm} of the Appendix.})

\rahul{As Theorem \ref{thm:symmetry} in Section {sec:app:symmetries} of the Appendix} shows, in
addition to the symmetry in the roadmap, there are symmetries within active
constraint regions themselves. For many active constraint regions it is
sufficient to sample a portion whose symmetry orbits complete the region.\\

\begin{figure*}[htpb]
\centering
\includegraphics[scale=0.2]{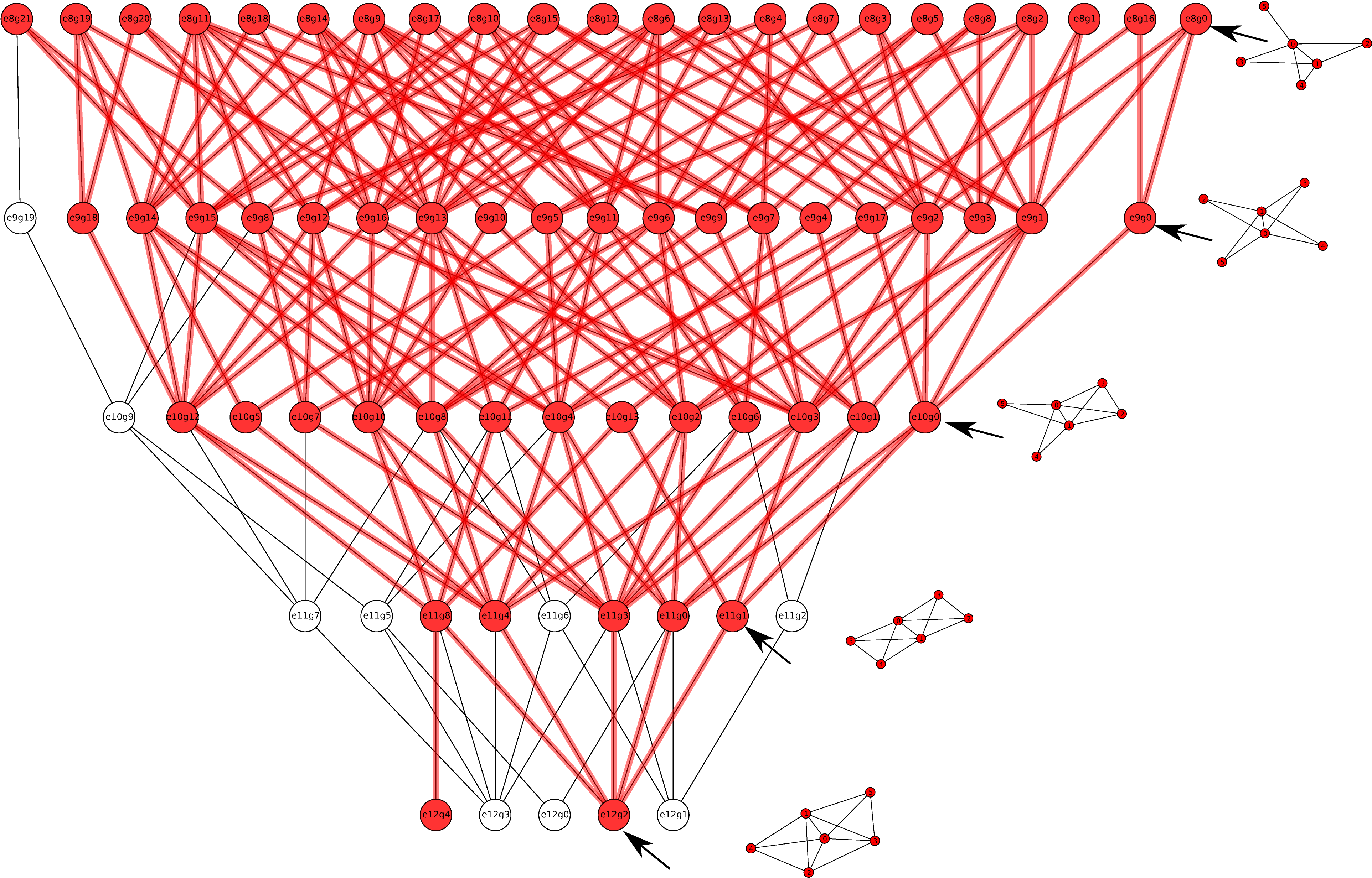}
\caption{\scriptsize \textbf{Atlas for $n=1$ and $k=6$ Assembly System}: 
The fundamental domain of the roadmap (up to 4D regions) for 
a $k=6,~n=1$ assembly system. Each node represents 
the \acr\ labeled by an \acgW. All the nodes in the same level have the same 
dimension. The red nodes have a non-empty \acr, the white nodes
have no feasible configurations. The graphs on the 
right of each level are the \acgW s of the nodes marked by arrows
		(see text in Section \ref{sec:AlgorithmVariantn1}).}
\label{fig:atlas}
\end{figure*}

\begin{figure}[htpb]
\centering
\includegraphics[scale=0.3]{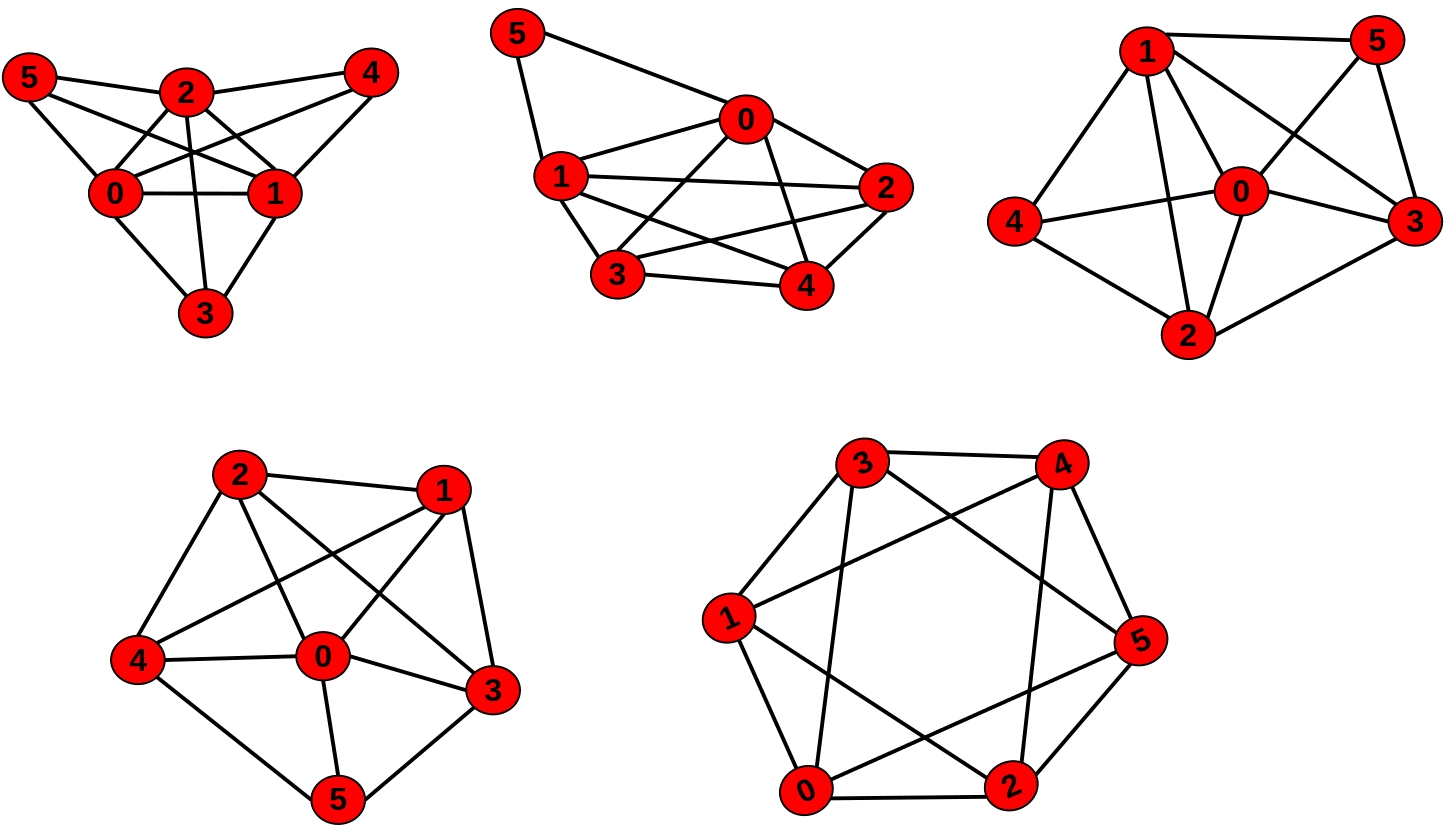}
\caption{ \scriptsize 
\textbf{Non Isomorphic Rigid Active Constraints Graphs for $n=1$ and $k=6$:} Six identical singleton spheres and 12 active constraints 
(see text in Section \ref{sec:AlgorithmVariantn1}).}
\label{fig:v6e12}
\end{figure}

\noindent\textbf{Illustrative Example:} Consider an assembly system of $k = 6$
identical hard spheres with pairwise distances equal to the sum of the
hard-sphere radii. Since all particles and all interaction distances are
identical, two isomorphic \acgW s represent the same \acr\ and only one
representative is needed in the atlas.  \figref{fig:atlas} shows the
fundamental domain of the atlas (up to 4D regions). The red nodes have
non-empty \acr s while white nodes have no feasible configurations. On the
right, we show the active constraint graphs of some active constraint regions.
The roadmap has exactly five 0D regions, which correspond to the 5
non-isomorphic rigid active constraint graphs shown in \figref{fig:v6e12}. 

\subsubsection{Algorithm Variant for arbitrary $n$ and $k$}
\label{sec:AlgorithmVariantnk}
\begin{figure}[htpb]
\centering
\includegraphics[scale=0.3]{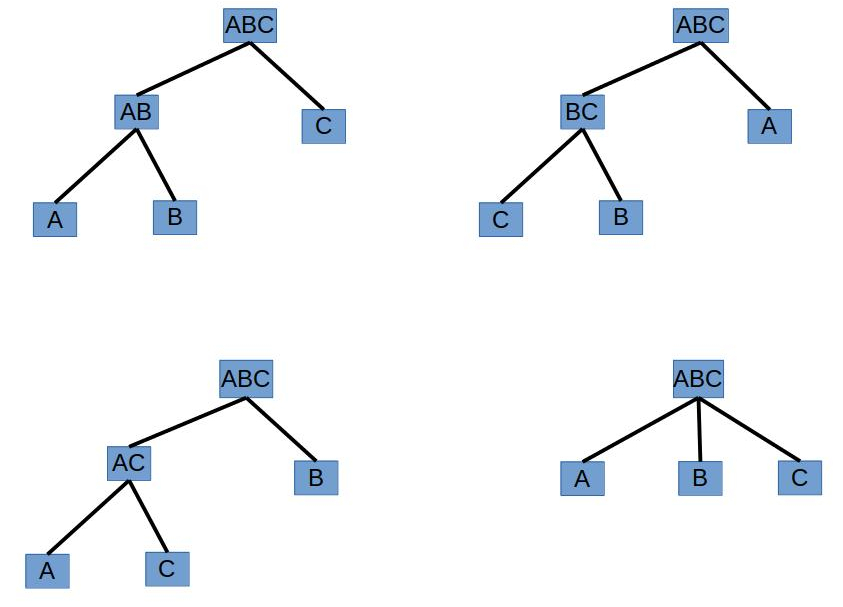}
\caption{\scriptsize \textbf{DR Plans for arbitrary $n$ and $k=3$}: 
All possible ways to combine
solutions of smaller assemblies to solve larger assemblies with three input
\rmc s. $A$, $B$, and $C$ are input \rmc s. $AB$, $AC$, $BC$, and
$ABC$ denote the output of assembly for respective input \rmc s
(see text in Section \ref{sec:AlgorithmVariantnk}).}
\label{fig:DRPlans}
\end{figure}

For arbitrarily large $n$ and $k$, some decomposition based method is required
to tractably deal with the size of the assembly landscape. The concept of
decomposition and recombination plans or \emph{DR-plans} is useful in this
context \cite{bib:HoLoSi98b,bib:HoLoSi98c,mvs2006}. Generally there are
several ways to combine the solutions to sub-assemblies. The overall complexity of the
recombination process depends primarily on the maximum number of sub-assemblies combined at 
a stage. \figref{fig:DRPlans} shows all the different ways of combining
sub-assemblies into larger assemblies, when we are given $k=3$ \rmc s $A$, $B$,
and $C$ as input.  In general, we choose a DR-plan where the number of
sub-assemblies being combined at any given stage is at most 12.  Now, the third
method from Section \ref{sec:AlgorithmVariantkg2} can be used at each stage.
In particular, if the number of sub-assemblies being combined at any given stage is
at most 2, as in the first three cases of \figref{fig:DRPlans}, then one of the
first two methods of Section \ref{sec:AlgorithmVariantkg2} can be used.

\begin{figure*}[htpb]
\centering
\subfigure[]{\label{fig:SimplicialComplex}\includegraphics[scale=0.25]{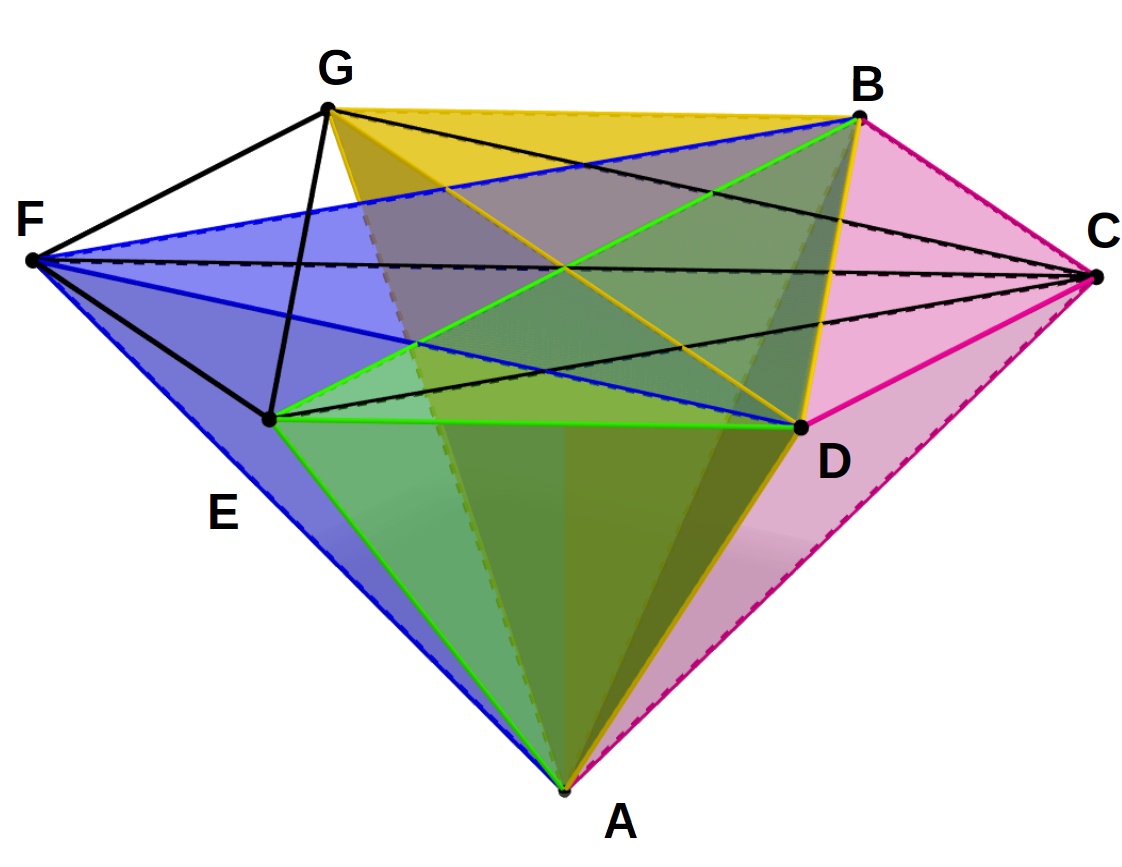}}
\subfigure[]{\label{fig:hyperstaticScheme}\includegraphics[scale=0.25]{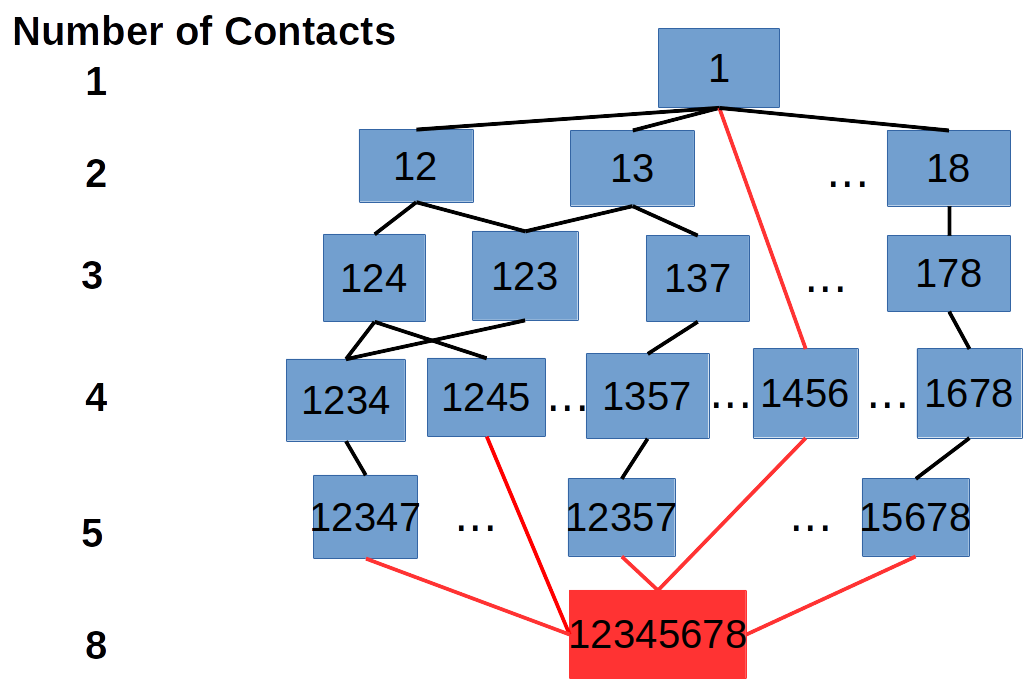}}
\caption{\scriptsize \textbf{Potential Energy Basin Structure}:
(a) Illustration of the structure (up to 3D regions) of a basin in an assembly
landscape.  Vertices B through F represent constraints between atoms
and the vertex A is a dummy vertex used for visualization. The basin
corresponds to the 0D region whose constraints are BCDEFG.  Tetrahedra
ABCD (resp. ABDG, ABDF, and ABDE) represents the 3D active
constraint region with constraints EFG (resp.  CEF, CEG, and CFG).  The
triangle ABD is a common 2D boundary region of these 3D regions with
one extra constraint. Its constraint set is CEFG. The 1D region AD with
constraint set BCEFG is a boundary of five 2D regions, all of which
have one fewer constraint (see text in Section \ref{sec:methods:BasinStructure}).
(b) Schematic illustration of different types of non-generic hyperstaticity.
Each box represents an active constraint region and the numbers inside the
boxes represent the set of active constraints. Nodes are shown for
active constraint regions whose graphs have constraint `1' in them.
The red box represents a hyperstatic node with $>d_\mathcal{A}$ constraints.
The red edges indicate hyperstatic edges, connecting nodes
that differ by more than one constraint.
See Section \ref{sec:methods:BasinStructure}.
}
\label{fig:Basin}
\end{figure*}

\subsection{Algorithm Extensions}
\label{sec:methods:immediateResults}
This section describes algorithms to compute several direct byproducts of the atlas:
(1) structure and neighborhood relationships of potential energy basins, 
(2) paths between active constraint regions and basins,
(3) approximate basin volumes, (4) path probabilities 
and (5) entropy integrals.

\subsubsection{Algorithm for Potential Energy Basin Structure}
\label{sec:methods:BasinStructure}
The lowest potential energy region of a basin or basin bottom is typically, but
not always, an effectively 0-dimensional active constraint region $R$, with
graph $G$, that splits into finitely many basins whose bottoms are rigid configurations 
satisfying the constraints in $G$. The potential
energy \emph{basin} (basins) corresponding to $R$ include portions of all ancestor active
constraint regions whose graphs are non-trivial subgraphs of $G$.
\emph{Hypostatic basins} have higher dimensional active constraint regions as
`bottoms', i.e., their active constraint graphs are maximal with no additional
constraints possible, but their configurations are flexible.  In this manner the
assembly landscape is partitioned into potential energy basins with clear
neighborhood relationships.  Free energy of a configuration depends both on the
potential energy and the weighted relative volume (entropy) of its potential
energy basin.

\figref{fig:SimplicialComplex} schematically uses a 6-simplex to illustrate the
structure of a potential energy basin (up to 3D regions). Specifically, it illustrates how
multiple active constraint regions (3D-0D) interact to form the basin. In the
figure, vertices B through F represent constraints between atoms and A is a
dummy vertex used for visualization. Each of the 20 tetrahedra represents a 3D
active constraint region, with Tetrahedron ABCD representing the region in
which the constraints E, F, and G are active. Each of the four colored
tetrahedra share a common triangle ABD, whose active constraints are C, E, F,
and G, i.e., the triangle 2D region ABD is a boundary of each of the colored
tetrahedral 3D regions formed by adding one more constraint to each of their
active constraint graphs.

\figref{fig:SimplicialComplex} shows how active constraint regions interact in
a generic setting. \figref{fig:hyperstaticScheme} shows a non-generic setting.
In addition to non-generic hypostatic basins previously discussed, there is a
further type of non-genericity where more than one
constraint becomes active at the same time. 

\figref{fig:hyperstaticScheme} 
schematically illustrates a hyperstatic regions and potential energy basins.
Each box represents an active constraint region and the numbers inside the
boxes represent the set of active constraints. The figure shows a portion of
the roadmap for nodes that have the active constraint `1' in them. Generically,
when going from a parent to child regions, only one more constraint gets added
to the set of active constraints.
However, in some
non-generic cases (shown as red edges), multiple constraints get added 
simultaneously. For example, the edge between 1 and
1456, causes a drastic reduction in energy/effective dimension from parent to
child when compared to a generic case. It also creates a higher potential
energy barrier when going from a region to its parent as multiple bonds need to
be broken simultaneously. Recalling $d_\mathcal{A}$, the ambient 
dimension of the assembly problem, generically, basins correspond to active constraint
regions with $d_\mathcal{A} =  6(k-1)$) constraints with rigid configurations,
but in hyperstatic basins more than $d_\mathcal{A}$ constraints could become
active (e.g., red node at the bottom with 8 constraints).

Crucially, since the core algorithm recursively generates the roadmap from
interior to boundary (higher to immediately lower energy/dimension), the
hypostatic basins are guaranteed to be found.  Hyperstaticity - when multiple
constraints become active simultaneously - is easily handled by the core
algorithm by creating clearly marked dummy nodes that fill out the intermediate
energy/dimension levels of the stratification, ensuring that only one extra
constraint is added in any parent child relationship.  These dummy nodes are
not included in the count of active constraint regions (a landscape design
variable). Results for potential energy basin structure are presented in
Section \ref{sec:results:potentialEnergy}.

\subsubsection{Finding Paths between Active Constraint Regions}
\label{sec:methods:ConfPaths}
\begin{figure}[htpb]
\centering
\includegraphics[scale=0.4]{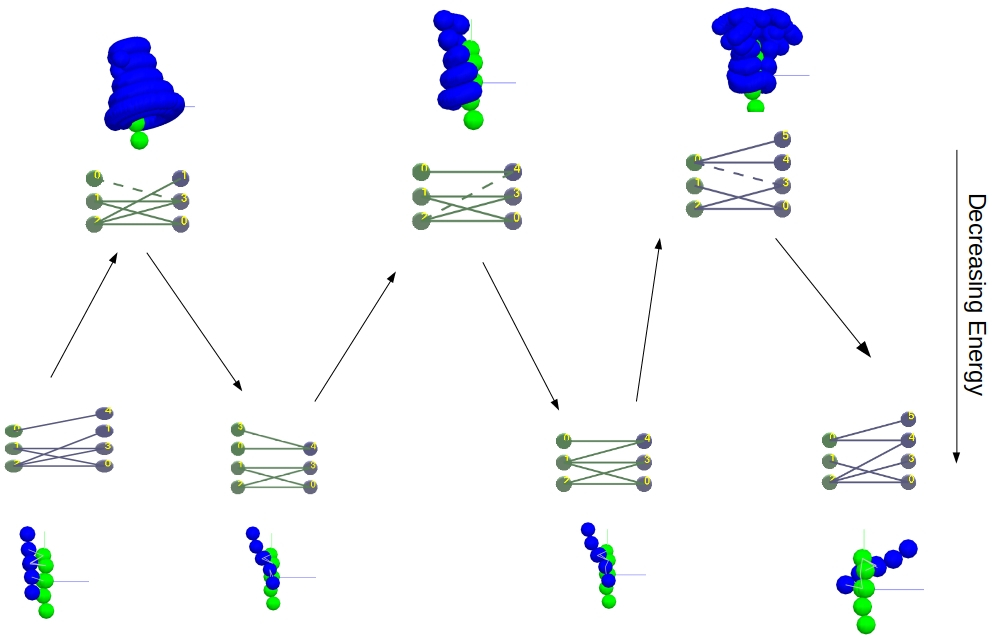}
\caption{\scriptsize \textbf{Paths between Active Constraint Regions}: A path 
in a toy atlas of two \rmc s shown at the bottom, connecting two 0D active 
constraint regions (left to right, active constraint graphs shown). 
The path only traverses regions with at least 5 constraints.
The arrows form a path, losing or gaining an active constraint, from the source 
to the destination regions. The sweep of configurations of the flip containing 
the configurations below, is shown next to each region (see text in Section 
\ref{sec:methods:ConfPaths}.)}
\label{fig:paths}
\end{figure}

From the generated roadmap, it is extremely fast to find (shortest) paths (or
number of paths of a given length) below a given energy level, between any two
active constraint regions.  Of particular interest is finding paths between two
0D active constraint regions that generically correspond to basin bottoms.
Specifically, we are interested in paths through 1D regions (with one higher
degree of freedom and one fewer constraint). These regions represent a generic
one degree of freedom motion path (see \figref{fig:paths}). 

Moving from one 0D region to another via a 1D region involves releasing a
constraint and making the same or a different constraint active. This
corresponds to crossing a local energy barrier. Breaking more than one bond at
a time means crossing a larger energy barrier. Either one (shortest) path or
all of them between two given active constraint regions are found by performing
a standard breadth first search through the roadmap DAG. Alternatively, the
number of paths of a given length can be found by performing a matrix
multiplication on the adjacency matrix of the DAG restricted to the relevant
energy/dimensional regions (see results in \ref{sec:results:ConfPaths}).

\subsubsection{Approximate Basin Volume Computation}
\label{sec:methods:approximateVolume}
To compute the volume of a potential energy basin for a given `bottom'
active constraint region $R_G$ with constraint graph $G$, we first identify all
active constraint regions $R_{G'}$ with $G' \subset G$. Not all configurations
in $R_{G'}$ contribute to this basin since $R_{G'}$ is part of all basins with
bottom regions $R_H$ with $G' \subset H$. Hence we locate those configurations
in $R_{G'}$ that are `close' to some boundary region $R_{G_i}$ with $G'
\subset G_i \subseteq G$.  For all $G' \subset G$, the total number of these
configurations, is the approximate volume of the potential energy basin.

To find the configurations in $R_{G'}$ that are close to a boundary region
$R_{G_i}$, we use a method of partitioning the set of configurations of any
parent region into sub-regions one for each child boundary region.  For
example, a given 5D region can be partitioned into sub-regions, one for each of
its 4D children.  The sizes of the sub-regions are proportional to the relative
sizes of the boundaries corresponding to children, since the size of the
boundary is a proxy for the probability that the new constraint corresponding
to that boundary becomes active. 

Let $b_i$ be the size of the $i^{th}$ boundary region $R_{G_i}$ in the parent
$R_{G'}$, based on the number of sampled configurations in $R_{G'}$. Then, the
relative size of the $i^{th}$ boundary  is given by $p(i) = b_i/\Sigma_i b_i$
and the number of configurations that lead to a boundary region $R_{G_i}$ is
then the product of $p(i)$ and the number of configurations in $R_{G'}$.

While different Cayley parameters choices for $R_{G'}$ could lead to different
sizes of boundaries, the relative sizes of the boundaries are independent of
the choice of Cayley parameters.  Nevertheless, using Cayley configurations, as
opposed to Cartesian configurations, to compute these quantities only provides
an approximate basin volumes (see results in Section \ref{sec:results:Volumes}).

\subsubsection{Computation of Path Probabilities}
\label{sec:methods:PathProb}
The probability of a path between two active constraint
regions is the product of the probabilities of each directed edge in the path.
The directed edges are either edges in the roadmap DAG or their reverse.
For a directed edge from a region $R$ to $R'$ in the roadmap DAG, the probability
is the relative size of the boundary corresponding to $R'$ within $R$. This is
computed as described in Section \ref{sec:methods:approximateVolume}.

The following example shows how to compute the probability on the reverse edges.
Consider a simplified atlas which has two active constraint regions $R_1$ and $R_2$
that are root nodes in the roadmap DAG, i.e., they have no ancestors. 
Let $\gamma$ (resp. $1-\gamma$) be the probabilities of a configuration being in $R_1$ (resp. $R_2$);
$\gamma$ can be computed as the relative volumes of $R_1$ and $R_2$ using the same Cayley parameters.
If $R_1$ and $R_2$ are not root node regions, $\gamma$ can be computed by treating $R_1$ and $R_2$ 
as bottoms of basins and computing their relative volumes as described in the previous section, but
using common Cayley parameters.

Let $R_3$ and $R_4$ be children of both $R_1$ and $R_2$.
Let $\alpha$ be the probability of the roadmap DAG edge $R_1R_3$ and let $\beta$ be the probability
of the roadmap DAG edge $R_2R_3$, computed as described above. 
Since $R_3$ and $R_4$ are the only children of $R_1$ and
$R_2$, the probabilities on the edges $p(R_1R_4) = (1 - \alpha)$ and $p(R_2R_4) = (1 -
\beta)$. Thus, the probabilities of a configuration being in $R_3$ and $R_4$ are given as follows.

\begin{align*}
p(R_3) &= \gamma \cdot \alpha + (1 - \gamma) \beta\\
p(R_4) &= \gamma \cdot (1 - \alpha) + (1 - \gamma) (1 - \beta)
\end{align*}

Let $a$ be the unknown the probability of a configuration in $R_3$ having come from $R_1$ on 
the addition of an active constraint, which is treated as the probability of ending up in $R_1$ after a constraint is released.
Therefore the probabilities $p(R_3R_1) = a$, and $p(R_3R_2) = (1-a)$.
Similarly, take $p(R_4R_1) = b$ and $p(R_4R_2) = (1-b)$. Now we can solve for unknowns $a$ and $b$
using the following two equations.

\begin{equation*}
\begin{aligned}
[\gamma \cdot \alpha + (1 - \gamma) \cdot \beta] \cdot a + &\\
[ \gamma(1 - \alpha) + (1-\gamma) (1 - \beta)]\cdot b &= \gamma\\
\end{aligned}
\end{equation*}

\begin{equation*}
\begin{aligned}
[\gamma \cdot \alpha + (1 - \gamma) \cdot \beta] (1 - a) +&\\ 
[\gamma(1 - \alpha) + (1-\gamma) (1 - \beta)] (1-b) &= (1 - \gamma)
\end{aligned}
\end{equation*}

Once $a$ and $b$ are solved for, the probability of the path $R_3R_1R_2R_4$, for example, 
can be computed as $a\beta b$ and
the probability of path $R_4R_3R_1R_2$ can be computed as $b\alpha a$.

Using the computation of the partitioning probabilities described above ($\alpha$, $\beta$ and
$\gamma$, $a$, $b$ in the above example), and the algorithm for
finding paths between active constraint regions (described in Section
\ref{sec:methods:ConfPaths}), probabilities for atlas paths can be computed. 

\subsubsection{Computing Cartesian Integrals using the Cayley Region}
\label{sec:methods:exactVolume}
\noindent\textbf{Metric on modulus space} A single flip $f$ of a Cartesian active
constraint region of dimension $d$ is a $(d+6)$-dimensional manifold $\Mn^d_f$ in
$R^{3k}$. Six of the dimensions represent rigid body motions that should not
enter the metric. Fortunately, every Cartesian configuration $\bx \in R^{3k}$
output by our methodology represents a unique orbit $\overline{\bx}$ of the
manifold modulo the special Euclidean group $SE(3)$ of rigid body motions 
(for those readers familiar with algebraic topology, we note that the
branched covering map gives a bijection between each Cartesian flip and its
corresponding Cayley image).

The distance of two orbits $\overline{\bx}_i, \overline{\bx}_j$ with orbit
representatives $\bx_i, \bx_j$ can then be defined as the minimal Euclidean
distance between Cartesian configurations from the orbits:

\[
   g(\overline{\bx}_i, \overline{\bx}_j) 
   := \argmin_{T_1, T_2 \in SE(3)} \left\|T_1 \bx_i - T_2 \bx_j\right\|.
\]

For two Cartesian configurations $\bx_a, \bx_b$, the projection $\bx^a_b$
of $\bx_b$ on the local tangent space of $\bx_a$ is

\[
   \bx^a_b := T^a \bx_b \text{ where } 
   T^a := \argmin_{T \in SE(3)}
   \left\|T \bx_b - \bx_a\right\|.
\]

\noindent\textbf{Entropy Integral} The \emph{partition function} for a manifold can be expressed as the
integral\cite{Holmes-Cerfon2013}.

\begin{equation}
   z_f := \int_{\overline{\Mn^d_f}}
   h(\bx) I(\bx) \sqrt{\left\| \overline{g}\right\|} d\bx
\label{eq:entropy_int}
\end{equation}

\noindent where $h(x)$ is called vibration factor, $I(x)$ rotation factor and
$\overline{g}$ is the metric on $\overline{\Mn^d_f}$. To compute the integral using the Cayley parameter samples output by our
methodology, we convert Eq. \eqref{eq:entropy_int} into an integral on the Cayley parameters
$\yy$:

\begin{equation}
\label{eq:entropy_int_2}
z_f = \int_{\overline{\Mn^k_f}} h(f(\yy)) I(f(\yy)) \sqrt{det(J^T J)} d\yy
\end{equation}

\noindent where $J$ where is the Jacobian matrix when changing variables from Cartesian
coordinates $\bx$ to Cayley parameters $\yy$. To approximately calculate $J$
for each $i$, we find a Cartesian configuration $\bx_c$ with Cayley parameters $(\yy^c_i)$, we find the
Cartesian configurations $\bx_i$ of the Cayley parameters $(\yy^c_1, \dots, \yy^c_i +
\Delta \yy_i, \dots, \yy^c_k)$ and project the $\bx_i$ into the tangent space
of $\bx_c$ as $\bx^c_i$. Then 

\[
   J := \left[ 
   \frac{\bx^c_1-\bx_c}{\Delta \yy_1}, \frac{\bx^c_2-\bx_c}{\Delta \yy_2}, 
   \dots, \frac{\bx^c_k-\bx_c}{\Delta \yy_k} \right].
\]

To capture multiple flips and to account for $p$ different
isomorphs of the same \acgW (for example when the \rmc s are identical), 
the integral over the Cartesian active constraint region in
the fundamental domain of the roadmap is 
determined as $ z = p \sum_f z_f $. 
The Jacobian
matrix has been used to develop a sophisticated adaptive sampling method,
different from the one presented in this paper, that samples the Cayley region
to get a guaranteed uniform coverage of samples in the Cartesian region. This 
adaptive Jacobian algorithm, is the subject of a companion paper \cite{Ozkan2014Jacobian}.

\section{Results}
\label{sec:results}

Results demonstrating the core algorithm's capabilities and high computational
efficiency have been generated by a resource-light, curated opensource software
implementation EASAL \cite{Ozkan:toms} (Efficient Atlasing and Search of
Assembly Landscapes, see software \cite{easalSoftware}, video \cite{easalVideo}
and user guide \cite{easalUserGuide}). 

We present 4 types of results demonstrating features of the new methodology
that were described respectively in Sections \ref{sec:stratification},
\ref{sec:recursiveBoundarySearch}, \ref{sec:convexification}, \ref{sec:CA},
\ref{sec:methods:BasinStructure}, \ref{sec:methods:ConfPaths},
\ref{sec:methods:approximateVolume}, and \ref{sec:methods:exactVolume}.

(1) Landscape Design and Path Finding. Section \ref{sec:results:DesignVolume}
demonstrates the connection between input shape variables and the two landscape
design variables described in Section \ref{sec:designVariable}, namely the
number of active constraint regions and the average Cayley parameter range. 

Section \ref{sec:results:DesignVolume} shows how the volumes of active
constraint regions can be designed by varying the concavity of the \rmc s. 

The independence of the two landscape design variables and their precise effect
on the total assembly landscape volume is demonstrated in Section
\ref{sec:results:design}. 

Section
\ref{sec:geometricBarCode} demonstrates the effect of the two landscape design
variables on the volumes of the different dimensional (energy) strata, both of
the atlas and individual basins. Together the landscape design variables
provide a bar-code summarizing the atlas. In addition the effect of the input
shape variables on the bar-code is shown. 

Section \ref{sec:results:ConfPaths} shows the
efficiency of computing paths between active constraint regions once the roadmap
has been found.

(2) Recursive Search and Decoupling of roadmap generation from sampling
density. Section \ref{sec:results:decoupling} first shows the diminishing
returns of number of samples in finding the regions of the roadmap.

Section
\ref{sec:results:potentialEnergy} shows the use of recursive boundary search
for finding diverse types of basins and their bar-codes.

Section
\ref{sec:results:interiorPoint} quantifies how recursive boundary search
decouples both efficiency and accuracy of roadmap generation from sampling
density.

Section \ref{sec:results:RegionSpecific} quantifies how
region-specific parametrization aids in decoupling.

(3) Volume Computation for Potential Energy Basins. Section
\ref{sec:results:approximateVolume} 
demonstrates the efficiency of computing approximate basin free energy
i.e., volumes stratified by their dimension or energy.

Section
\ref{sec:results:ExactVolume} gives results on accurate volume and entropy
computations for the case of $n=1$ and $2<k<24$, i.e., cluster assembly from
identical spheres.

(4) Verifying the Theoretical Time Complexity from Section \ref{sec:CA}.
Section \ref{sec:results:complexity} tabulates computational experiments
verifying the time complexity analysis for the single threaded version of
EASAL, in particular showing the dependence of the sampling time on the 
landscape design variables and the input shape variables.

\subsection{Details of the Experiments}
We describe the experimental setup in Section \ref{sec:expSetup},
the main computations in Section \ref{sec:keyComputations} and the key measurements
in Section \ref{sec:keyMeasurements}.

\subsubsection{Experimental Setup}
\label{sec:expSetup}
The experiments were broadly run on two types of input.

\begin{figure}[htpb] \centering
\subfigure[]{\label{fig:6Straight}\includegraphics[scale=0.08]{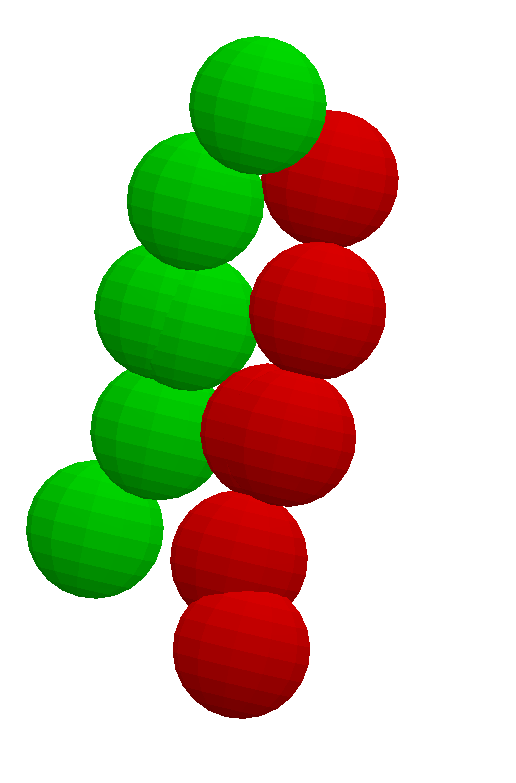}}
\subfigure[]{\label{fig:6Pocketed}\includegraphics[scale=0.08]{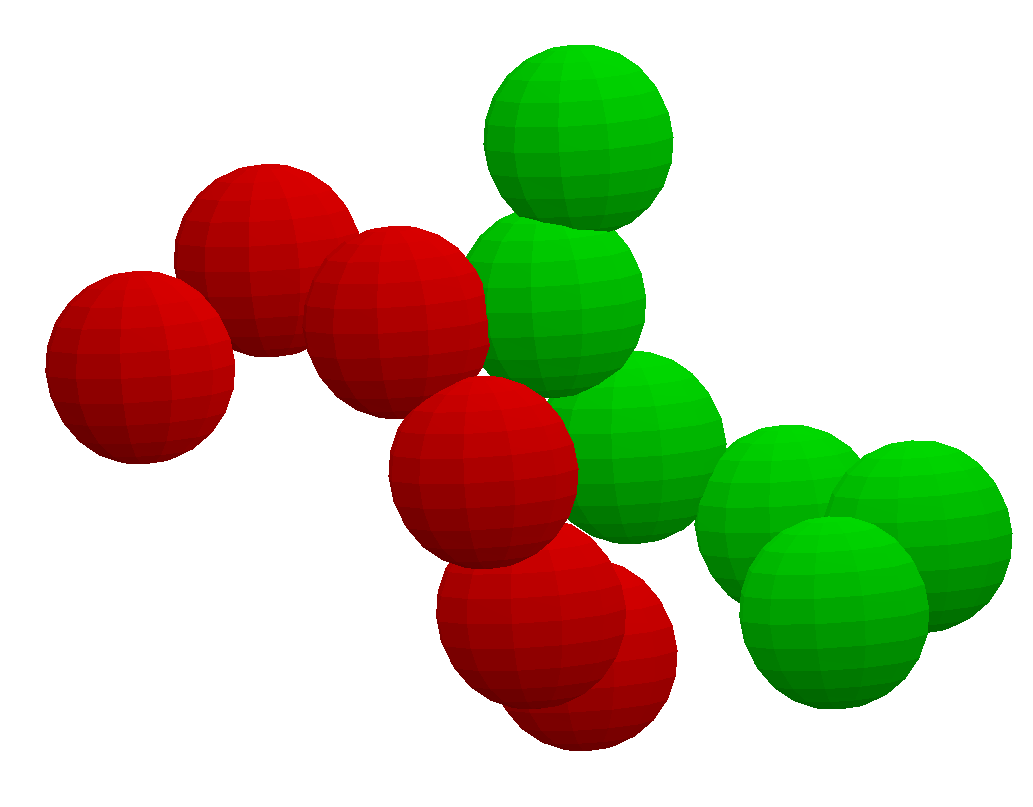}}
\subfigure[]{\label{fig:10Pocketed}\includegraphics[scale=0.08]{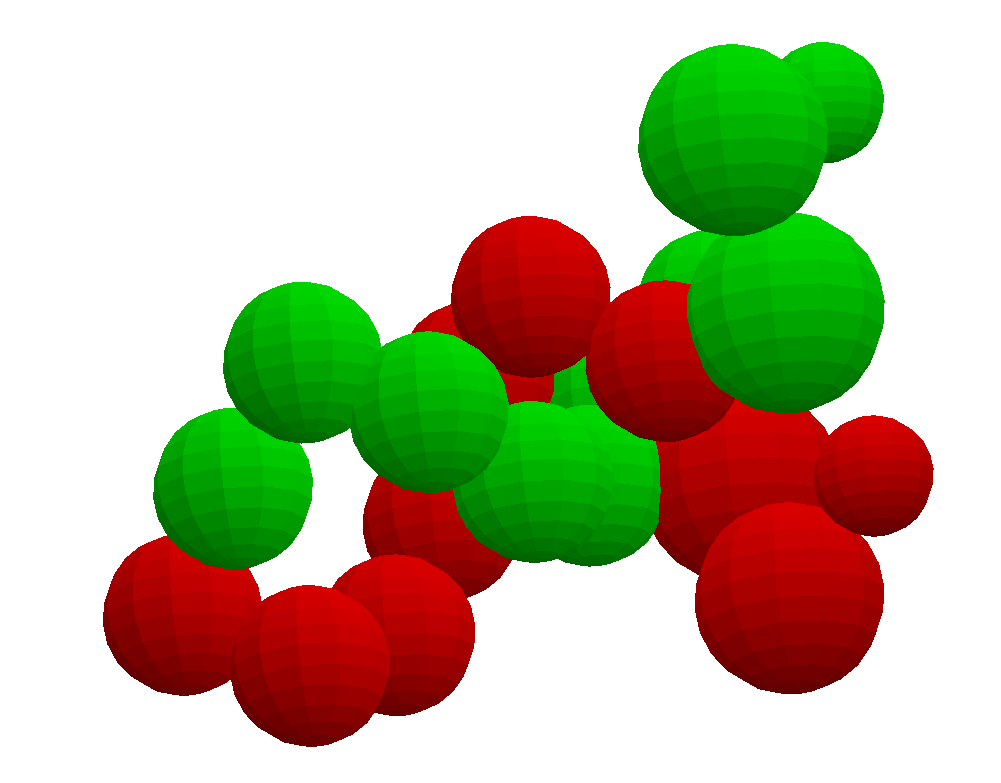}}
\subfigure[]{\label{fig:20Pocketed}\includegraphics[scale=0.08]{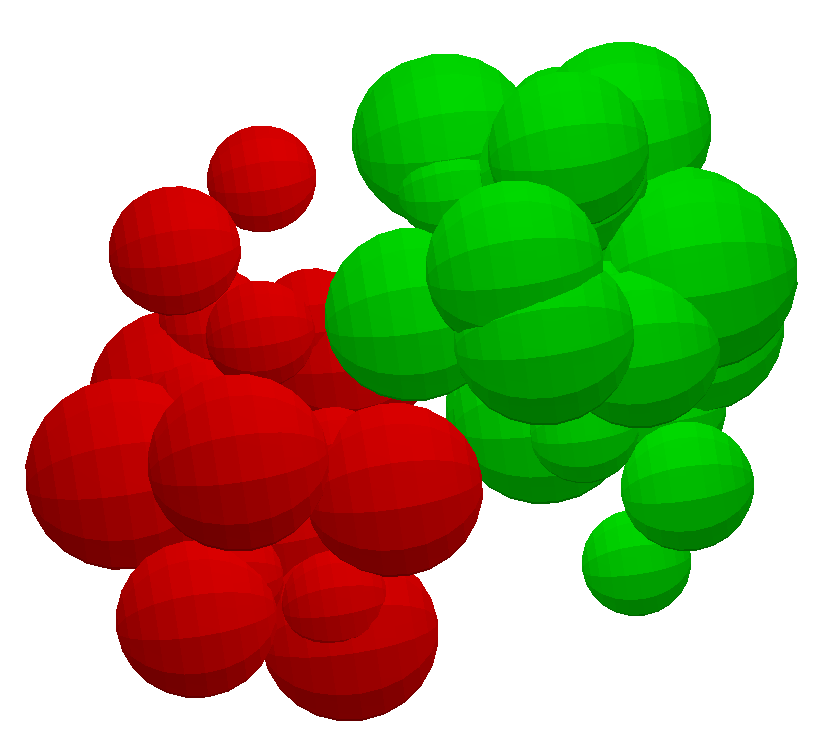}}
\subfigure[]{\label{fig:20Straight}\includegraphics[scale=0.08]{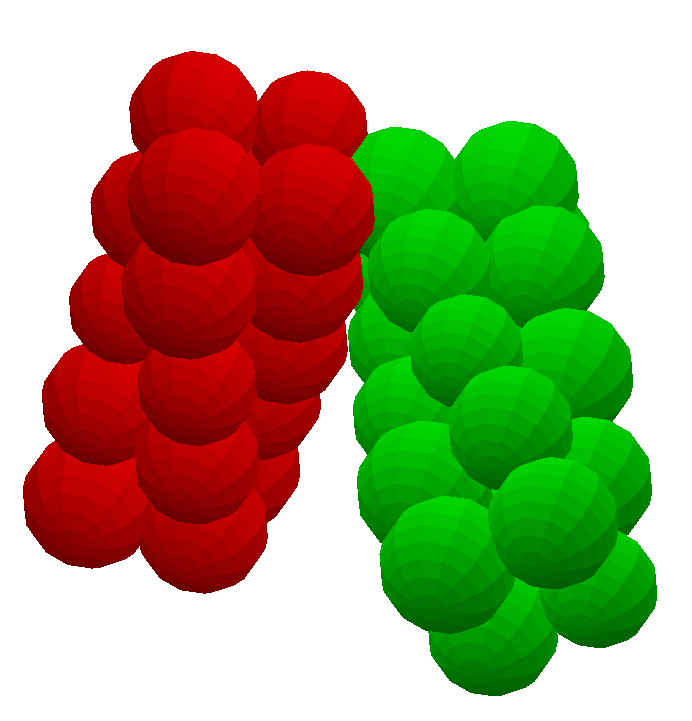}}
\subfigure[]{\label{fig:42Pocketed}\includegraphics[scale=0.08]{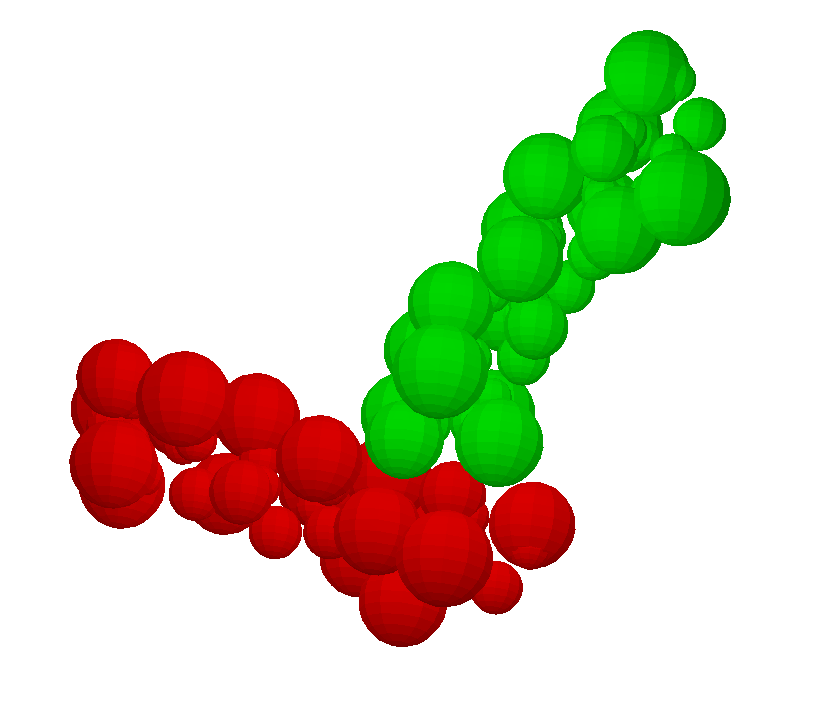}}
\caption{\scriptsize \textbf{List of \rmc s used in the experiments, illustrating different input 
shape variables}: (a) Narrow Convex (6 Atoms). (b) Narrow Concave (6
Atoms). (c) Narrow Concave (10 Atoms) (d) Wide Convex (20 Atoms). (e) Wide
		Concave (20 Atoms). (f) Wide Concave (42 Atoms) (see text in Section \ref{sec:expSetup}).}
\label{fig:inputMolecules}
\end{figure}

(1) The results for $k=2$ input \rmc s (macromolecular assembly,
Sections
\ref{sec:results:DesignVolume}, %\ref{sec:results:decoupling},
\ref{sec:results:ConfPaths},
\ref{sec:results:potentialEnergy}, 
\ref{sec:results:interiorPoint}, \ref{sec:results:RegionSpecific},
\ref{sec:results:approximateVolume}, and \ref{sec:results:complexity}) are set
up to cover a wide variety of geometric shapes (input shape variables,
including number of atoms, concavity and width, see
\figref{fig:inputMolecules}).  Although the proof of concept results here, are
restricted to $n=42$ atoms, EASAL has been used to analyze assemblies of \rmc s
with $n\approx 5000$ atoms \cite{Wu2012}.

The width of input \rmc s \figref{fig:6Straight} - \figref{fig:10Pocketed} is
comparable to the average diameter of their constituent atoms and hence
considered \emph{narrow}. The width of the rest of the input \rmc s
(\figref{fig:20Pocketed} - \figref{fig:42Pocketed}) are much greater than the
average atomic diameter and are hence considered \emph{wide}. 

The experiments were run on a laptop with Intel(R) Core(TM) i7-7700 @ 3.60GHz
CPU with 16GB of RAM. Suitable g++ compiler optimizations were used on the
EASAL code \rahul{(open source software available on Bitbucket at
https://bitbucket.org/geoplexity/easal) } to enhance its performance.  On average
52 million configurations were generated per hour.

(2) The results $n=1$ (spherical cluster assembly, Section
\ref{sec:results:ExactVolume}) are setup as follows. Each assembly system
consisted of $k$ identical singleton spheres. The experiments were run on a
laptop with Intel(R) Core(TM) i5-2500K @ 3.2GHz CPU with 8GB RAM. For $k\le12$,
on average, 4 million Cartesian configurations in the underlying fundamental
region of the roadmap (up to 3D active constraint regions) were generated per
hour.

\subsubsection{Main Computations}
\label{sec:keyComputations}
We perform 3 main computations. 

(1) Atlas Generation. For $k=2$ (shown in \figref{fig:inputMolecules}), we
generate the atlas for 10 typical randomly chosen 5D active constraint regions
and all their descendants. To be able to compare the results across \rmc s, we
fix the ratio of the average atom radius to the sampling step size $t$.  We
additionally fix the width of the Lennard-Jones' well, $\dhi_\ijx -\dlo_\ijx$
in Assembly Problem (\cone, \ctwo) for an atom pair $(a, b)$, with radii
$\rho_a$ and $\rho_b$, to $0.25 * ( \rho_a + \rho_b)$. Each of these input
assembly systems was sampled with 3 different values of $t$, to analyze the
effects of step size on the sampling time. The core algorithm in Section
\ref{sec:coreAlgorithm} is used to generate the atlas.

For $n=1$, the fundamental domain of the atlas was generated for 0D to 3D
active constraint regions. The algorithm variant in Section
\ref{sec:AlgorithmVariantn1} is used. 

(2) Basin Mapping and Volume Computation. The input is a previously generated
atlas and an active constraint region whose corresponding basin is to be
mapped. All ancestor regions that contribute to the basin are generated (by
taking subgraphs of the active constraint graph of the input region). To
compute the volume of the basin, the algorithm described in Section
\ref{sec:methods:approximateVolume} is used.

(3) Path Finding. The input is a previously generated atlas and 100 randomly
chosen pairs of active constraint regions. To find the shortest paths and the
number of paths of a given length, the algorithm in Section
\ref{sec:methods:ConfPaths} is used.

(4) Entropy integral. The input is a previously generated atlas ($n=1$). The
algorithm described in Section \ref{sec:methods:exactVolume} is used.

\subsubsection{Key Measurements}
\label{sec:keyMeasurements}
(1) The number of {\sl active constraint regions}, a 
landscape design variable, is measured both for atlases and basins.

(2) The {\sl average Cayley parameter range}, a second landscape design variable 
(defined in Section \ref{sec:designVariable}), is measured for 
atlases regions.

(3) The {\sl weighted samples} are the number of Cayley configurations sampled,
weighted by the number of their corresponding Cartesian configurations. This
is measured both for atlases and basins. 

(4) We measure the {\sl time} required for atlas, basin and path computation.

\subsection{Results on Designing Landscapes and Finding Paths} 
\label{sec:results:design}
To validate the discussion in Section \ref{sec:stratification}, we show how to
design assembly landscapes, by using the input shape variables, to alter the
landscape design variables.  We analyze the independence of the two landscape
design variables and their precise effect on assembly landscape volume.  We
show how volumes of active constraint regions are designed by changing the
average Cayley parameter range for the region, a landscape design variable.
Time required to compute both the shortest path and the number of paths of a
given length are tabulated. 

\subsubsection{Independence of the two Landscape Design Variables}
\label{sec:results:LandscapeVariables}
The goal of this experiment is to verify that (a) the independence of the two
landscape design variables, namely the number of active constraint regions and
the average Cayley parameter range; and (b) that the number of weighted
samples, a proxy for volume, is linear in the number of regions where the
linear factor depends only on the average Cayley parameter range and the
sampling step size.  This validates the discussion in Section
\ref{sec:designVariable}.
\begin{figure*}[htpb]
\centering
\subfigure[]{\label{fig:CRSa}\includegraphics[width=0.3\columnwidth]{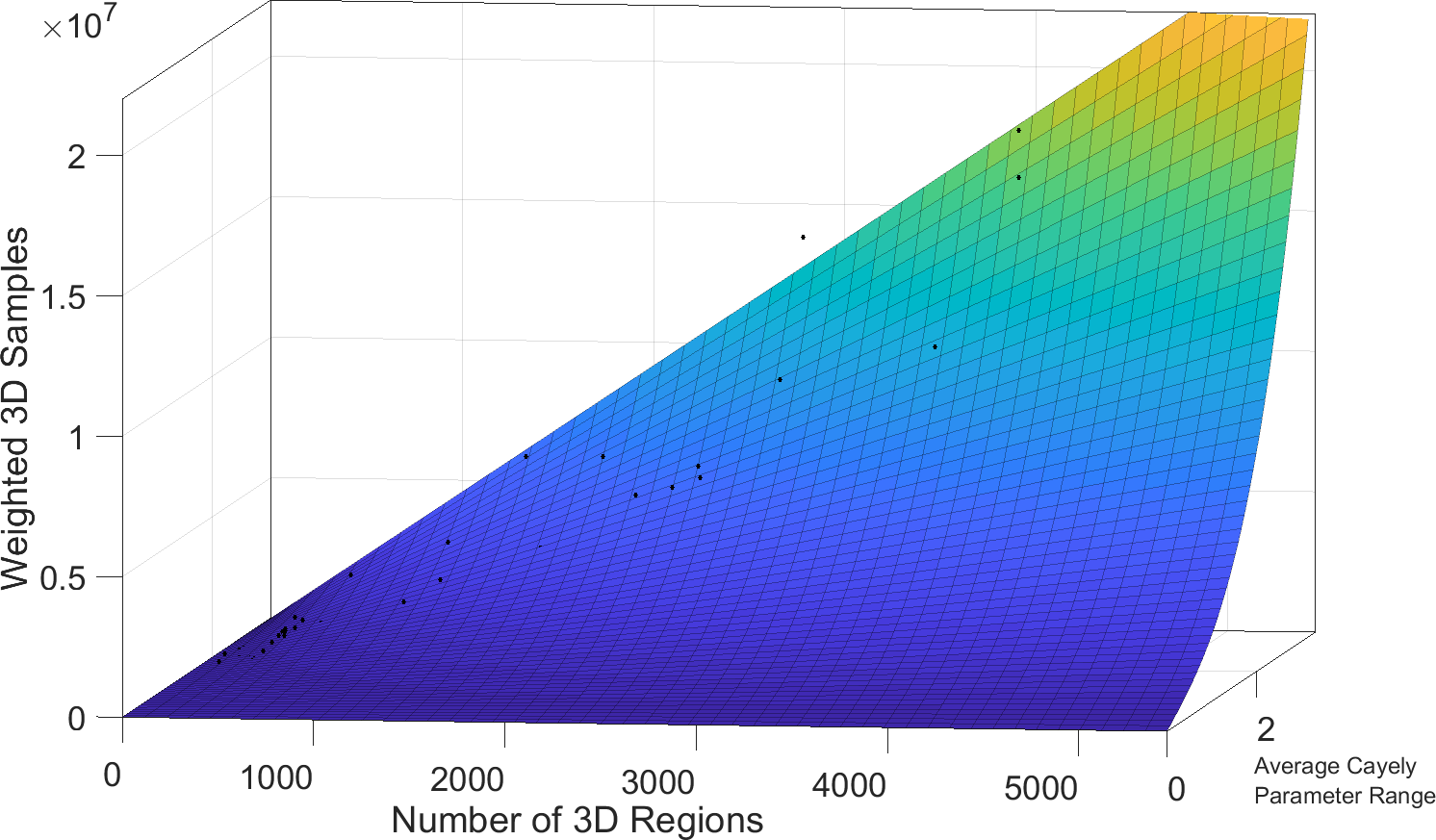}}
\subfigure[]{\label{fig:CRSb}\includegraphics[width=0.3\columnwidth]{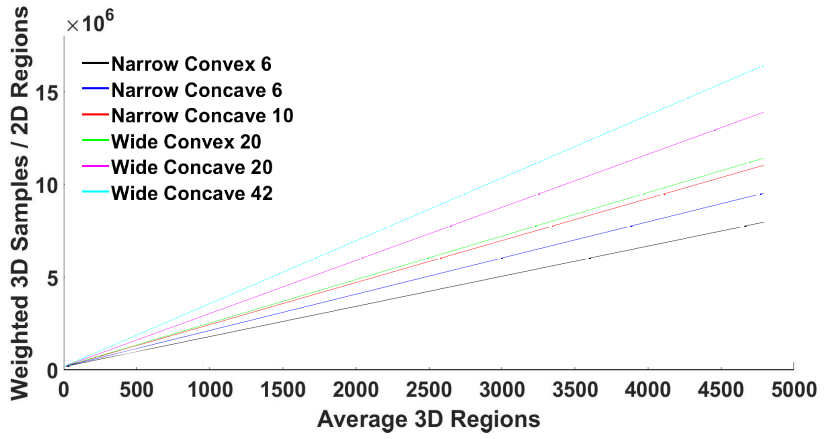}}
\subfigure[]{\label{fig:CRSc}\includegraphics[width=0.3\columnwidth]{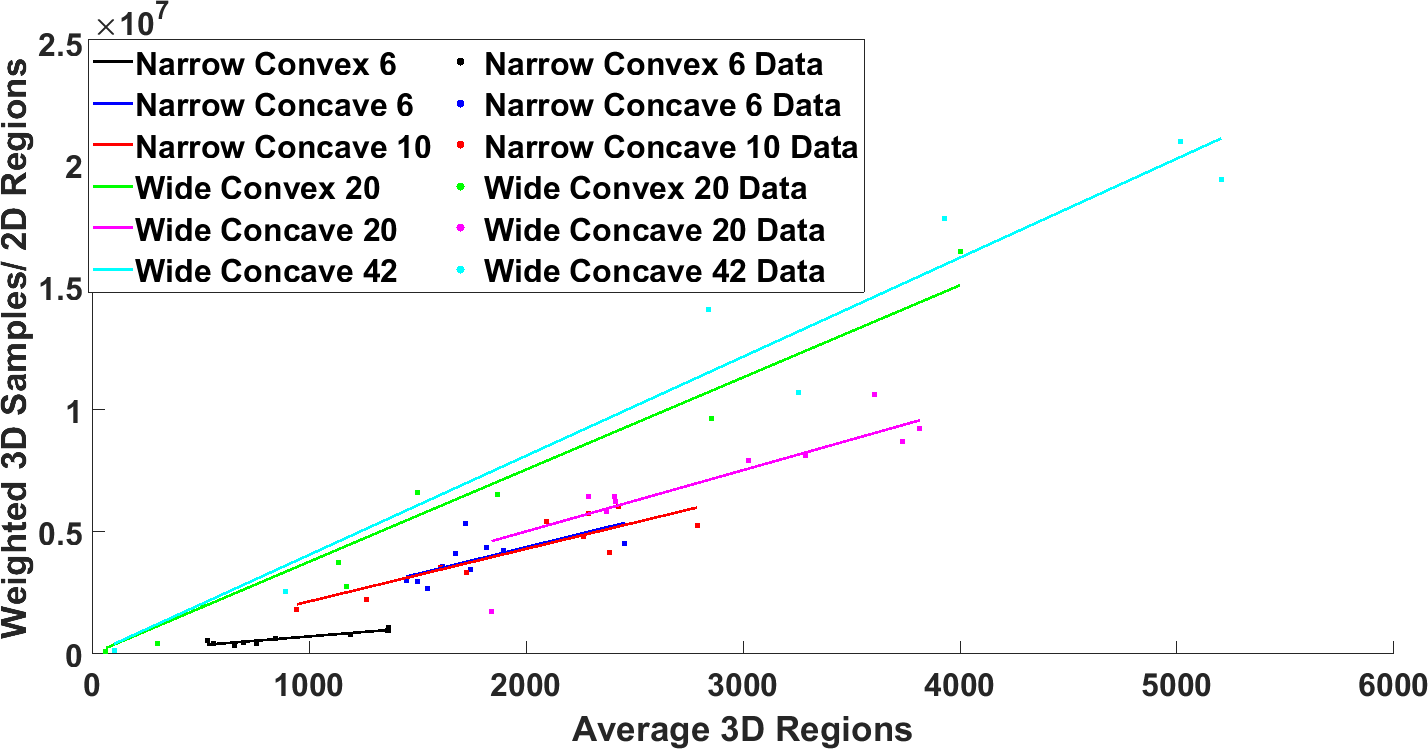}}
\caption{\scriptsize \textbf{Results Illustrating the Effects of Landscape Design Variables on the Atlas}:
(a) Plot of the average Cayley parameter value and the number of 
regions against weighted samples for 3D regions. These regions 
are the descendants of the 10 5D regions for the 6 input \rmc\ 
pairs described in Section \ref{sec:expSetup}. A bivariate 
function fit conforms to the prediction that the number of 
samples is cubic in average Cayley parameter range and linear 
in the number of regions. (b) Slices of the bivariate function 
in (a) at Cayley parameter values corresponding to the average 
3D Cayley parameter values of the different input \rmc\ pairs.
(c) Independently plots the number of 3D regions against the 
number of weighted 3D samples for the 6 input \rmc\ pairs, 
showing a match between observation and prediction. 
See text in Section \ref{sec:results:LandscapeVariables} for details.}
\label{fig:CRS}
\end{figure*}

\figref{fig:CRSa} plots the average Cayley parameter value and the number of 3D
regions against the number of weighted samples in those 3D regions. These
regions are the descendants of the 10 5D regions for the 6 input \rmc\ pairs
described in Section \ref{sec:expSetup}. Each point represents a single atlas
for one of the input \rmc\ pairs.

We expect the number of weighted samples to vary as the third power of the
average Cayley parameter range and linearly in the number of regions. As
predicted, \rahul{a } bivariate function that is cubic in the average Cayley parameter
range and linear in the number of active constraint regions shows a good fit to
the points.

\figref{fig:CRSb} shows 6 slices of the plot in \figref{fig:CRSa} taken at values 
corresponding to average 3D regions' Cayley parameter ranges, for each of the
6 input \rmc\ pairs.

Independently, for each input \rmc\ pair, \figref{fig:CRSc} plots the number
of 3D regions against the number of weighted samples in those regions.
As can be seen, the predicted plots (\figref{fig:CRSb}) and the
observed plots (\figref{fig:CRSc}) match except for the 20 atom wide concave and the
20 atom wide convex systems, which have switched order. We expect this is because
of the highly non-generic nature of the wide convex \rmc\ which has several
atoms with similar radii.

\subsubsection{Designing Volumes of Regions via Average Cayley Parameter Range}
\label{sec:results:DesignVolume}
\begin{figure*}[htpb]
\centering
\includegraphics[width=0.7\textwidth]{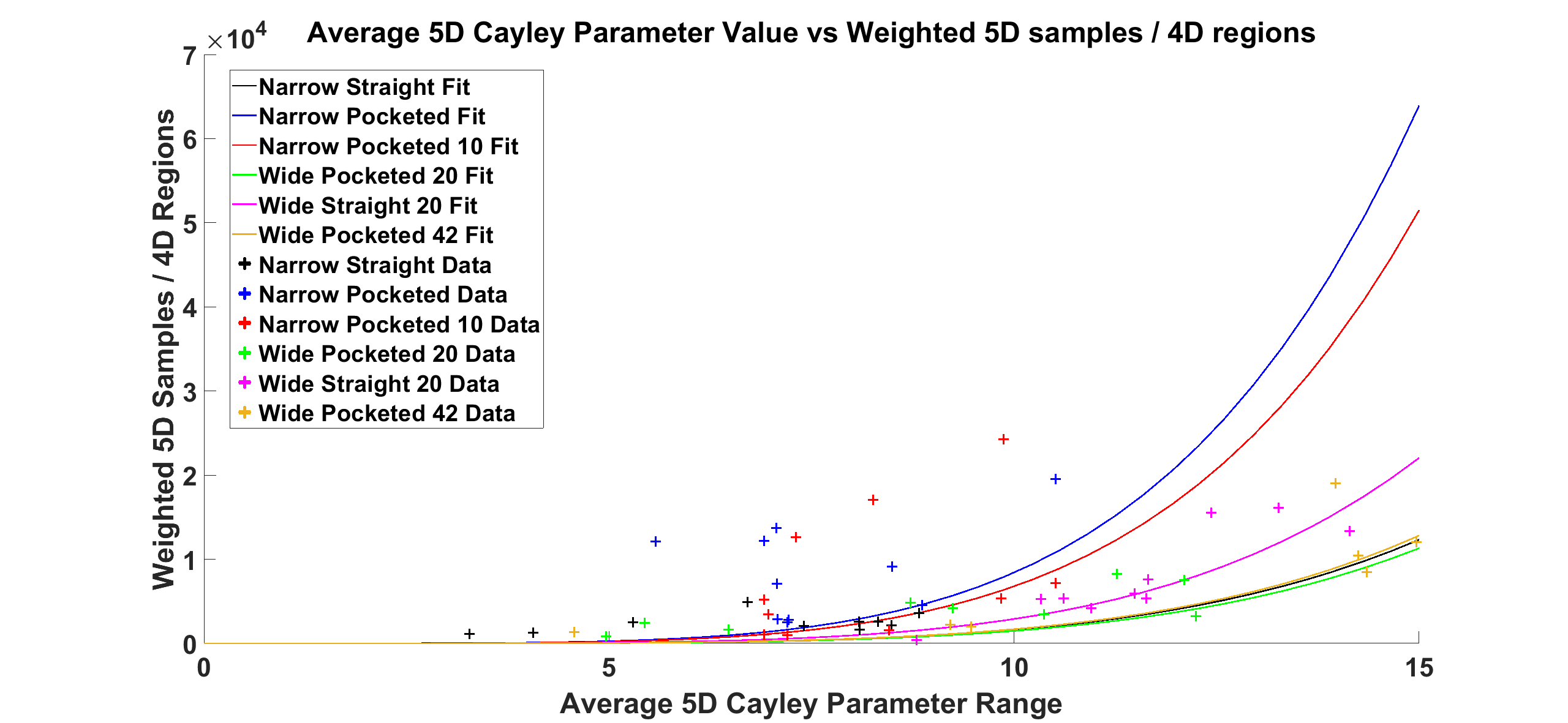}
\caption{\scriptsize \textbf{Effect of the Average Cayley Parameter Range on Volume}: 
Each point of a single color represents one of the 10, 5D active constraint 
regions; each color represents one of the six input molecular pairs shown in
\figref{fig:inputMolecules}.
Trend lines fit a $5^{th}$-degree polynomial curve
(see text in Section \ref{sec:results:DesignVolume}).
}
\label{fig:CayleyVsSamples}
\end{figure*}
%%%
The goal of the experiment is to verify that the volumes of active constraint
regions can be designed using the average Cayley parameter range, a landscape
design variable, which in turn can be tuned using the input shape variables.

However, once the Cayley parameter range is fixed, we expect the number of
weighted samples to be the same across all \rmc s regardless of their size and
their geometry.

We present results validating the discussion in Section
\ref{sec:designVariable}, by demonstrating that the number
of weighted Cayley samples (a proxy for volume) in an active constraint region
is proportional to its average Cayley parameter range raised to its dimension.

To be able to compare the effect of the average Cayley parameter range on
weighted samples across different \rmc s, we divide the average Cayley
parameter range by the average radius of the atoms in the input \rmc s.  In
addition, we correct for the extra samples created by boundary search.  While
finding the boundaries of an active constraint region, the EASAL software
performs binary search (described in Section \ref{sec:algorithms}), which leads
to the creation of more samples. The number of extra samples created in an
active constraint region due to binary search is proportional to the number of
children of that region. Hence we divide the number of weighted samples in the
active constraint region by its children.

Therefore in \figref{fig:CayleyVsSamples}, the x-coordinate represents the
normalized average Cayley parameter range value for that active constraint
region obtained by dividing it by the average radius of atoms in the input \rmc
s. The y-coordinate represents the number of weighted samples in the active
constraint region divided by the number of its children.

\figref{fig:CayleyVsSamples} shows trend lines of the variation of weighted
samples with the normalized average Cayley parameter range. It is clear that
the trend lines match the a $d^{th}$ degree polynomial fit from MATLAB, where
$d$ is the dimension of the active constraint region. The results are shown for
$d=5$.

\subsubsection{Using the Bar-Code of the Atlas}
\label{sec:geometricBarCode}
\begin{figure*}[htpb]
\centering
\subfigure[]{\label{fig:RVDNS6}\includegraphics[scale=0.3]{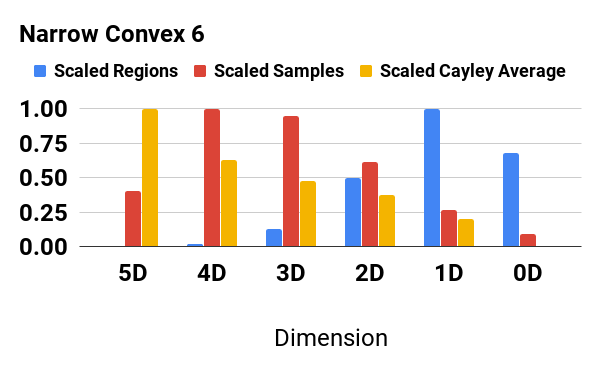}}
\subfigure[]{\label{fig:RVDNP6}\includegraphics[scale=0.3]{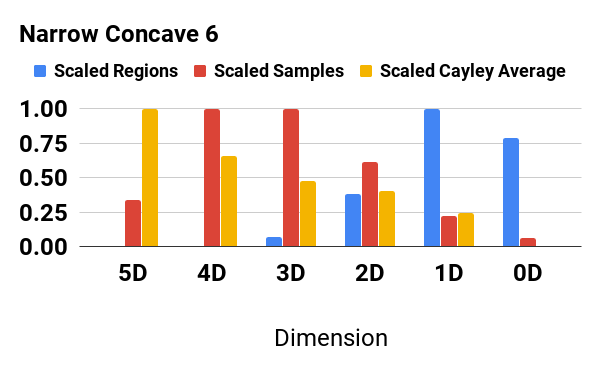}}
\subfigure[]{\label{fig:RVDNP10}\includegraphics[scale=0.3]{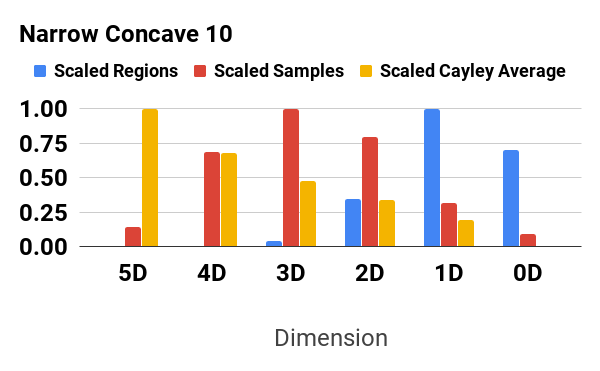}}
\subfigure[]{\label{fig:RVDWP20}\includegraphics[scale=0.3]{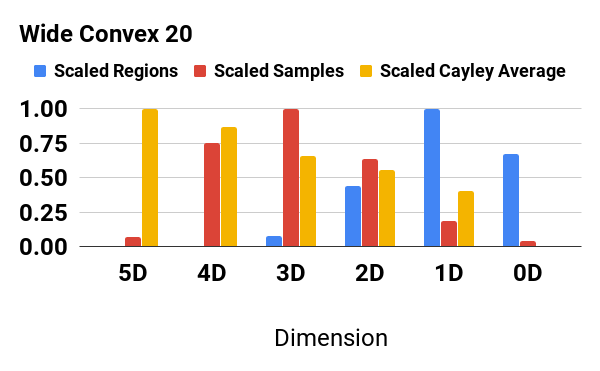}}
\subfigure[]{\label{fig:RVDWS20}\includegraphics[scale=0.3]{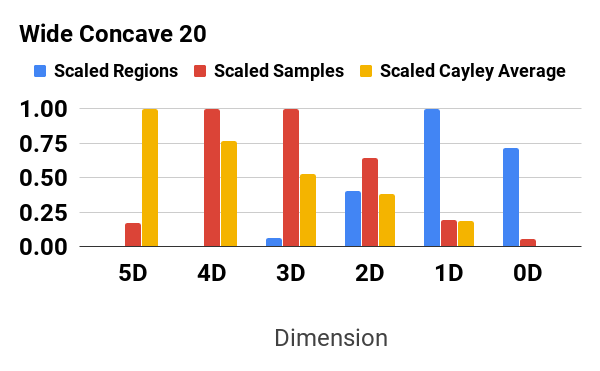}}
\subfigure[]{\label{fig:RVDWP42}\includegraphics[scale=0.3]{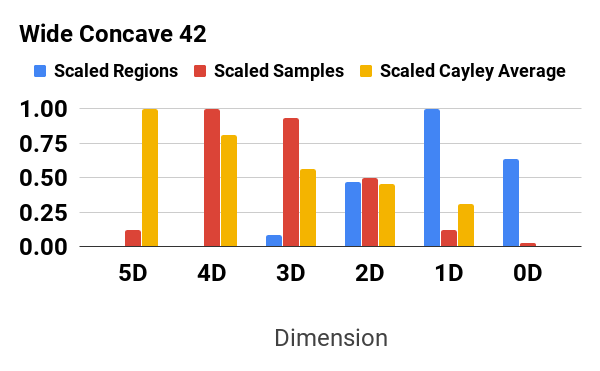}}
\subfigure[]{\label{fig:RVDWS20}\includegraphics[scale=0.18]{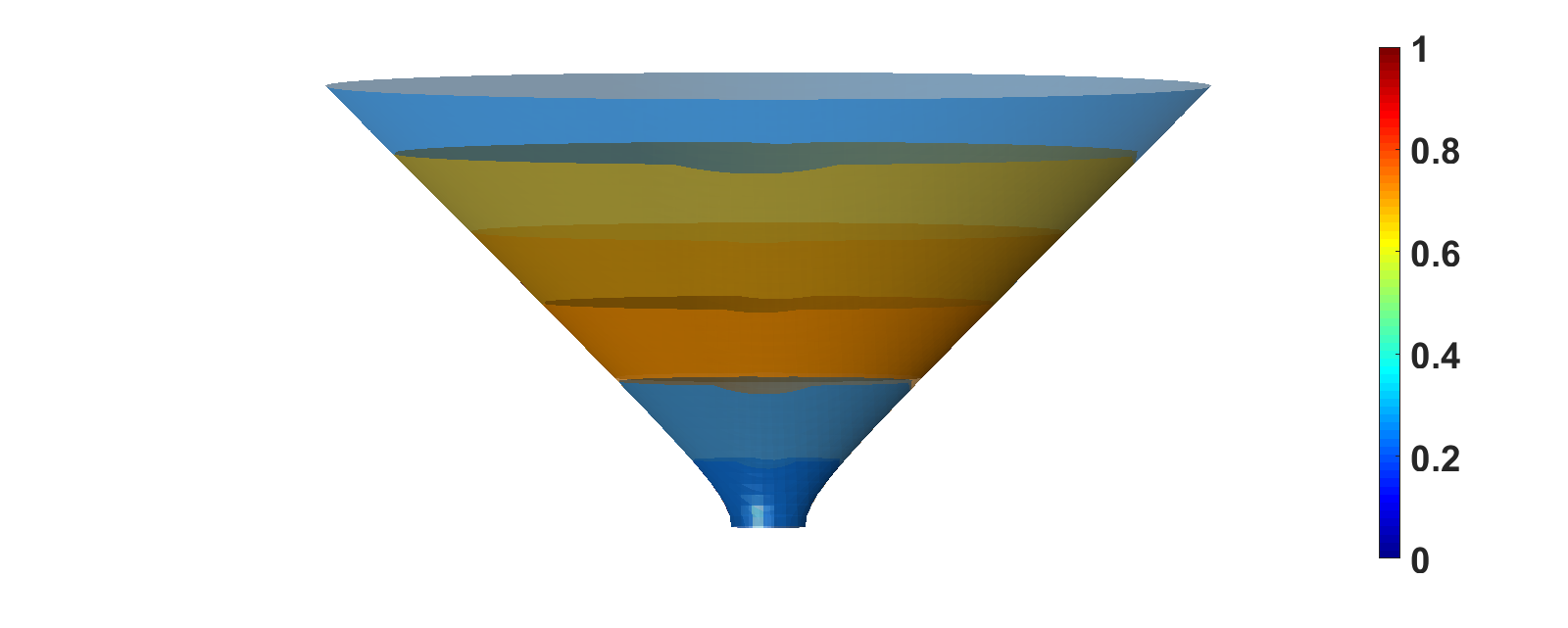}}
\subfigure[]{\label{fig:RVDWP42}\includegraphics[scale=0.18]{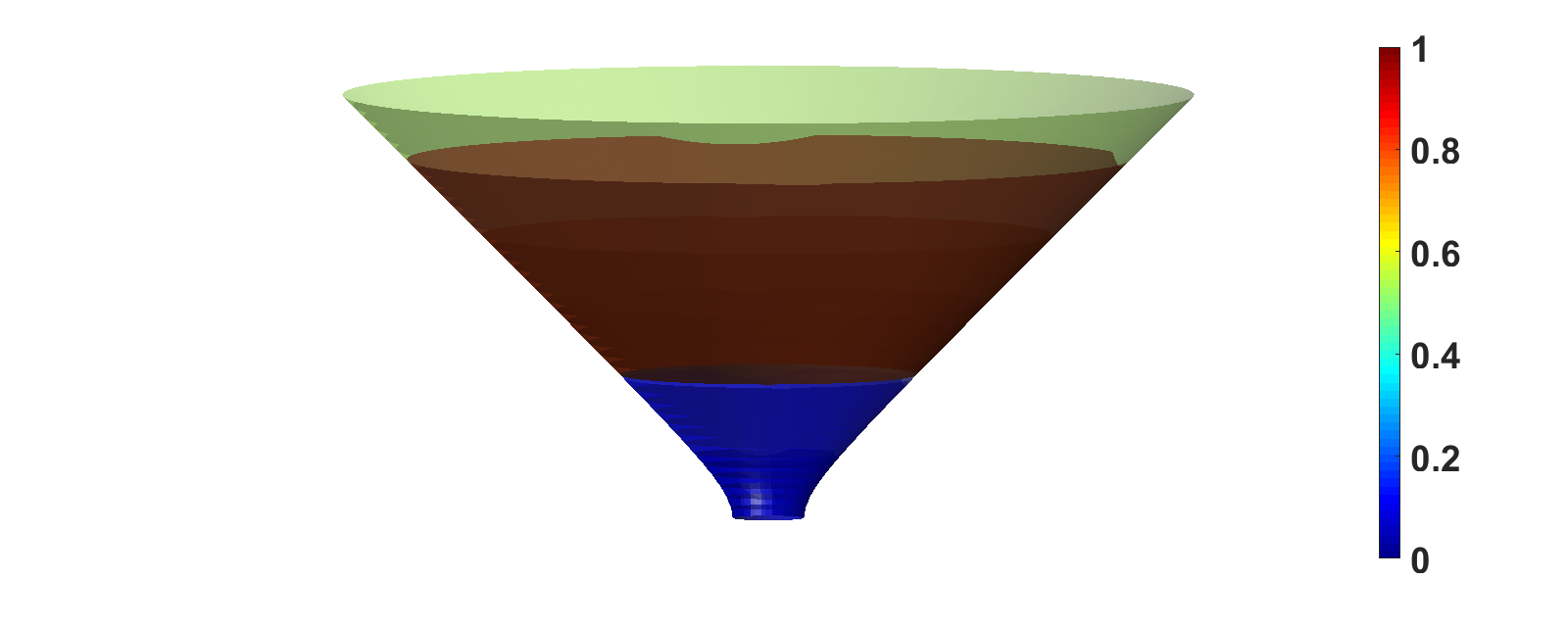}}
\caption{\scriptsize \textbf{Bar-Code of Output Atlas}: 
Figures	(a) through (f) compare the number of regions (scaled and shown in blue), the 
number of weighted samples (scaled and shown in red) and the average Cayley 
parameter range (scaled and shown in yellow)
in each dimension for different input \rmc s. (a) Narrow Convex 6, 
(b) Narrow Concave 6, (c) Narrow Concave 10, (d), Wide Convex 20 (e) 
Wide Concave 20, and (f) Wide Concave 42. As can be seen from the figure,
the profile of the histograms shown look similar.
Figures (g) and (h) show an analogous dimensional profile for a typical basin
of the 6 atom narrow convex system, i.e., the number of regions and the number of weighted samples
respectively. Each concentric section of a paraboloid depicts the number of regions (resp. 
the number of weighted samples) in different dimensions (0D at the bottom).
See text in Section \ref{sec:geometricBarCode}.}
\label{fig:RVsD}
\end{figure*}

This experiment demonstrates a surprising behavior of the succinct bar-code
(shown in \figref{fig:RVsD}) of Section \ref{sec:designVariable}. The bar-code
consists of (1) the normalized profiles of the two landscape design variables
and (2) the weighted samples of the output atlas or basin, both stratified by
dimension. The experiment shows that the bar-code does not change appreciably
with variations in the input shape variables of the 6 input \rmc\ pairs. This
is {\sl despite} the strong influence of input shape variables on the {\sl
absolute values} of the landscape design variables of the atlas as a whole, of
individual regions, and of dimensional strata, as demonstrated in the previous
two experiments.

In other ways, however, the bar-code behaves as expected. The highest number
of regions belong in the lower dimensional strata (0D, 1D, 2D) and a negligible
number belong in the higher dimensional strata (5D, 4D, and 3D). Despite this,
as expected, the total number of samples in higher dimensional regions is
significantly higher than those in lower dimensional regions, due to
exponential dependence on dimension. In addition, the higher dimensional
regions also have higher values of the average Cayley parameter range, which
decreases exponentially by dimension. The bar-code for the basin behaves
similarly to the bar-code of the atlas.

\subsubsection{Finding Paths from Roadmap}
\label{sec:results:ConfPaths}
The goal of this experiment is to tabulate the time that the algorithm of
Section \ref{sec:methods:ConfPaths} takes to find shortest paths and number of
paths between active constraint regions of previously generated atlas.  The
experiment was run on the atlases of the narrow convex 6 and the narrow convex
20 \rmc s. The results are summarized in Tables \ref{table:paths} and 
\ref{table:numPaths}.

\begin{table}
\centering
\begin{tabular}{|cccc|}\hline
$n$ & $r$ & Path Length& Time \\\hline
6& 176 & 7 & 1.9 ms \\\hline
6& 145 & 6 & 2.2 ms \\\hline
20& 787 & 18 & 119ms\\\hline
\end{tabular}
\caption{\scriptsize \textbf{Time to Find the Shortest Path between Active 
Constraint Regions}: 
The time on a standard laptop (see text), to find the 
shortest path between 100 pairs of randomly chosen 0D active constraint regions 
through other 1D and 0D active constraint regions (see Section \ref{sec:results:ConfPaths}).}
\label{table:paths}
\end{table}

\begin{table}
\centering
\begin{tabular}{|cccc|}\hline
$n$ & $r$ & $l$ & Time \\\hline
\multirow{3}{*}{6}
				&176 & 2 & 2.02 s\\
				&176 & 4 & 4 s\\
				&176 & 8 & 6.04 s\\
				&176 & 10 & 8.08 s\\\hline
\multirow{3}{*}{20}
				& 787 & 2 & 6 min\\
				& 787 & 4 & 11.58 min\\
				& 787 & 8 & 18.04 min\\
				& 787 & 10 & 27.44 min\\\hline
\end{tabular}
\caption{\scriptsize \textbf{Time to Find the Number of Paths between Active 
Constraint Regions}: The 
time on a standard laptop, to find the number of paths of 
length $l$, between all pairs of 0D active constraint regions 
in a toy atlas with $r$ total 0D active constraint regions (see Section \ref{sec:results:ConfPaths}).}
\label{table:numPaths}
\end{table}

\subsection{Results on Recursive Search and Decoupling}
\label{sec:results:decoupling}
We demonstrate the core algorithm's effective decoupling
of roadmap generation from sampling discussed in Section \ref{sec:advantagesRS}.
\begin{figure}[htpb]
\centering
\includegraphics[width=\columnwidth]{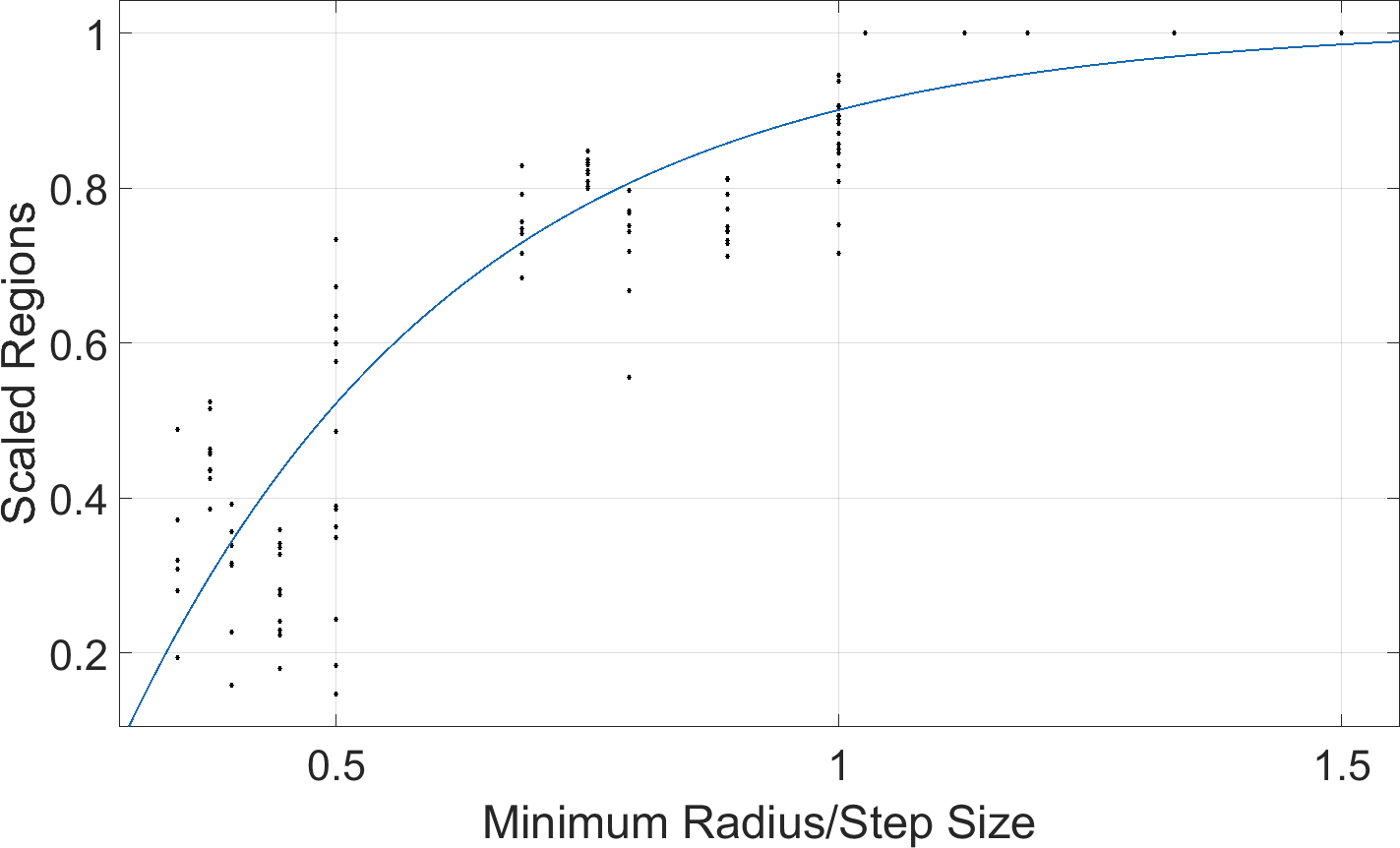}
\caption{\scriptsize \textbf{Decoupling Exploration from Sampling}: 
Plot of the ratio of the radius of the smallest atom in
the assembly system to the step size on the x-axis and on the normalized number of regions
discovered using this step size on the y-axis.
As the step size decreases, the number of regions discovered 
converges (see text in Section \ref{sec:results:decoupling}
for details).}
\label{fig:Convergence}
\end{figure}

\begin{figure*}[htpb]
\centering
\begin{tabular}{ccc}
\begin{minipage}{0.32\textwidth}
\subfigure[\scriptsize Generic Basin]{\bottominset{\label{fig:NormalBasin}\includegraphics[scale=0.06]{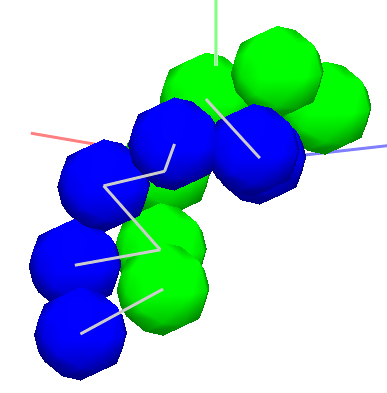}}
{\includegraphics[scale=0.18]{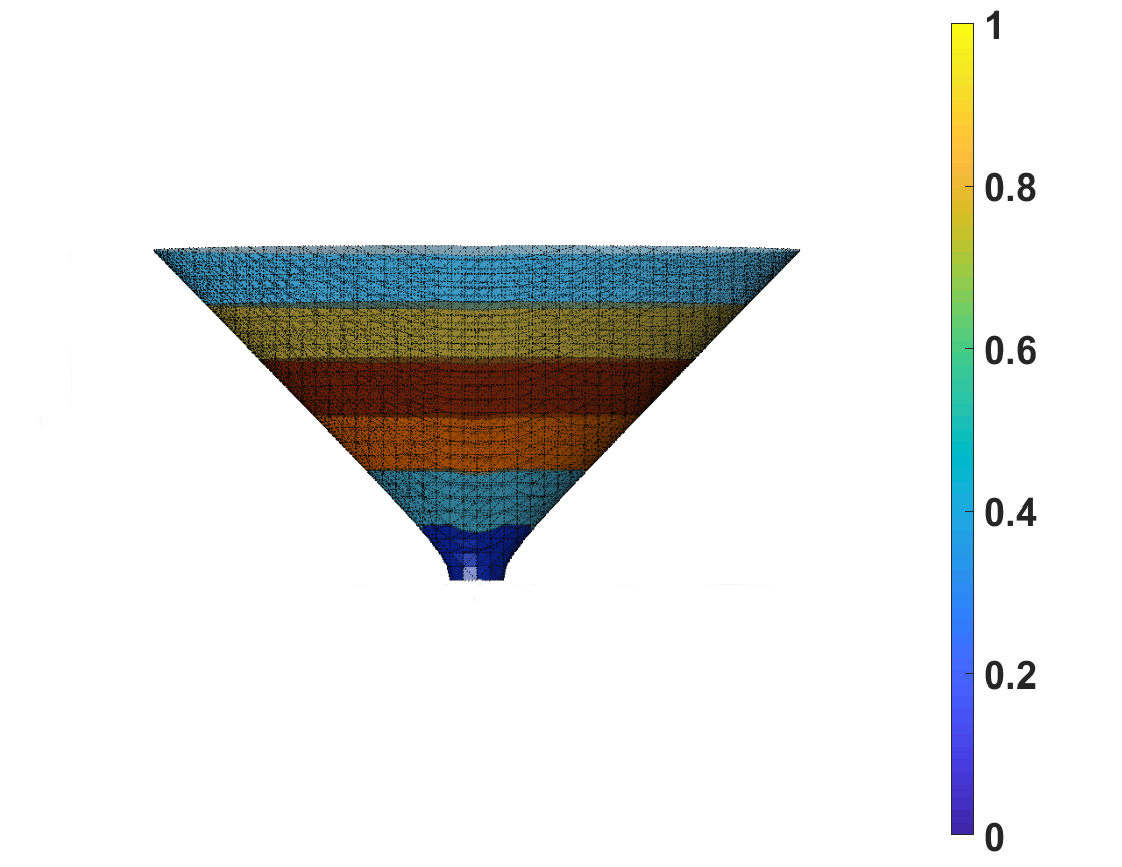}}{30pt}{18pt}}
\end{minipage}
\begin{minipage}{0.32\textwidth}
\subfigure[\scriptsize Hypostatic Basin]{\label{fig:HypoBasin}\bottominset{\includegraphics[scale=0.05]{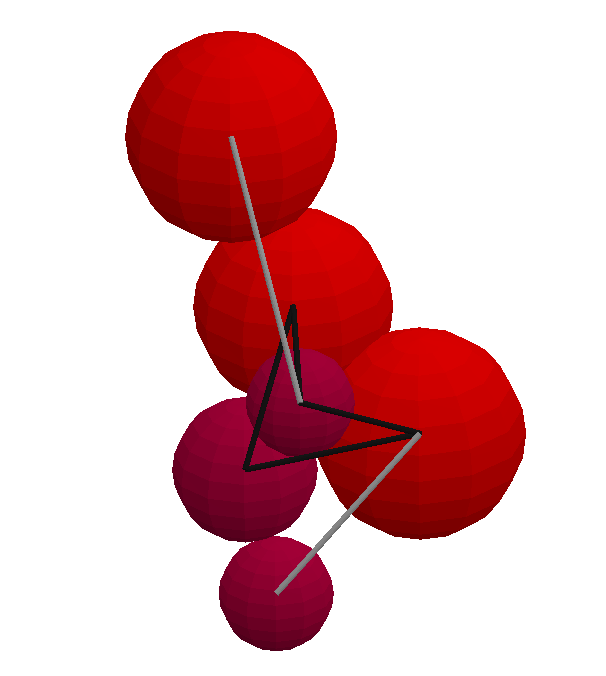}}
{\includegraphics[scale = 0.18]{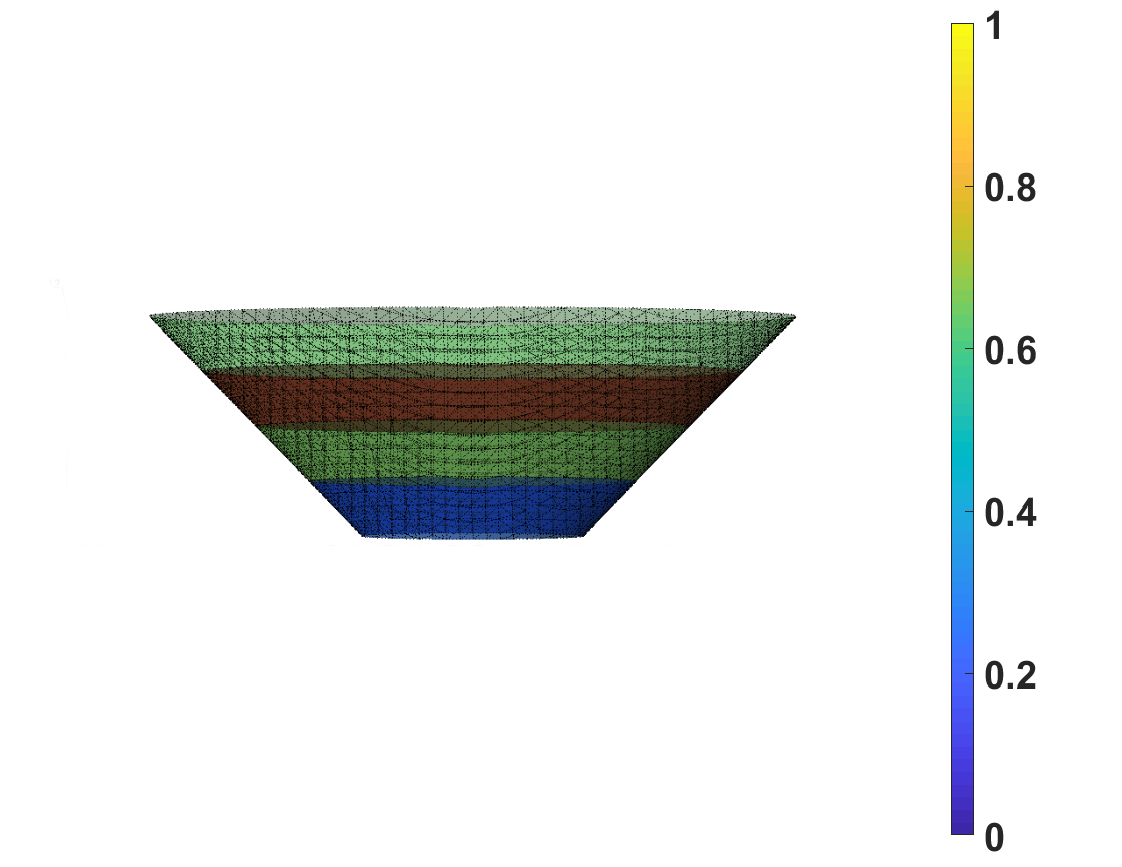}}{20pt}{30pt}}
\end{minipage}
\begin{minipage}{0.32\textwidth}
\subfigure[\scriptsize Hyperstatic Basin]{\label{fig:HyperBasin}\bottominset{\includegraphics[scale=0.04]{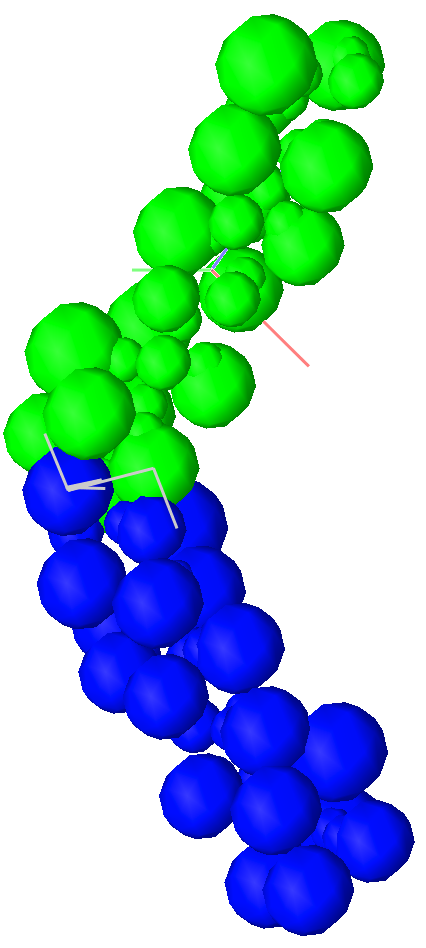}}
{\includegraphics[scale = 0.18]{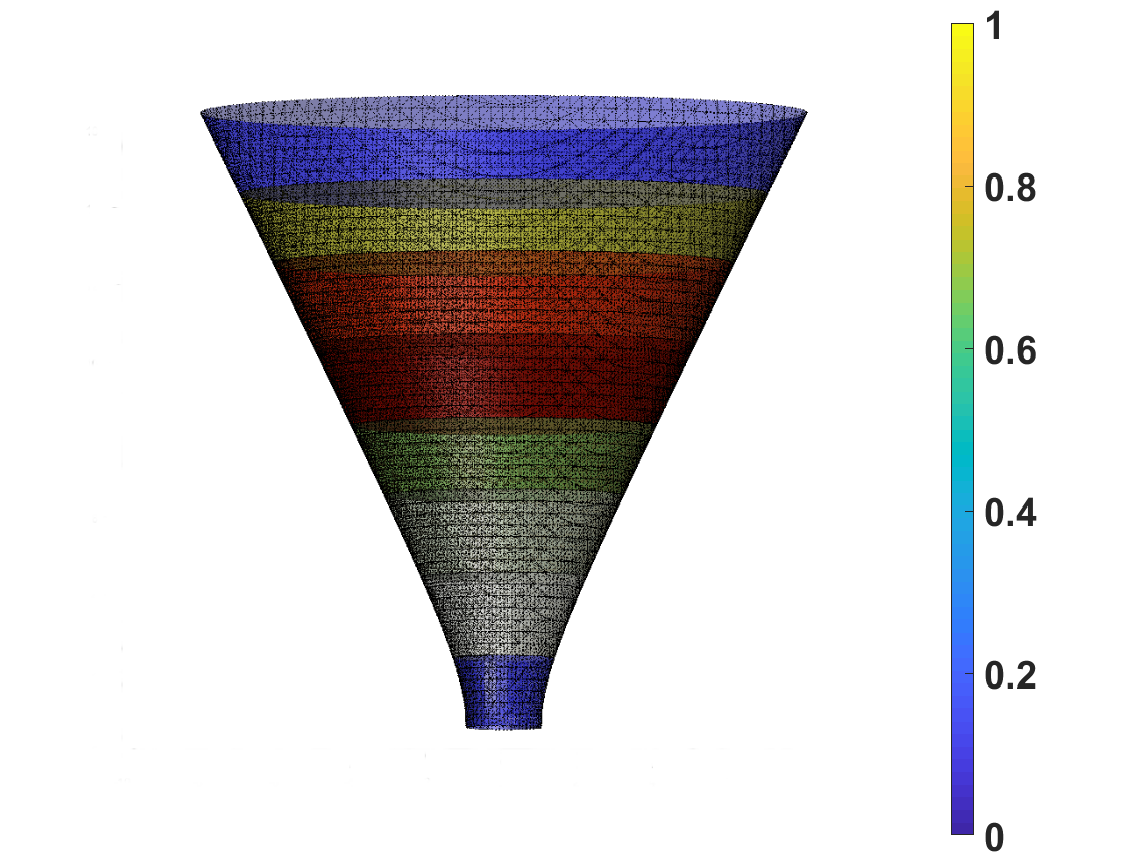}}{10pt}{20pt}}
\end{minipage}
\end{tabular}
\caption{\scriptsize \textbf{Bar-codes for 3 types of Potential Energy Basins}:
Bar-code showing the number of active constraint regions of different dimensions in
the potential energy basins of 3 different input \rmc\ pairs. 
The top most portion of the paraboloid shows the number of 5D
regions and each concentric layer shows the number of regions one dimension
lower. The inset figures show one Cartesian configuration at the bottom
of each potential energy basin. (a) A generic potential energy basin.
(b) A non-generic, hypostatic basin. (c) A non-generic, hyperstatic basin. 
See text in Section \ref{sec:results:potentialEnergy}.}
\label{fig:BasinRings}
\end{figure*}

The first experiment shows that decreasing the step size (a way to get better 
coverage of the configurational region), gives diminishing returns in terms of
the number of regions discovered. To be able to compare the phenomenon
across the 6 different input \rmc\ pairs, we normalized the step size by dividing
it by the radius of the smallest atom in the input \rmc\ pair.

In \figref{fig:Convergence}, we plot the ratio of the radius of the smallest
atom in the assembly system to the step size on the x-axis and on the
normalized number of regions discovered using this step size on the y-axis.
Each point represents one run starting from one of the 10 5D regions, for one
of the 6 input \rmc\ pairs, for one of the step sizes, as described in Section
\ref{sec:expSetup}. As the step size decreases the number of regions discovered
converges. Specifically, as the step size approaches the radius of the smallest
atom in the assembly system, the number of regions discovered does not
significantly increase. This supports the effectiveness of our methodology in
generating the roadmap with minimal sampling and thus decoupling roadmap
generation from sampling.

\subsubsection{Recursive Search for Locating and Mapping Basins}
\label{sec:results:potentialEnergy}
This experiment demonstrates that the recursive approach to roadmap generation
easily locates all three types of basins (normal, hypostatic, and hyperstatic)
and maps their structure using the algorithm in Section
\ref{sec:methods:BasinStructure}.

\figref{fig:BasinRings} schematically shows a portion of the basin bar-code, i.e.,
the number of ancestor regions of different dimensions that contribute to different
types of basins occurring in the assembly landscapes of 
of 3 different input \rmc\ pairs.

\figref{fig:NormalBasin} shows the basin bar-code for the 6 atom narrow concave
system, a generic basin.

Unlike the generic basins shown in \figref{fig:NormalBasin}, the basin bar-code
in \figref{fig:HypoBasin} does not have any 0D or 1D regions. The inset
\rmc\ shows a 2D configuration with 4 active constraints that is at the
bottom of its potential energy basin. Despite the configurations in this region
having 2 degrees of freedom, more constraints cannot become active due to the
special geometry of the \rmc s involved. 

\figref{fig:HyperBasin} shows the bar-code of a hyperstatic potential energy
basin. Notice that there are no 0D regions, and that the bottom of the basin
has more than 6 contacts.
\subsubsection{Decoupling aided by Recursive Boundary Search}
\label{sec:results:interiorPoint}
\begin{figure}[htbp]
\centering
\includegraphics[scale=0.4]{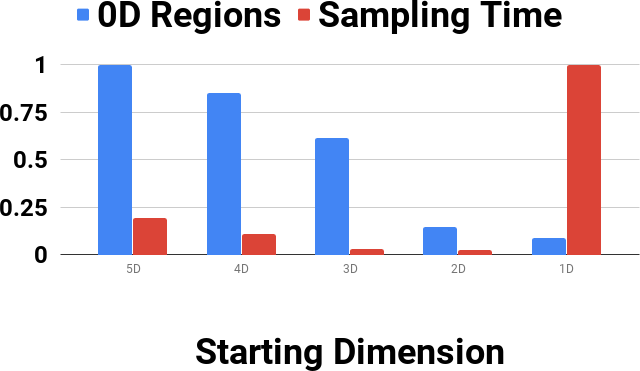}
\caption{\scriptsize \textbf{Results Quantifying Accuracy and 
Complexity of Decoupled Recursive Boundary Search}:
Comparing the scaled number of 0D regions discovered and the scaled time 
required to enumerate them in an assembly system with the 6 atom narrow 
concave \rmc\ as input. The x-axis shows the starting dimension of sampling. 
See text in Section \ref{sec:results:interiorPoint}. 
}
\label{fig:0D} 
\end{figure}
This experiment illustrates that when recursive search - starting from higher
dimensional (energy) interior to lower dimensional boundary - is decoupled from
sampling, it has a huge advantage over typical methods that remain in lower
dimensional (energy) regions, validating the discussion in Section
\ref{sec:advantagesRS}. 

We use the EASAL software to enumerate all the lowest potential energy 0D
regions in the assembly system with a pair of narrow concave (6 atoms)
\rmc\ as input. While EASAL typically starts sampling from the interior of
5D regions, in this experiment, we compare its time complexity and accuracy in
discovering 0D regions to alternatively starting the sampling from the interior
of 4D, 3D, 2D and 1D regions.

\figref{fig:0D} shows that when EASAL starts from 5D regions it discovers the
most number of 0D regions and takes a reasonable amount of time. The number of
0D regions found decreases as the starting dimension for sampling decreases.
The time required initially decreases with dimension as well, but dramatically
increases, once the starting dimension decreases to a turning point. This is
due to the combinatorial blow up of the number of possible starting regions,
most of which do not have valid configurations, being low dimensional. Thus,
starting from the interior of a higher dimensional (energy) region and
recursively finding lower dimensional boundary regions is most effective at
finding all minimum potential energy regions. 
\subsubsection{Region Specific Parametrization for Decoupling and Volumes}
\label{sec:results:RegionSpecific}
\begin{figure*}[htpb]
\centering
\subfigure[]{\label{fig:E1vsE3Samples}\includegraphics[width=0.45\columnwidth]{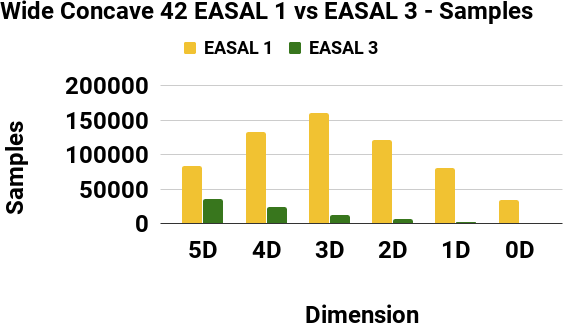}}
\subfigure[]{\label{fig:E1vsE3Regions}\includegraphics[width=0.45\columnwidth]{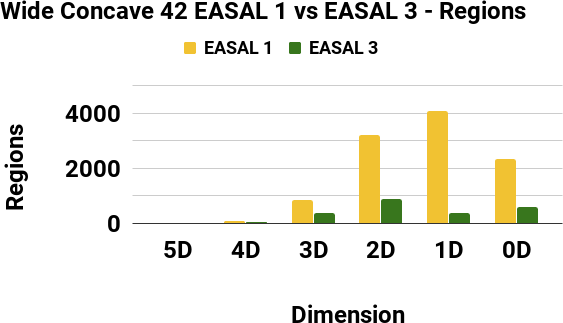}}
\caption{\scriptsize \textbf{Results Quantifying the Advantages of Region-Specific Parametrization}: 
Comparing the output of EASAL-1 (with region-specific parametrization) and 
EASAL-3 (without region-specific parametrization) for 42 atom wide concave system.
(a) EASAL-1 discovers more regions than EASAL-3. (b) EASAL-1 discovers more 
configurations than EASAL-3. See text in Section \ref{sec:results:RegionSpecific}.}
\label{fig:E1vsE3}
\end{figure*}
The results in this section demonstrate the advantages of region-specific
parametrization both in finding active constraint regions and in computing
volumes, validating the discussion in Section \ref{sec:boundaryDetection}.

The first experiment quantifies the advantage of sampling each lower
dimensional boundary region using a region-specific parametrization instead of
just using a single parametrization of a region to sample its lower dimensional 
boundary regions. 

\begin{figure*}[htpb]
\centering
\subfigure[]{\label{fig:pSpaceLL} \includegraphics[scale=0.35]{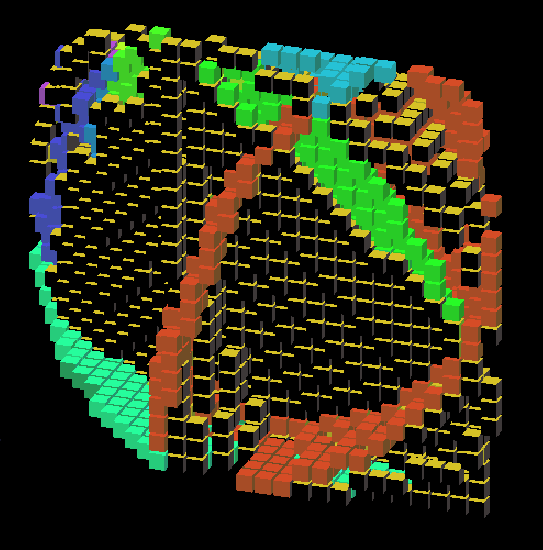}}
\subfigure[]{\label{fig:pSpaceLR} \includegraphics[scale=0.35]{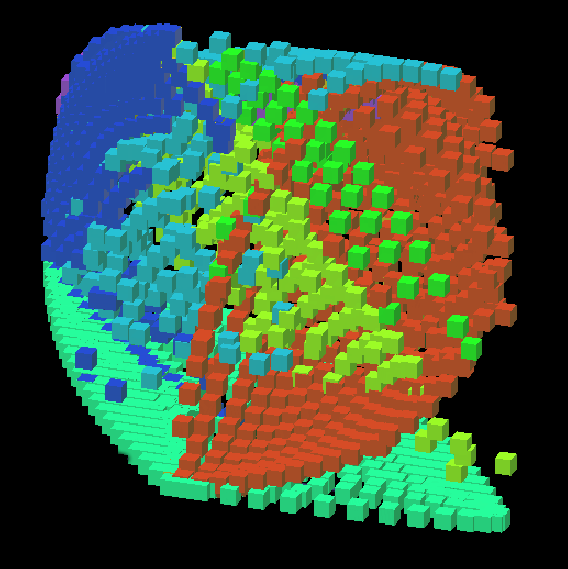}}
\subfigure[]{\label{fig:childInParentRealizations} \includegraphics[scale=0.3]{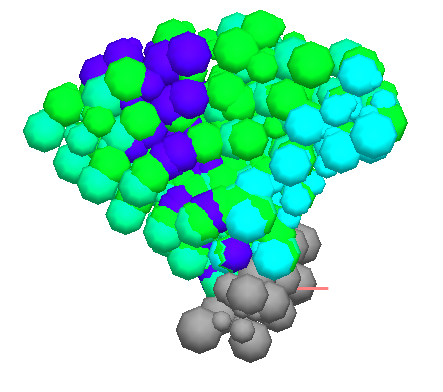}}
\caption{\scriptsize \textbf{Results showing the advantage of Region Specific Parametrization}: 
(a) and (b) show two different screenshots (from the EASAL software),
of the Cayley region of a typical 3D active constraint region for the input 
\rmc s shown in (c).
(a) shows the Cayley region with the transparent mustard colored points 
representing interior points and the points in other colors representing 
points close to the lower dimensional boundaries.
(b) The configurations in the children regions have been sampled using their 
own parametrization, but are shown in the parent parametrization.
As can be seen, using region specific parametrization for children regions, 
discovers more Cayley configurations. 
(c) The sweep of configurations in a different 2D Cayley region for the same
input \rmc. The gray \rmc\ is held fixed, and the green portions of the
sweep represent configurations that were discovered while sampling the parent.
The light blue and purple configurations were additionally discovered while
sampling children regions with region-specific parametrization.
See text in Section \ref{sec:results:RegionSpecific}.}
\label{fig:reParametrizationExample}
\end{figure*}

We sample the assembly landscape of the 42 atom concave system using two
different methods with the same starting step size. The first method, EASAL-1
with region-specific parametrization, starts by sampling a 5D Cayley region.
When a child region is discovered, it creates a different Cayley
parametrization for the child region and recursively samples the child region
until all its descendants are explored.
\begin{figure*}[htpb]
\subfigure[]{\bottominset{\label{fig:6StraightMid}\includegraphics[scale=0.07]{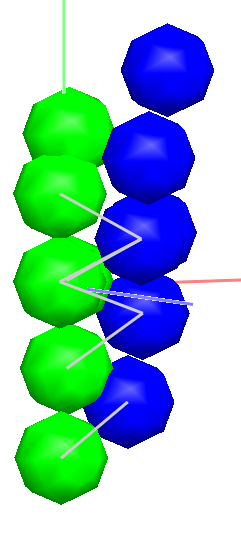}}
{\includegraphics[scale=0.15]{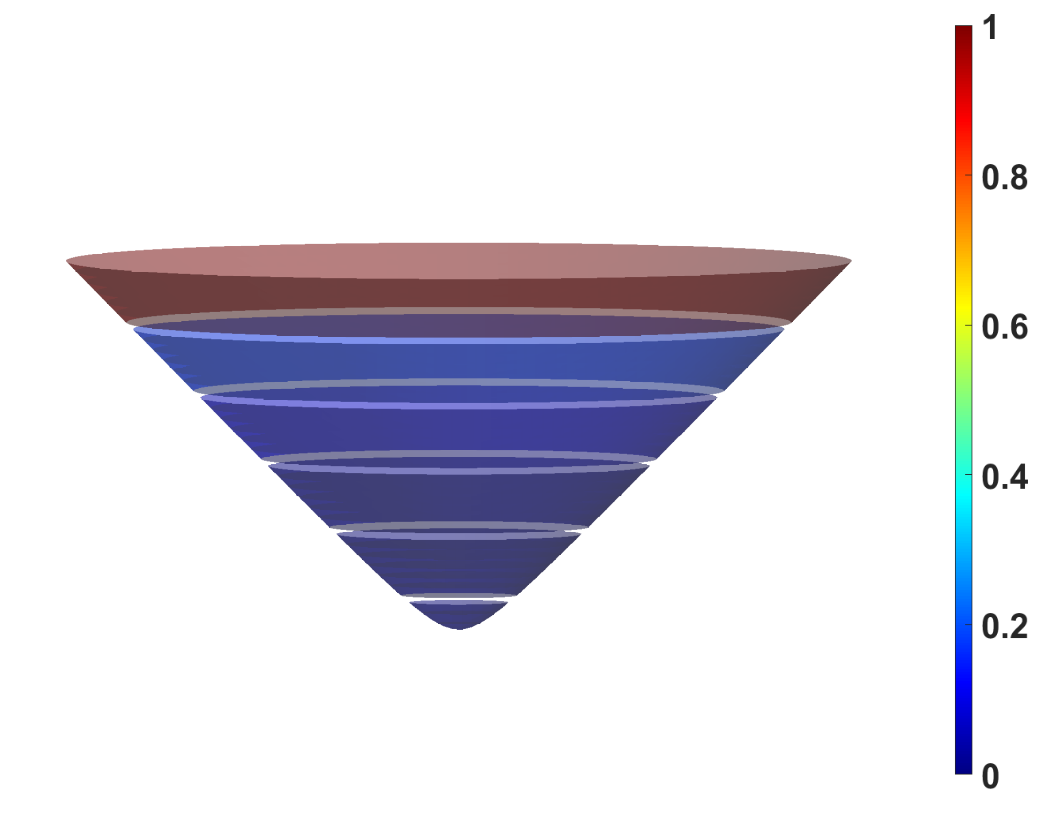}}{10pt}{23pt}}
\subfigure[]{\bottominset{\label{fig:Basin6P}\includegraphics[scale=0.035]{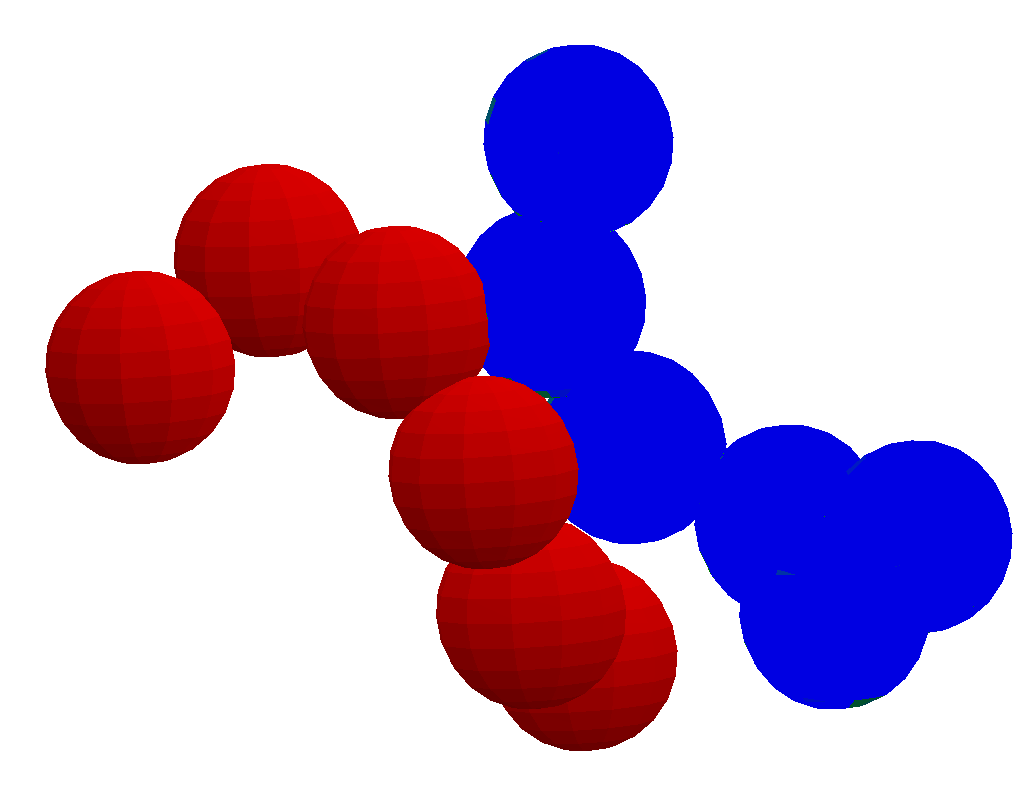}}
{\includegraphics[scale=0.15]{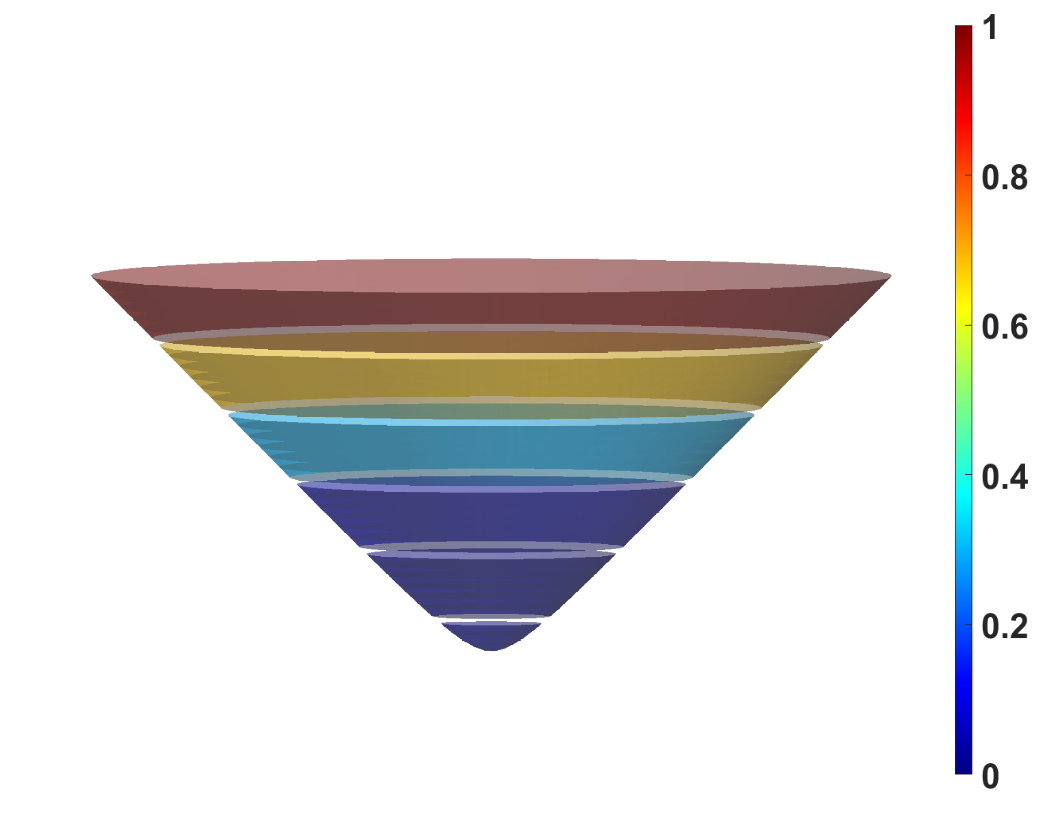}}{23pt}{43pt}}
\subfigure[]{\bottominset{\label{fig:10PocketedMid}\includegraphics[scale=0.08]{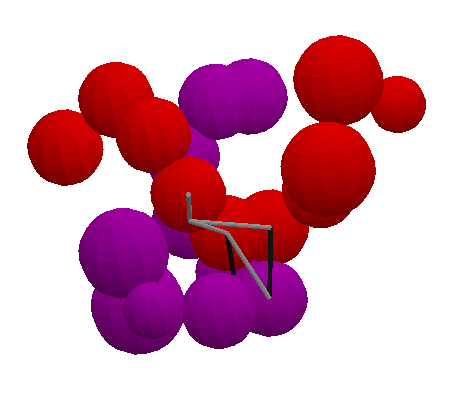}}
{\includegraphics[scale=0.15]{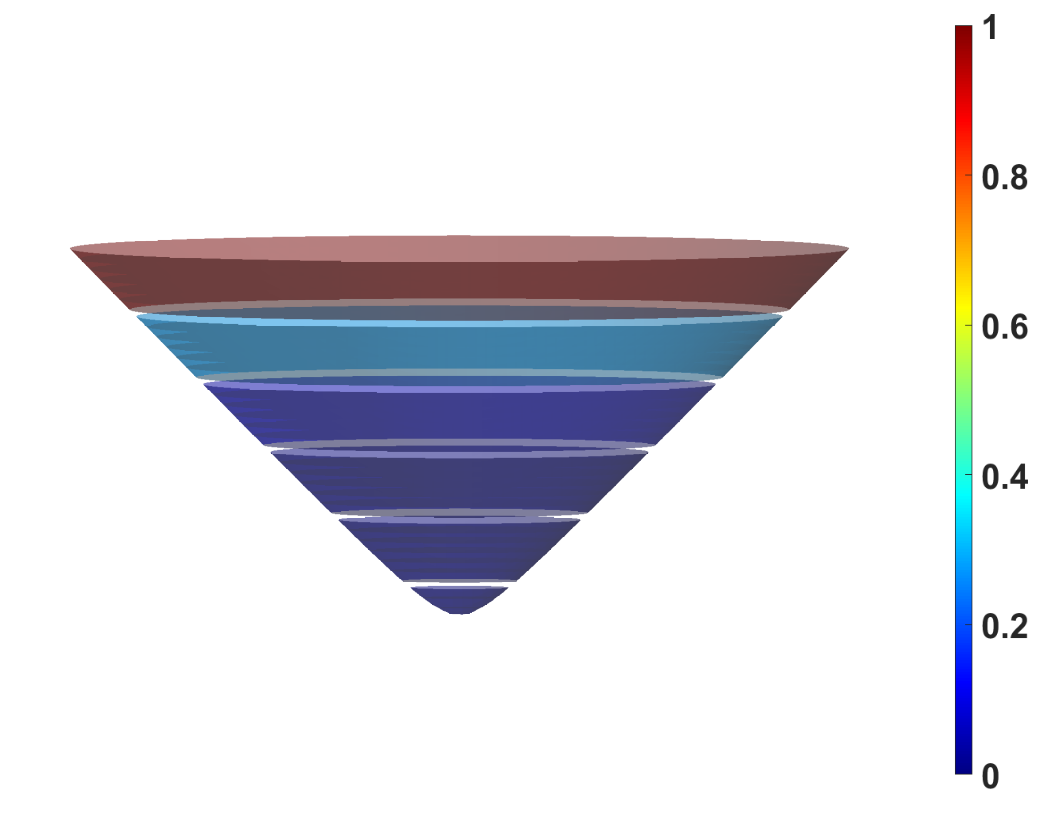}}{18pt}{33pt}}

\subfigure[]{\bottominset{\label{fig:20PocketedMid}\includegraphics[scale=0.05]{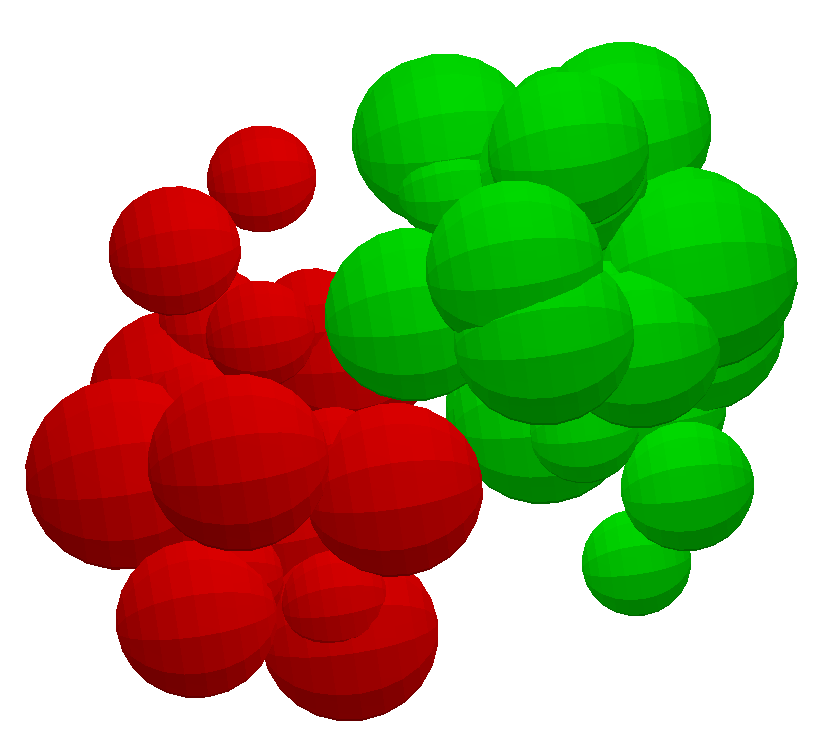}}
{\includegraphics[scale=0.15]{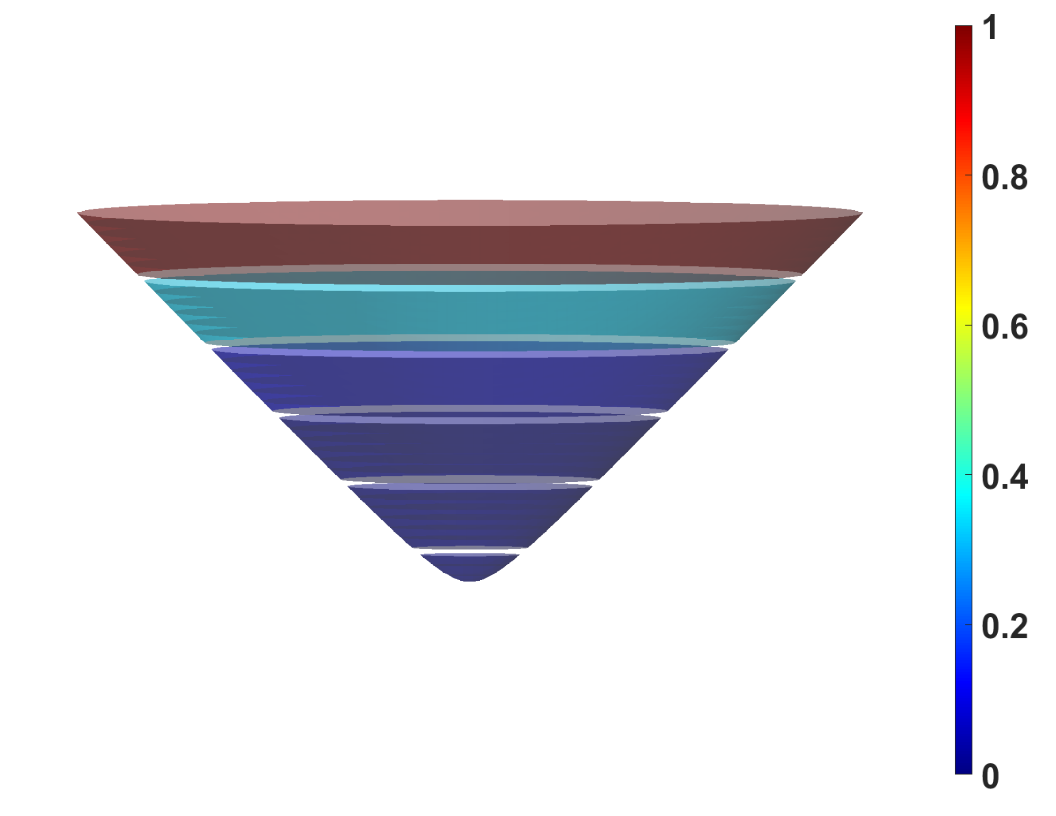}}{18pt}{38pt}}
\subfigure[]{\bottominset{\label{fig:20StraightMid}\includegraphics[scale=0.05]{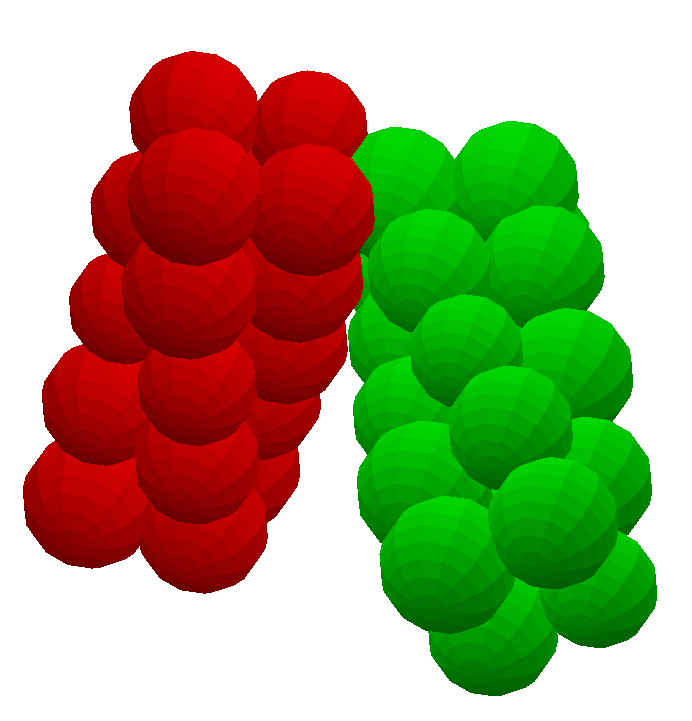}}
{\includegraphics[scale=0.15]{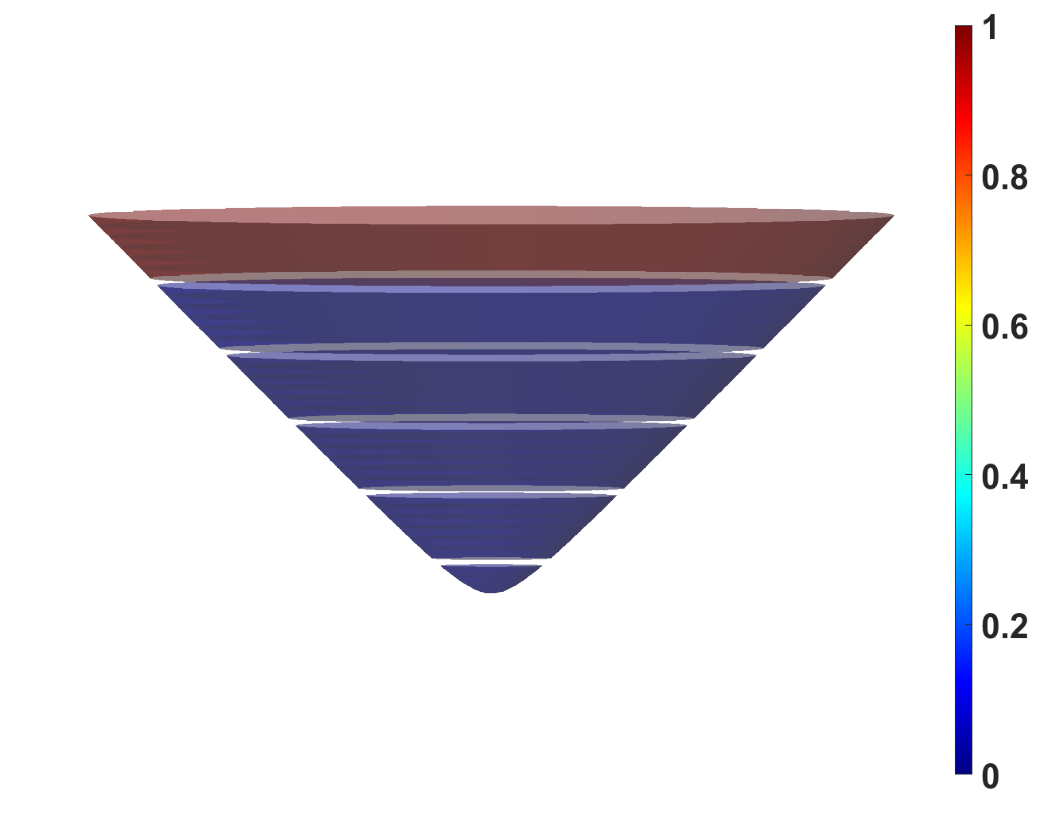}}{18pt}{38pt}}
\subfigure[]{\bottominset{\label{fig:42PocketedMid}\includegraphics[scale=0.07]{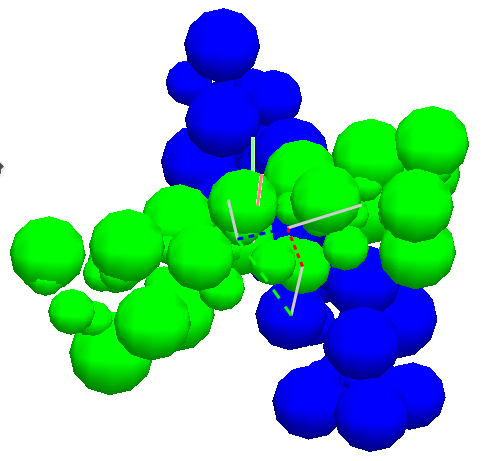}}
{\includegraphics[scale=0.15]{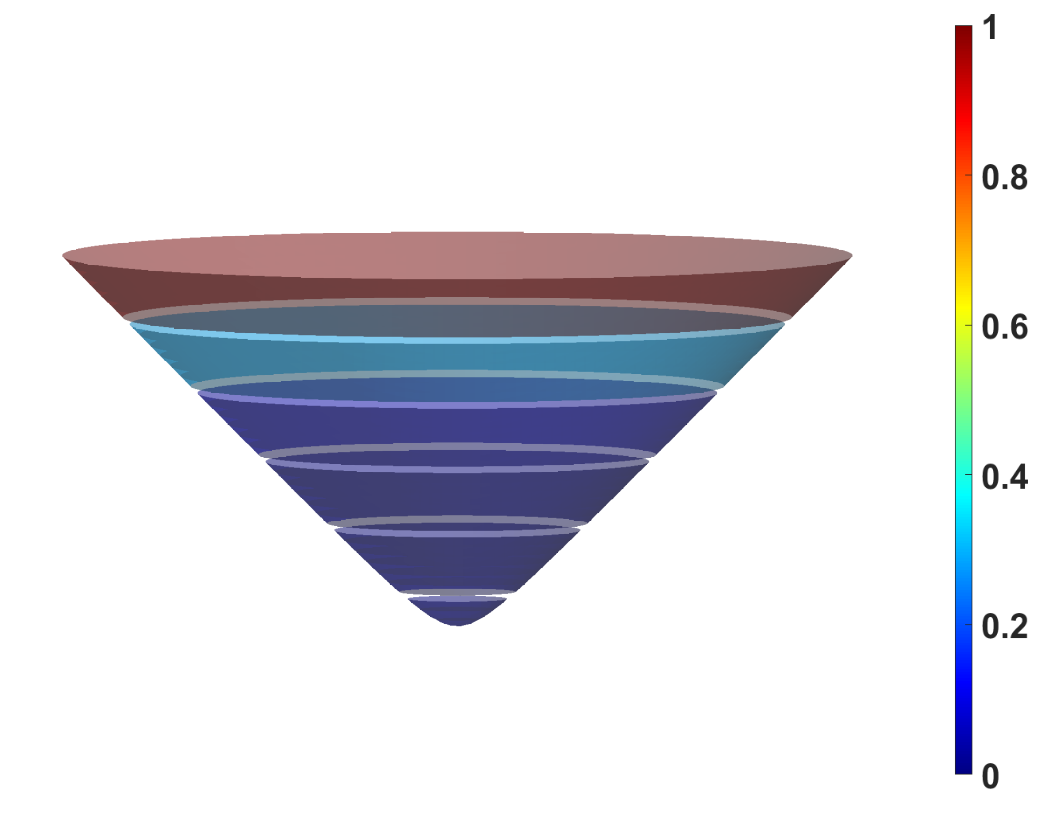}}{18pt}{33pt}}
\caption{\scriptsize \textbf{Bar-Codes for Approximate Basin Volumes:}
Volumes of potential
energy basins in the assembly landscape of the 6 input \rmc\ pairs described in
Section \ref{sec:expSetup}.
The bottom most portion of the paraboloid shows the volume of all the 0D regions
and each successive layer shows the volume of regions 1 dimension higher. 
Higher volumes are shown
in red and lower volumes are shown in blue (see scale beside each figure). 
See text in Section \ref{sec:results:approximateVolume}.}
\label{fig:BasinVolume}
\end{figure*}
The second method, EASAL-3 without region-specific parametrization, samples a
5D region densely close to the boundaries and sparsely in the interior, thereby
sampling the descendant regions, in a single Cayley parametrization.

\figref{fig:E1vsE3} compares the number of regions and the number of samples
found using the two methods. \figref{fig:E1vsE3Regions} shows that EASAL-1
finds more configurations as compared to EASAL-3. In addition,
\figref{fig:E1vsE3Regions} shows that EASAL-1 finds more lower dimensional
active constraint regions as well. Thus, region-specific parametrization
further helps to decouple roadmap generation from sampling by reducing the number of samples
required to find a given number of regions.

The second experiment shows that using region-specific parametrization provides
better coverage sampling of active constraint regions.
\figref{fig:reParametrizationExample} shows, using a concrete example in the
EASAL software, how region-specific parametrization discovers more
configurations.

\subsection{Results on Volume Computation}
\label{sec:results:Volumes}
\begin{table}[htpb]
\centering
\begin{tabular}{cc}\hline
Rigid Molecular Component&Volume Computation Time (seconds)\\\hline
6 Convex&	1.4261\\\hline
6 Concave& 4.4712\\\hline
10 Concave& 9.0646\\\hline
20 Convex& 22.6449\\\hline
20 Concave& 20.1107\\\hline
42 Concave& 106.8745\\ \hline
\end{tabular}
\caption{\scriptsize \textbf{Time to Compute Approximate Basin Volumes}: 
The average time required to compute the volumes of potential energy basins. 
See text in Section \ref{sec:results:approximateVolume}.
}
\label{table:BasinVolume}
\end{table}
%%%%%%%%%%%%%
\begin{figure*}[htpb]
   \centering
\subfigure[\scriptsize 1D regions ($k=7$)]{\label{fig:result_v7:a}\includegraphics[width=0.45\columnwidth]{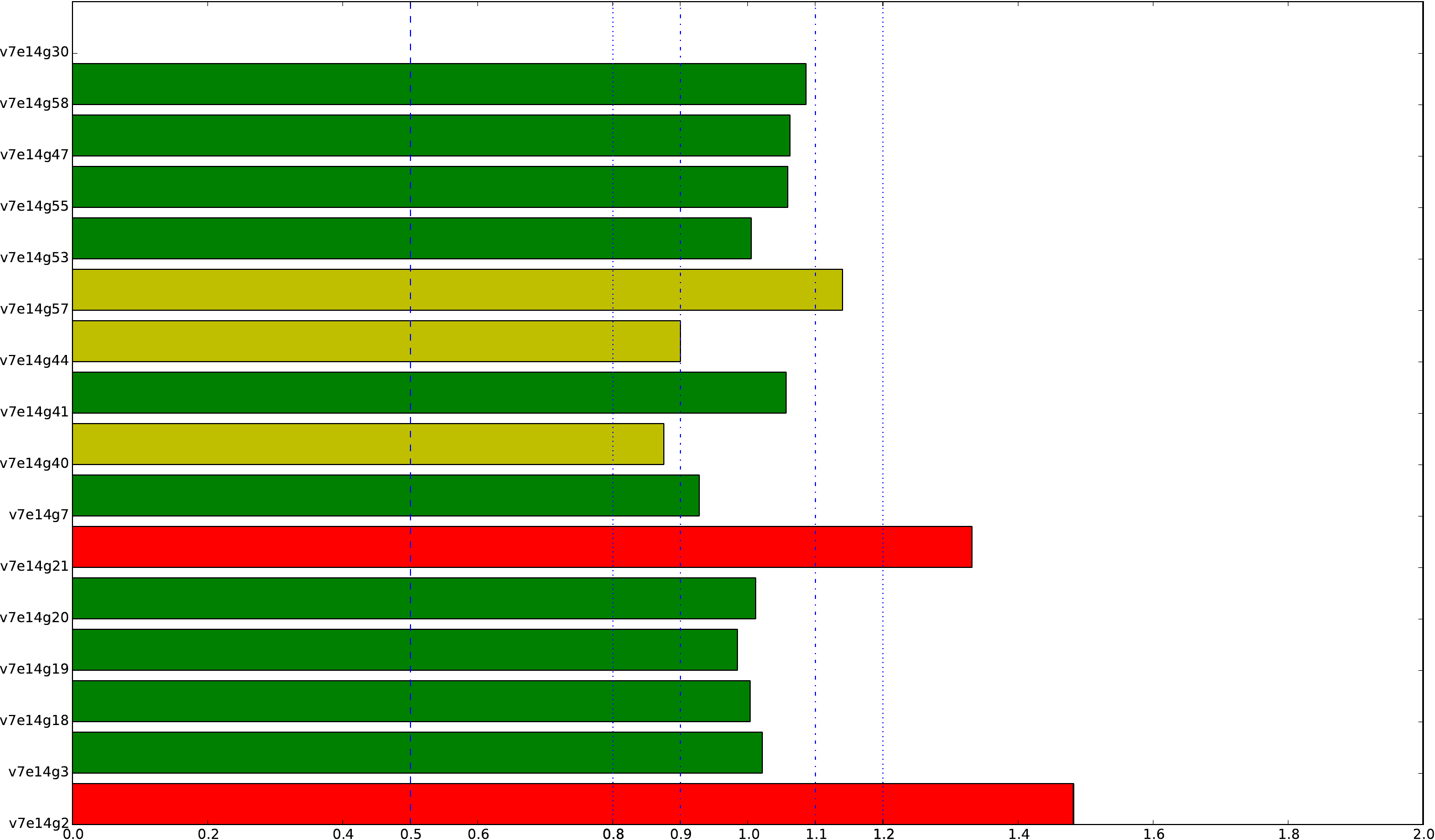}}
\subfigure[\scriptsize 2D regions ($k=7$)]{\label{fig:result_v7:b}\includegraphics[width=0.45\columnwidth]{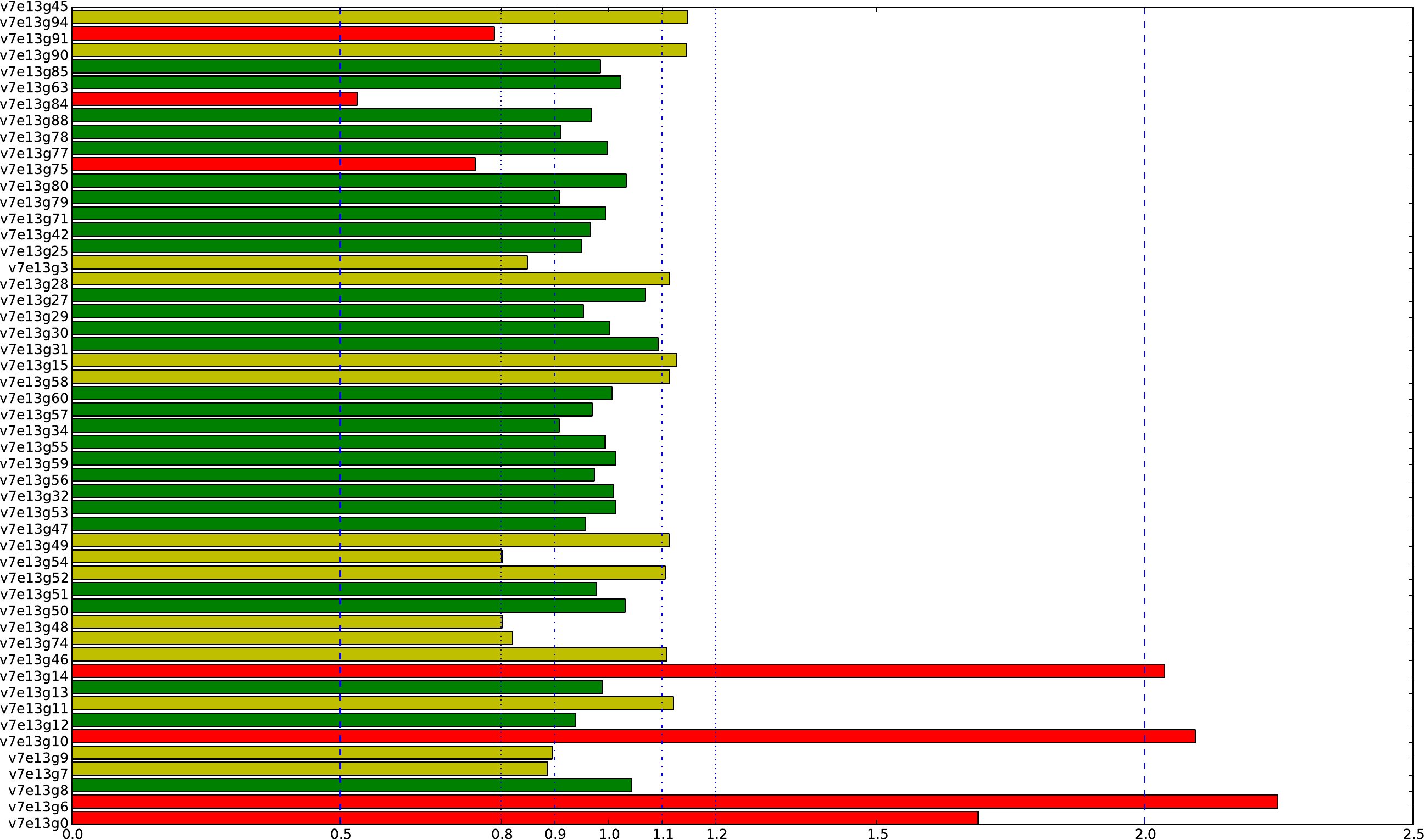}}
\subfigure[\scriptsize 1D regions ($k=8$)]{\label{fig:result_v8:a}\includegraphics[width=0.45\columnwidth]{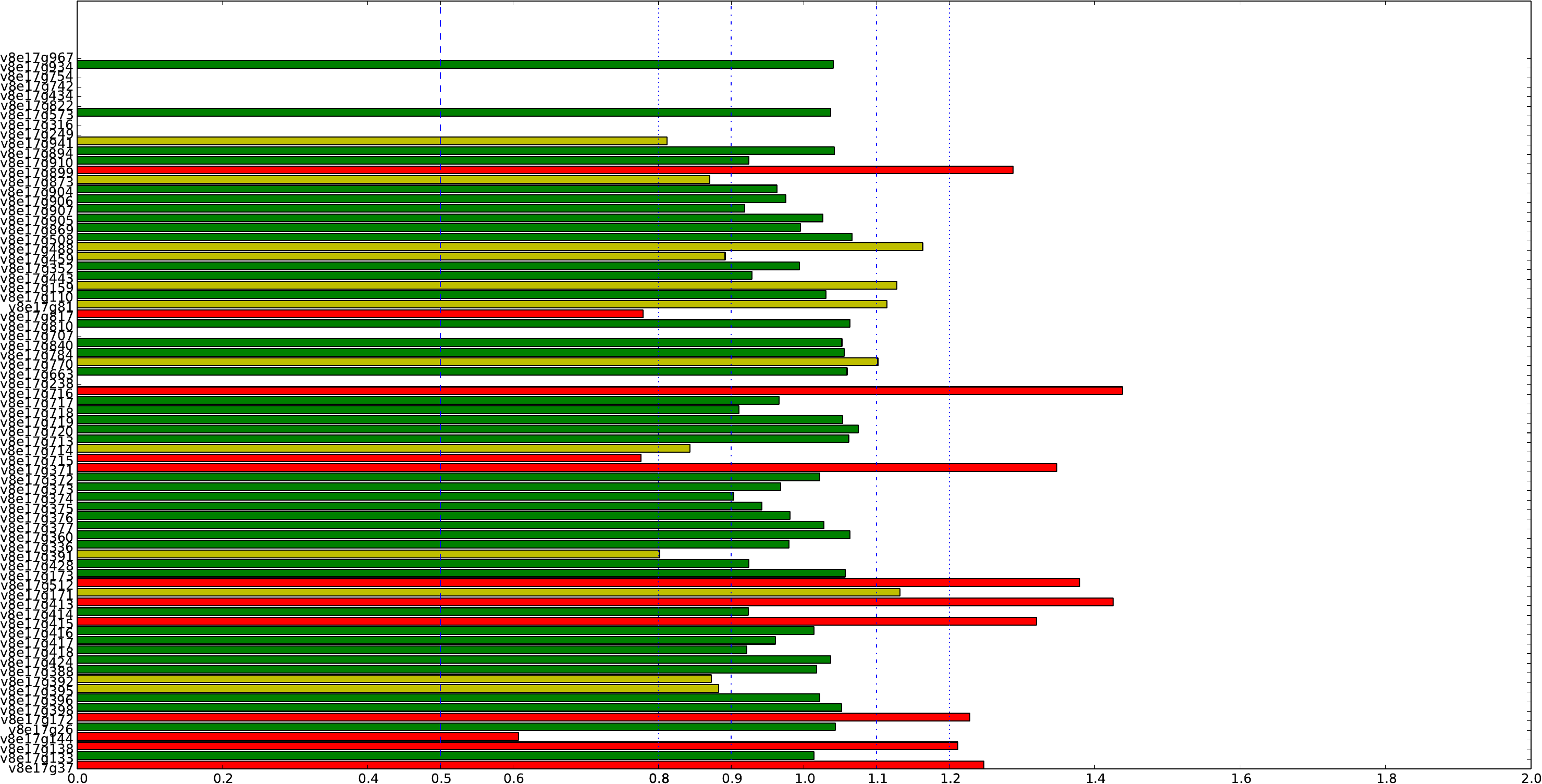}}
\subfigure[\scriptsize 2D regions ($k=8$)]{\label{fig:result_v8:b}\includegraphics[width=0.45\columnwidth]{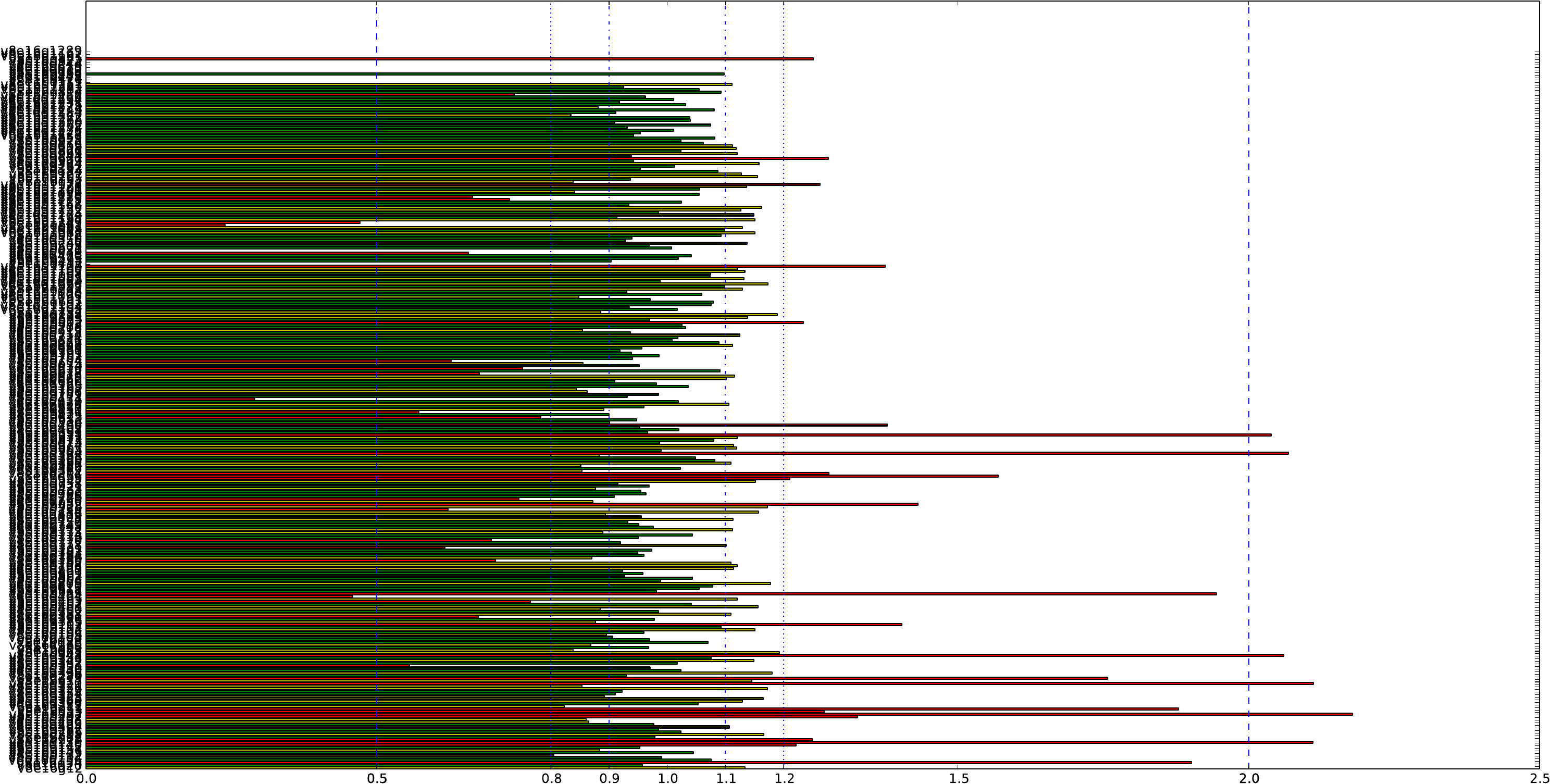}}
\caption{\scriptsize \textbf{Entropy Calculations}: (a) and (b) show the ratio of integral values computed by our 
method (using an extremely coarse Cayley sampling) against that of paper \cite{Holmes-Cerfon2013}, for 
multiple 1D and 2D region of a $n=1$ assembly landscape with $k=7$ identical spheres.
Green bars indicate a ratio in the range $[0.9, 1.1]$, yellow bars
indicate a ratio in the range $[0.8,1.2]$ but not in $[0.9, 1.1]$ and red bars indicate a ratio not in the range $[0.8, 1.2]$.
About 82\% of ratios are green or yellow, i.e., show at most 20\% error.
(c) and (d) Similar comparison for an $k=8$ system.
See text in Section \ref {sec:results:ExactVolume}.}
\label{fig:result_v7}
\end{figure*}
%%%%%%%%%%%%%
Results in this section demonstrate the performance of two algorithms
for finding approximate and accurate volumes given in 
Sections \ref{sec:methods:approximateVolume} and
\ref{sec:methods:exactVolume}.
\begin{figure*}[htpb]
\centering
\subfigure[\scriptsize 2D nodes ($k=8$)]{\label{fig:v7d1_converge}\includegraphics[scale=0.3]{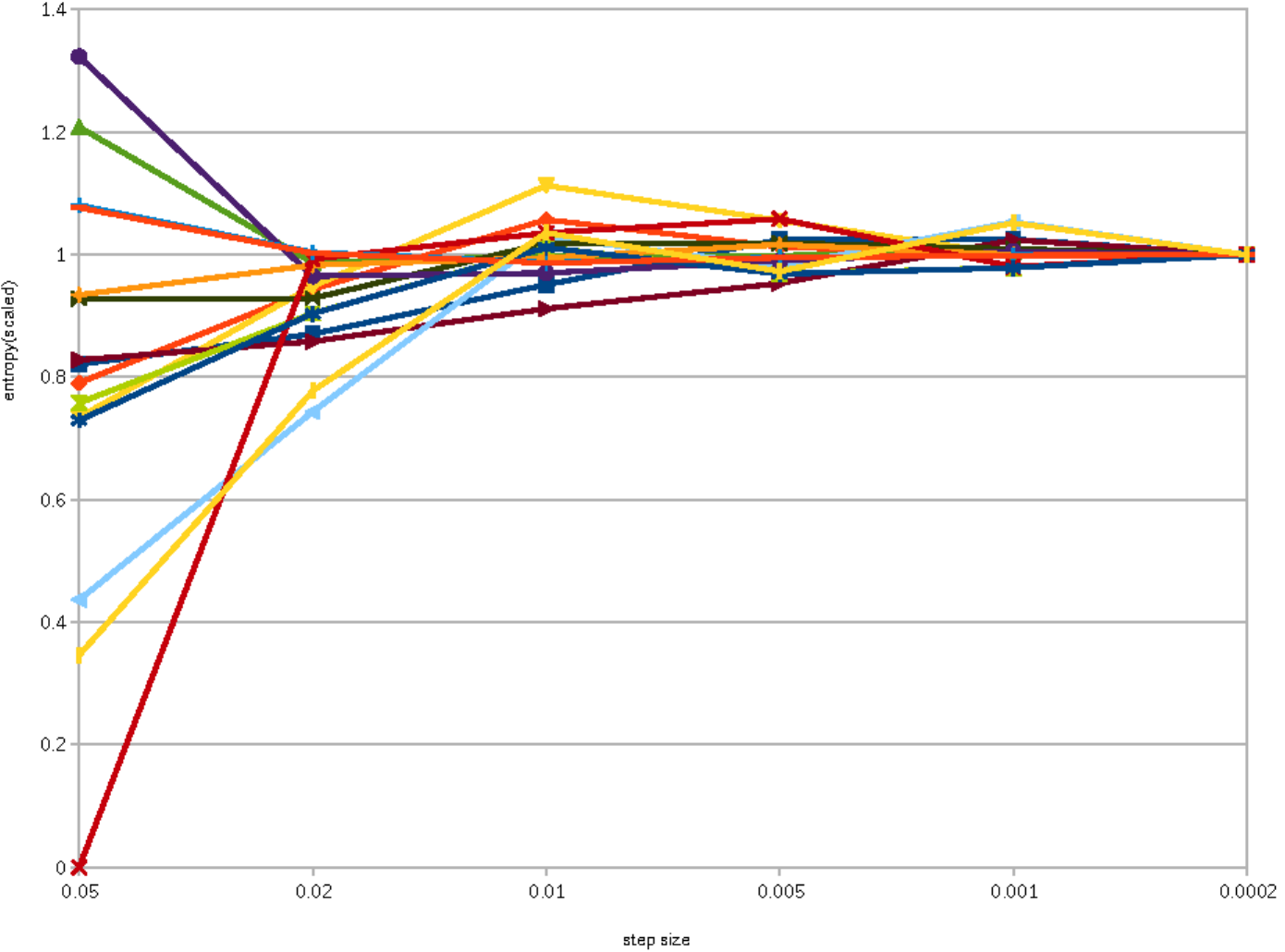}}
\subfigure[\scriptsize 3D nodes ($k=6$)]{\label{fig:result_3d}\includegraphics[scale=0.5]{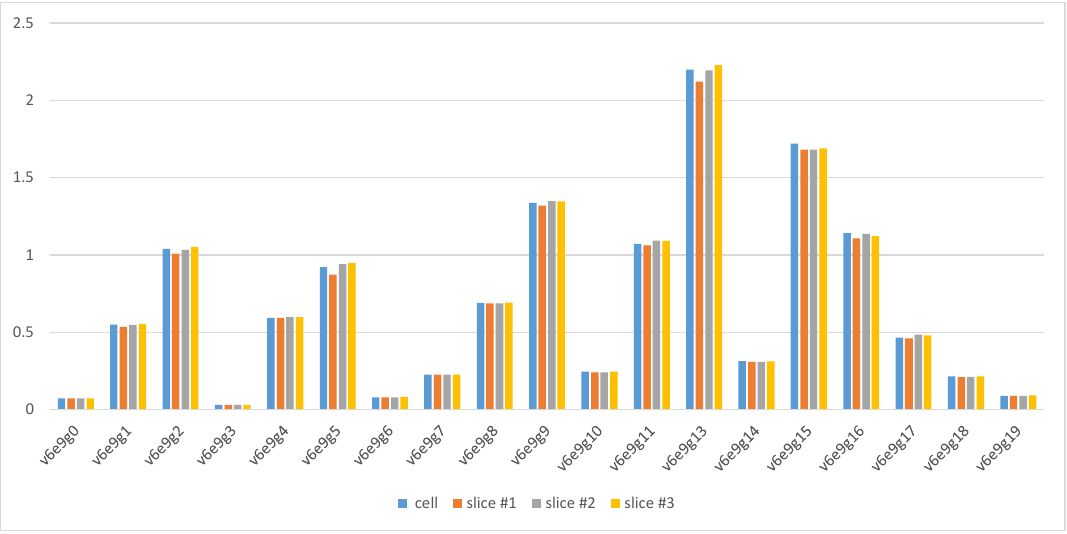}}
\caption{\scriptsize \textbf{Entropy Calculations}: (a) Convergence of entropy calculation as the step size decreases, 
		for multiple 1D active constraint regions (shown in different colors) in a $n=1$ assembly landscape
with $k=8$ identical spheres. The entropy integral values are scaled by the one with the smallest step size. 
		(b) \scriptsize Comparison of integral values computed by 4 different methods (shown in different colors) on multiple 
		3D active constraint regions for
a $n=1$ assembly landscape with $k=6$ identical spheres. See text in Section \ref{sec:results:ExactVolume}.
}
\end{figure*}
\subsubsection{Approximate Basin Volume Computation}
\label{sec:results:approximateVolume}
This experiment starts from a previously generated atlas
and finds the volumes of potential energy basins
extremely fast using the algorithm in Section \ref{sec:methods:approximateVolume}.
Results are tabulated in Table \ref{table:BasinVolume}. 

\figref{fig:BasinVolume} shows portions of the basin bar-codes, namely
the weighted samples or approximate volumes stratified by dimension.
\subsubsection{Comparison and Convergence of Entropy Integrals} 
\label{sec:results:ExactVolume}

This section demonstrates the performance of the algorithm in Section
\ref{sec:methods:exactVolume}.

\figref{fig:result_v7} compares, for $n=1$ assembly systems with $k=7$ and
$k=8$ identical spheres, 1D and 2D region integrals (See Equation
\eqref{eq:entropy_int_2} in Section \ref{sec:methods:exactVolume}) computed by
our method to the results of the paper \cite{Holmes-Cerfon2013}. Already with
coarse sampling, most of the ratios of values returned are in the range $1\pm
0.1$. Although some integrals differ by more than 50\%, increasing sampling density has
these integrals converge rapidly as shown in \figref{fig:v7d1_converge}.

The samples are uniformly distributed in Cayley region; the Jacobian relating
Cayley to Cartesian parametrization (of Section \ref{sec:methods:exactVolume})
was not used to adjust either the sampling
step size or direction, but only to scale the integral.

Our method extends to regions of any dimension. But since the paper
\cite{Holmes-Cerfon2013} does not provide 3D region integrals we use two
different methods to compute the integral and compare against each other.

The first `cell based' method is the generalization of the 2D integral in 3D. At
each grid point in Cayley region, the function value is computed and multiplied
by the volume scaling factor calculated from the local Jacobian. The sum of
this weighted function value at all the grid points is used as the numerical integral.
By calculating the local numerical Jacobian we linearized the space locally, which
could contribute to the error when the dimension is high and sampling
is coarse.

The other ``slice based'' method is based on the integral of a 2D slice of the
space. We partition the 3D Cayley region into a collection of 2D regions, or
slices, by fixing one of the Cayley parameters. For each slice, the 2D integral
is calculated as before. The sum of the slice integrals weighted by
their thickness is used as the numerical 3D integral. The thickness for each
2D region is calculated as the average distance between neighboring slices. Our
previous comparison shows that the error of local linearization is
negligible for the 2D integral. So this result can be credibly used to check
the magnitude of linearization error for the 3D integral.

\figref{fig:result_3d} shows the comparison between these two methods for all
3D regions of the $n=1$ assembly system with $k=6$ identical spheres. For the slice
based method, we can partition the region in 3 different directions depending
on the Cayley parameter to be fixed for each slice. So there are 3 different
slice based results. As shown in the figure, the result from the 4 different
methods are very similar, implying that the linearization error for the cell-based
3D integral is negligible.

\subsection{Results on Verifying Time Complexity}
\label{sec:results:complexity}
We first demonstrate the performance of the core algorithm in 
Section \ref{sec:results:coreComplexity} and in Section \ref{sec:results:variantComplexity}
we demonstrate the performance of the algorithm variant for $2<k<12$ and arbitrary $n$.

\subsubsection{Verifying the Core Algorithm's Time Complexity}
\label{sec:results:coreComplexity}
We demonstrate the performance of the core algorithm in Section
\ref{sec:coreAlgorithm} and verify the time complexity analysis in Section
\ref{sec:CA}. Table \ref{tab:performance} shows the sampling time, averaged
across the 10 5D regions and their descendants for each of the 6 input \rmc\
pairs, varying linearly in the number of weighted samples. As expected, the
sampling time varies linearly in the number of regions but the factor of
proportionality depends on the input \rmc 's shape variables and is captured
by the second landscape design variable as demonstrated in Section
\ref{sec:results:DesignVolume}.

\begin{table*}[htpb]
\centering
		\begin{tabular}{ccccc}
\hline
Rigid Molecular Component&n&Regions&Weighted Samples& Sampling Time(seconds)\\\hline\hline
Narrow Convex &6&3270&419k&20\\\hline
Narrow Concave &6&10128&2.2 million& 86\\\hline
Narrow Concave &10&110554&11.5 million& 451\\\hline
Wide Convex &20&112625&19.5 million&1190\\ \hline
Wide Concave &20&63835&12.8 million& 775\\ \hline
Wide Concave &42&90190&31.5 million& 8580\\ \hline
\end{tabular}
\caption{\scriptsize \textbf{Verifying the Core Algorithm's Time Complexity}:
Number of regions and sampling time for different input \rmc\ pairs. 
These are averages for sampling 10 randomly chosen initial 5D nodes
and all their descendants. In these results, the ratio of the radius to step size 
is set to 3. See Section \ref{sec:results:complexity}}
\label{tab:performance}
\end{table*}
\begin{figure*}[htpb]
\centering
\subfigure[]{\label{fig:BivariatePlot}\includegraphics[width=0.45\textwidth]{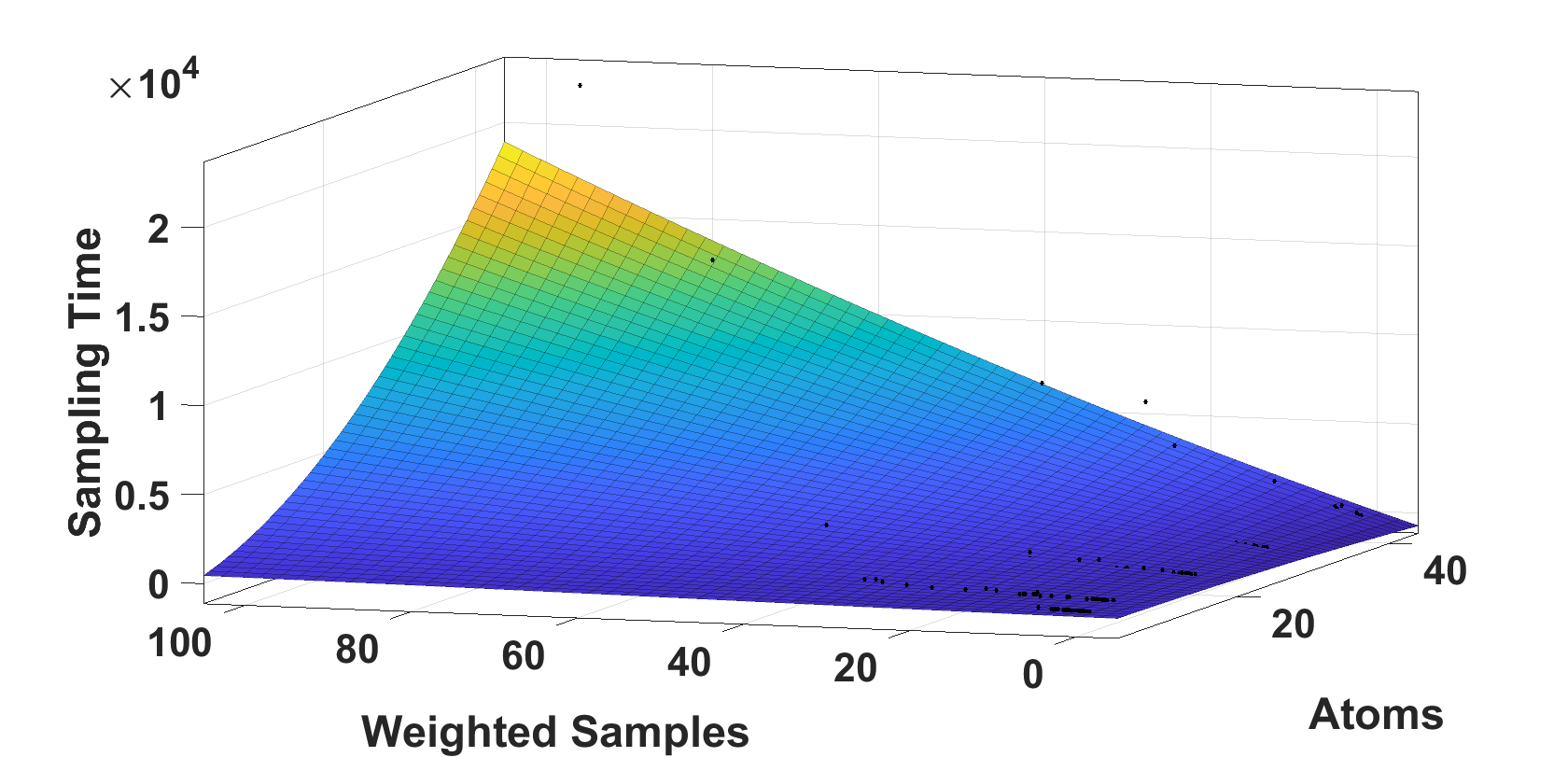}}
\subfigure[]{\label{fig:BivariatePlotSlice}\includegraphics[width=0.45\textwidth]{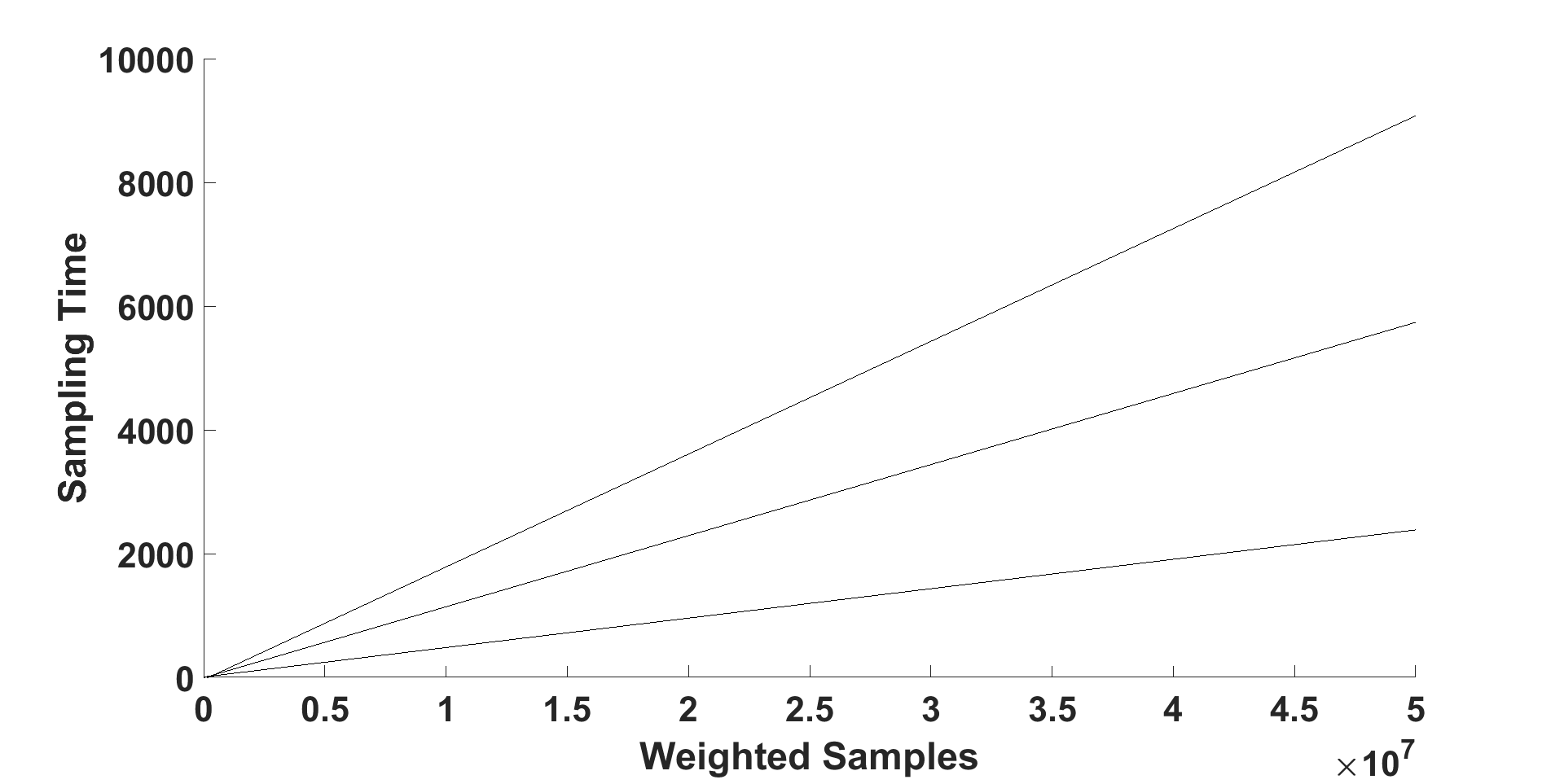}}
\subfigure[]{\label{fig:RVsSStep}\includegraphics[width=0.45\textwidth]{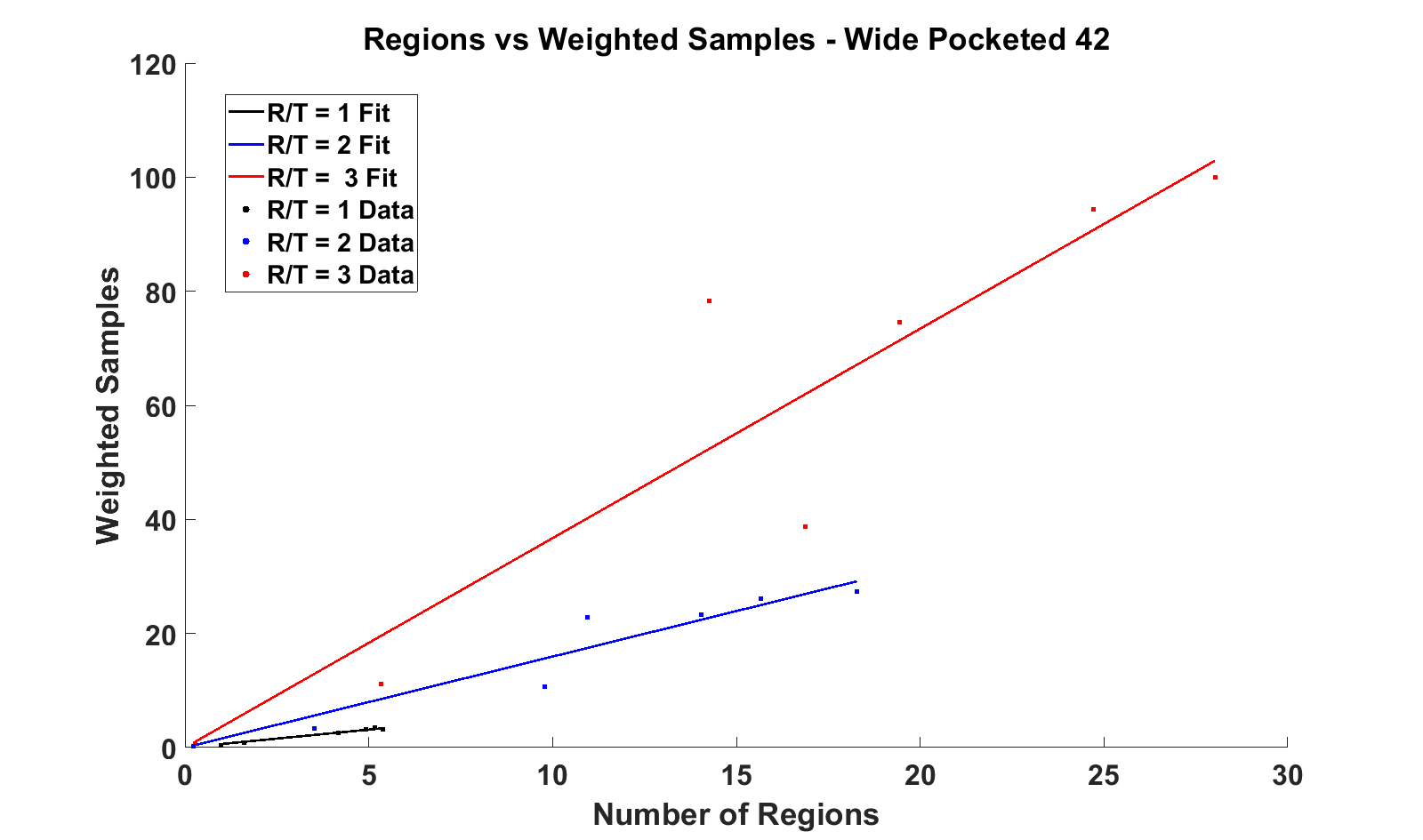}}
\caption{\scriptsize \textbf{Verifying Time Complexity}: (a) The plot of the number of atoms and the number
of samples against the sampling time. The plot also shows bivariate function fit to this
data. (b) shows slices of the bivariate function in (a) at $n =$ 15, 25 and 35. These
show a linear relationship between the number of samples and the sampling time
when $n$ is kept constant. (c) Plot of the number of weighted samples against
the number of regions for the wide concave \rmc\ system (42 Atoms). The plot shows
that the linear factor in regions vs weighted samples also depends on step
size. As the step-size increases, the slope of the line 
increases (see text in Section \ref{sec:results:complexity} for details).}
\end{figure*}

The next experiment verifies: (1) the expected linear dependence of sampling
time on the number of weighted samples and quadratic dependence on the number
of atoms (a constituent of the input shape variables); and (2) the influence of
the sampling step size on the constant of proportionality between weighted samples 
and the number of regions.

\figref{fig:BivariatePlot} plots the number of atoms and the number of weighted
samples against the sampling time. Each point corresponds to a single run
(using a 5D root node) described in Section \ref{sec:expSetup}. As expected, the plot also
shows a bivariate function fit to this data using MATLAB conforming to a
quadratic dependence on the number of atoms $n$. Also as expected, slices of
the function taken at different values of $n$ (see
\figref{fig:BivariatePlotSlice}) show a linear relationship between the number
of weighted samples and the sampling time. 

\figref{fig:RVsSStep} plots the number of samples against the number of regions
for the wide concave \rmc\ system (42 atoms). It shows three different
lines, one each for the different step size. As can be seen from the plots, the
slope of the line showing the number of samples against
the number of regions, increases as the step size increases.

\subsubsection{Verifying the Time Complexity of Algorithm Variant for arbitrary $n$ and $2<k<12$}
\label{sec:results:variantComplexity}
We demonstrate the performance of the second algorithm variant, described in
Section \ref{sec:AlgorithmVariantkg2}, at atlasing the configuration space of
the folding of the tryptophan zipper (trpzip 1). We decompose the tryptophan
zipper into smaller rigid sub-units as described in \cite{joseph2017exploring,
kusumaatmaja2012local}. We use the benchmark data available at \cite{optmin} as
input and generate the topology of the landscape 

The benchmark data for the tryptophan zipper consists as input a single
molecule with 147 atoms divided into 17 groups of rigid sub-units, each varying
in size from 4 atoms to 10 atoms. Using the second method in Section
\ref{sec:AlgorithmVariantkg2}, we generate atlases with pairs of the rigid
sub-units as input and give an estimate for the time required to find the
complete atlas using this data.

\begin{figure}[htpb]
\centering
\includegraphics[width=0.5\columnwidth]{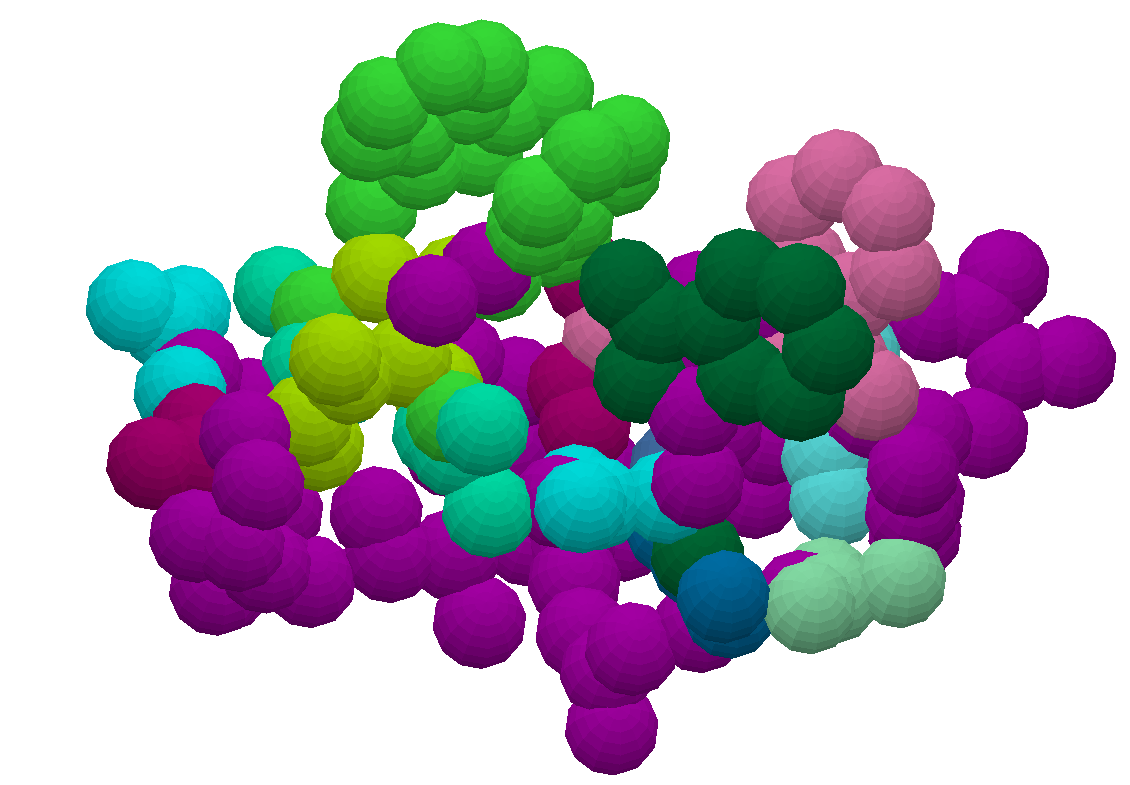}
\caption{EASAL screenshot of a configuration in the folding energy 
landscape of the tryptophan zipper (trpzip1). The different colors represent the 
different input rigid components.}
\label{fig:tryptophan}
\end{figure}

The output of the benchmark as described in the papers \cite{Jerelle2016,
joseph2017exploring} is a potential energy disconnectivity graph which, is
gives information about all the minima in the landscape and the energy barriers
separating them. The wealth of information in EASAL's atlas subsumes  this
information.  For example, the atlas provides the complete connectivity of
basins, including all  path lengths, basin volumes, and energy level
information \rahul{(see Figure \ref{fig:tryptophan} for an EASAL screenshot of a
configuration in the folding landscape)}. 

The following experiments were performed using the prototype parallel version
of EASAL running on the hipergator supercomputer. We used 10 cores of Intel(R)
Xeon(R) Gold 6142 CPU @ 2.60GHz processors and 10 GB of memory.  Table
\ref{tab:trypto} summarizes the atlasing time required by EASAL to generate the
atlases of pairs of rigid molecular components in the first step. With these,
we estimate the time required for the direct sum at each stage and finally
the time required for generating the entire atlas.

\begin{table*}
\resizebox{\textwidth}{!}{%
%\begin{tabular}{|>{\color{red}}c>{\color{red}}c>{\color{red}}c>{\color{red}}c>{\color{red}}c|}\hline
\begin{tabular}{ccccc}\hline
		Rigid Molecular Components & Macrostates & Number of Energy Minima & Atlasing Time & Time per Minima\\\hline
		10 - 10 & 2.6 million & 1.1 million & 6m 22s & 0.3 ms\\\hline
		10 - 4 & 200k & 80k & 1m 30s & 1.1 ms\\\hline
		4 - 4 & 7k & 2k & 50s & 25 ms\\\hline
\end{tabular}
}
		\caption{Atlasing time for the atlases of the rigid molecular components in
		the tryptophan zipper (trpzip1) molecule. See text in Section \ref{sec:results:variantComplexity}
		for details.}
\label{tab:trypto}
\end{table*}

Direct sum involves searching for active constraint regions of atlases. 
We assume a naive approach in the worst case,
taking the direct sum at each internal node of the tree, which gives us a
search time no worse than the product of the roadmap traversal time at the children.
With more sophisticated techniques for direct sum (such as the one described in
Section \ref{sec:AlgorithmVariantkg2}), we would in fact achieve significantly better search
times, hence this is a conservative estimate. 
Next, we use the sampling time of the 10-10 pair at first level of the 
direct sum. Since most of the nodes at this level are of size 8 (4-4 pairs),
this serves as a good approximation.

From Table \ref{tab:trypto}, we estimate the search time for an atlas of
size 10 as follows. On average, each node in the 10 atom atlas has 100
configurations, thus the time required to search for an active constraint
region in the roadmap should be a hundredth of the time required to generate all the samples.
In the case of the 10-10 molecular pair, this is 3.8 seconds. 
Propagating this product up the tree, we obtain a time of 12 hours at the root
node to complete the process.

The number of energy minima at the root node can be estimated conservatively as in a similar
way. Theoretically, every pair of minima at the child level produces a new
minima at the parent node, but of these only a fraction are valid.
Using the fraction we obtain from the 10-10 molecule (theoretical v/s
observed), and propagating the number of minima up the tree, we get 5.8 billion
minima in total at the root node, i.e., for the entire tryptophan zipper atlas. 
This gives us a sampling time per energy minimum of 7$\mu$s.

\section{Discussion}
\label{sec:discussion}
The new methodology presented in this paper combines many ideas e.g.
geometrization, stratified roadmaps, recursive search. But the unique
advantages (see Section \ref{sec:conclusion} for a summary) can be traced
directly or indirectly to the combination of these ideas with the use of Cayley
parametrization to convexify constrained regions of the energy landscape that
are topologically complex. Crucially, this parametrization permits traversal of
the constrained region without having to explicitly enforce the constraints
(which is usually done by minimizing constraint violation using gradient
descent). Moreover, the convexity brings clear advantages in boundary
detection, path and volume computations, ameliorating the curse of dimension.
To the best of our knowledge, our methodology is the only one that uses this
type of parametrization and convexification.

However, as pointed out, such convexification is only viable for active
constraint graphs that occur in assembly, but not more generally in folding and
other systems. Secondly, Cayley parameters are defined only for active
constraint graphs, which in turn assume a pair-potential energy model that can
be treated using geometric constraint systems. More global energy functions can
be incorporated into the methodology, but not in ways that leverage Cayley
parametrization.  These represent inherent limits to extending the advantages
of Cayley parametrization.

On the other hand, although the new methodology has not so far been leveraged
beyond combining with standard multiscale techniques
\cite{sitharam:Assembly,virus2019}, new methods are now being developed that
transmit the unique strengths of Cayley parametrization across scales, e.g. in
conjunction with the use of decomposition and recombination (DR) plans (as
sketched in the core algorithm's adaptation to large assemblies).

Overall, the new methodology can be used either as a resource-light,
stand-alone method, or any combination of its features can be combined
piecemeal with prevailing methods including MC and MD. Ongoing development of a
parallel EASAL version, using the C++ Actor Framework (CAF)
\cite{cshw-nassp-13,chs-rapc-16}, assigns different active constraint regions
to different threads to sample in parallel, with one thread dedicated to keep
track of the atlas. Results so far indicate no obstacles to linear speedup in
the number of processors. This version would permit comparisons with other
parallel methods \cite{cha2015accelerated,Griffiths2019}.

\section{Conclusions}
\label{sec:conclusion}
The paper has introduced a resource-light geometric method to generate and
sample assembly landscapes characterized by short-range, Lennard-Jones pair
potentials in an implicit solvent. The new methodology's key strengths are
robust and query-able generation of assembly landscape topology decoupled from
sampling; intuitive design of landscapes using succinct bar-codes that connect
input geometry with key landscape features; efficient sampling with refinable
accuracy towards path and volume computations; and generalizability to a
variety of assembly systems. Proof-of-concept results illustrating these
features have been provided using a curated opensource software implementation
\cite{Ozkan:toms}. 

The underlying theory is grounded in the classical concept of topological
stratification and, more crucially, a recent understanding of convexification
of configurational region using Cayley parameters. Consequently, the method
provides quantifiable accuracy, isolation and coverage of narrow, low potential
energy regions. Since the sampling is done by direct, region-specific
parametrization that avoids the need to enforce constraints, it minimizes
discarded samples and improves efficiency. Convexity additionally facilitates
path finding, boundary detection, and volume computation even for high
dimensional landscapes. While it can be used as a stand-alone method, its
individual features can be hybridized with prevailing methods - with
complementary strengths - promising an alternative approach to persistent
computational challenges including free energy, configurational entropy,
kinetics and multiscale modeling of significantly larger assemblies.

While the methodology formally characterizes and leverages the simplicity of
assembly as opposed to folding, this also represents its inherent limits in
that convex Cayley parametrization is only viable for assembly as opposed to
folding or more general macromolecular energetic processes. Moreover the
advantages of Cayley parametrization do not extend to more global energy
functions that are not based on pair-potentials.

Current and future work include comparing the parallel version of EASAL with
other methods \cite{Griffiths2019} on benchmarks
\cite{Chill2014,cha2015accelerated}. This would permit broader application of
the methodology towards modeling, prediction and design of landscapes for a
variety of poorly understood assembly scenarios.

\section*{Acknowledgement}
This research was supported in part by NSF Grants
DMS-0714912, CCF-1117695, DMS-1563234, and DMS-1564480.

\bibliographystyle{unsrt}
\bibliography{easal}

\section*{Appendix}
\appendix
\section{Rigidity Preliminaries}
\label{sec:app:rigidity}
The generic rigidity-based analysis of active constraint graphs in Section
\ref{sec:rigidity} uses the following concepts of combinatorial
rigidity (we additionally refer the reader to the works
\cite{SJS:Handbook,CombinatorialRigidity}).  A (Euclidean)
\emph{realization} in $\mathbb{R}^d$ of a graph $G = (V, E)$, with edge
lengths $\gamma: E \rightarrow \mathbb{R}$, is an assignment of points
in $\mathbb{R}^d$ to vertices such that the Euclidean distance between
pairs of points are the given edge lengths $\gamma$ (factoring out the
$d+1 \choose 2$ Euclidean rigid body motions namely, the rotations and
translations of $SE(d)$). A realization is said to be (locally)
\emph{rigid} if there is no other realization in its neighborhood that
has the same edge lengths. A graph is said to be rigid if a generic
realization of the graph with given edge lengths is rigid. Otherwise,
the graph is said to be flexible (not rigid).  A rigid graph,
generically has finitely many realizations for a given set of edge
lengths. A graph is said to be \emph{minimally rigid}, \emph{well
constrained} or \emph{isostatic} if it is rigid and the removal of any
edge causes it to be flexible.  All non-edges whose lengths are fixed by
fixing the edge lengths of the graph, are 
(locally) \emph{implied} or
\emph{dependent}. 

The degrees of freedom (dof) of a graph is the minimum number of
edges whose addition makes it rigid. Thus, the number of degrees of freedom is
the same as the generic (effective) dimension of the realization space of the
graph with given edge lengths. In $R^d$, Maxwell's theorem
\cite{maxwell} states that if a graph $G=(V,E)$ is rigid, then there is a
subset of edges $E'$, such that for every subset of vertices $S \subseteq V$,
$|E'(S)| \le d|S| - {d+1 \choose 2}$, and $|E'| = d|V| - {d+1 \choose 2}$.
For $d=2$, the right hand side is $2|V| -3$ and for $d=3$ the right hand
side is $3|V|-6$.
The converse is true for $d\le2$, but fails for $d\ge3$.
A graph is \emph{independent} if the removal of any edge increases the degrees 
of freedom. Thus, for an independent graph, the number of edges plus
the number of degrees of freedom is $d|V| - {d+1 \choose 2}$.

\section{Cayley Convexification}
\label{sec:app:convexity}

We define a class of graphs that have 
a convex Cayley parametrization used in Section \ref{sec:convexity}.

A \emph{complete 3-tree} is any graph obtained by starting with a triangle and
adding a new vertex adjacent to the vertices of a triangle in the current
graph. Alternatively, this amounts to successively pasting a complete graph on
4 vertices (a \emph{tetrahedron}) onto a triangle in the current graph. This
yields a natural ordering of vertices in a 3-tree (we drop `complete' when the
context is clear). A 3-tree has $3|V| -6$ edges and
is minimally rigid in $\mathbb{R}^3$. Therefore, a 3-tree generically has
finitely many realizations, and removing any edge gives a flexible
\emph{partial 3-tree}. 

In $\mathbb{R}^2$, the analogous graphs for 3-trees are 2-trees. A
\emph{complete 2-tree} is any graph obtained by starting with an edge and
successively pasting a triangle onto an edge in the current graph. 
A 2-tree is minimally rigid in $\mathbb{R}^2$ and has $2|V| -3$ edges.
A \emph{partial 2-tree} is any subgraph of complete 2-tree.

Theorem \ref{thm:SiGa} asserts that the length tuples of non-edges or \emph{Cayley parameters}, 
$F$ - that complete a partial 3-tree $G$ into a 3-tree $G\cup F$, or complete a partial 2-tree into a 2-tree - form a convex set
(see the example in Section \ref{sec:app:toytwodC}).
These length tuples are called \emph{Cayley configurations} of the partial 3-tree $G$. The \emph{chart} for a graph $G$ and 
Cayley parameters $F$ is a map that takes a Cartesian realization of $G$ (with given edge lengths)
to a Cayley configuration, i.e., the tuple of lengths of the non-edges $F$ in the Cartesian realization.
The chart is a branched covering map. Therefore the theorem states that the chart for a partial
3-tree maps the set of Cartesian realizations of $G$ (given edge lengths or edge length intervals) to a convex set.

In addition to proving Theorem \ref{thm:SiGa}, the paper \cite{SiGa:2010} shows
the existence of convex Cayley regions for a much larger class of
graphs (beyond the scope of this paper).

\begin{theorem}%{(\cite{SiGa:2010} Any partial 3-tree yields an exact convex chart)}
\label{thm:SiGa} 
Let the active constraint graph $G = (V, E)$  of a region $R_G$ be a partial
3-tree. Let $F$ be a set of non-edges such that $(V, E\cup F)$ is a 3-tree.
Then the chart for $G$ with Cayley parameters $F$ 
maps the set of Cartesian realizations of $G$ (given edge lengths or edge length intervals) to a convex set.
The convex set is bounded by $O(|G|)$ polynomial inequalities (typically linear or quadratic). 
There is a sequence in which the parameters in $F$ can be fixed, such that the exact bounds on each successive parameter can
be computed in time $O(|G|)^2$.
\end{theorem}

The quadratic and linear polynomials defined in Theorem \ref{thm:SiGa} arise from
simple edge-length (metric) relationships in triangles and tetrahedra
and are called \emph{triangle} and \emph{tetrahedral inequalities}. The exact bounds
mentioned in the theorem are called \emph{tetrahedral bounds}. The core algorithm 
of Section \ref{sec:coreAlgorithm} leverages
this efficient computation of the convex bounds enhanced by the Theorem 5.1.3
in the master's thesis \cite{ugandhar}. 

\subsection{Example \toytwodtwo}
\label{sec:app:toytwodC}
Here, we use an active constraint graph in the assembly system of two molecules
with 3 and 2 atoms respectively, to illustrate Cayley convexification in
$\mathbb{R}^2$. Since that example is in $\mathbb{R}^2$, \emph{2-trees}
(defined in Section \ref{sec:app:convexity} in the Appendix) serve the purpose
of 3-trees used in our methodology \cite{SiGa:2010}.

\begin{figure}[htpb]
\begin{center}
\includegraphics[width=.8\columnwidth]{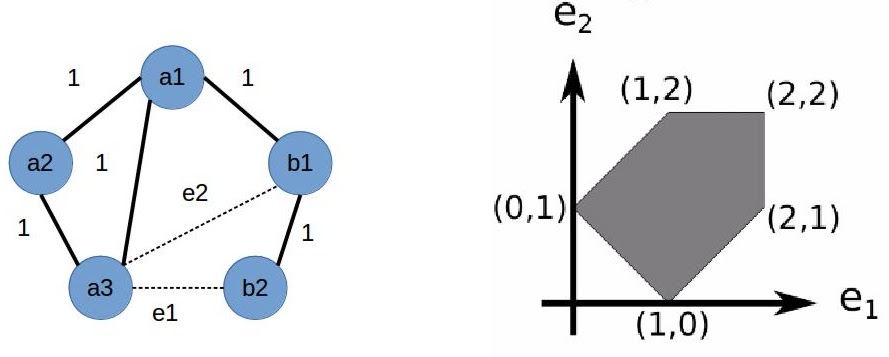}
\end{center}
\caption{\scriptsize \textbf{Cayley Convexification for Example {\toytwodtwo}}: (left) An
active constraint graph and the chosen Cayley parameters $e1$ and $e2$. 
(Right) the 2D convex Cayley region for the 
graph delineating the realizable lengths of $e1$ and $e2$, i.e., those lengths 
of $e1$ and $e2$ that admit a feasible configuration for the graph (see Section 
\ref{sec:app:toytwodC}).}
\label{fig:2DCayley}
\end{figure}

Consider the partial 2-tree graph shown in \figref{fig:2DCayley} (left). To
represent the configurational region of this flexible graph, we add the
non-edges $e1$ and $e2$, shown with dotted lines, to complete the 2-tree. This
not only makes the graph rigid, but finding its Cartesian configurations is easy by a
straightforward ruler and compass construction, solving two quadratics at a
time. The non-edges $e1$ and $e2$ are the Cayley parameters and correspond to
independent flexes. \figref{fig:2DCayley} shows the convex Cayley
region corresponding to this graph. 

If the edges in the graph in \figref{fig:2DCayley} (left) were assigned length
intervals instead of fixed lengths, yielding an active constraint graph, the
resulting configurational region would continue to be convex, but would be 7
dimensional. However, when these intervals are relatively small in comparison
to the edge lengths, the Cayley region remains effectively 2
dimensional.\hfill$\blacksquare$

Furthermore, in the context of active constraint graphs, 
\emph{generalized 3-trees} also yield convex Cayley parameters. This is because
each \rmc\ represents a unique realization of their underlying complete
graph. A generalized 3-tree is defined by construction similar to a 3-tree.
If 3 or more vertices in the already
constructed graph $G$ belong to the same \rmc\ $A$, when a new vertex
$v$ is added with edges to the vertices of a triangle $T$ in $G$, then
the $m \le 3$ vertices in $A \cap T$ can be replaced by any other $m$ distinct
vertices in $A$ to which $v$ is adjacent.  Moreover, generalized 3-trees, just
like 3-trees, have an underlying sequence of tetrahedra, and are rigid with
finitely many realizations.  In this paper, we simply refer to generalized
(partial) 3-trees as (partial) 3-trees.

\section{Cayley Configuration to Cartesian Configurations}
\label{sec:app:realization}

\begin{theorem}
\label{thm:realization}
In an assembly system with $k=2$ molecular units, each Cayley configuration
corresponds to up to 8 Cartesian configurations.
\end{theorem}
\begin{proof}
Consider the first tetrahedron in a 3-tree completion of an active constraint
graph that contains at least one edge (active constraint or Cayley parameter)
between the two \rmc s $A$ and $B$. The tetrahedron contains either (i) 2
vertices $a_1$ and $a_2$ in $A$ and 2 vertices $b_1$, and $b_2$ in $B$ 
or without loss of generality (ii) 3 vertices $a_1$, $a_2$, and $a_3$ in 
$A$ and 1 vertex $b_1$ in $B$.  In either case the 3-tree will contain
exactly 2 new (generalized) tetrahedra: In case (i) there are 4 edges between
$A$ and $B$ and at most 2 further edges can be added to attain minimal rigidity.
2 additional (generalized) tetrahedra can be added: one with a vertex $a_3$ 
from $A$ and one with a vertex $b_3$ from $B$ distinct from $a_1, b_1, a_2, b_2$. 
The vertex $a_3$ is connected by a new edge to $b_1$ or $b_2$.
The vertex $b_3$ is connected by a new edge to any vertex in $A$. In case (ii) there
are 3 edges between $A$ and $B$ and at most 3 further edges can be added to attain
minimal rigidity. Again, 2 additional (generalized) tetrahedra can be added. The first
containing $b_2 \ne b_1$ and the second with $b_3 \ne b_1, b_2$. 
The vertex $b_2$ is connected to two of the vertices among $a_1, a_2, a_3$,
and the vertex $b_3$ is connected to any one of the vertices $a_1, a_2, a_3$.
\end{proof}

\section{Core Algorithm Details}
\label{sec:app:algorithm}

\begin{algorithm} [htbp]
 \SetKwInOut{Input}{input}\SetKwInOut{Output}{output}
 {\bf sampleAtlasNode}\\
 \Input{atlasNode: node}
 \Output{Complete sampling of the atlasNode and all its children}
 \BlankLine

	$E$ = node.activeConstraints\\
	$G$ = node.activeConstraintGraph\\
	\If{ $G$ is minimally rigid}
		{stop;	}
	$F$ = complete3Tree($G$)\\
	
	$C$ = computeConvexChart($G$, $F$)\\

	\For{ each cayleyPoint $p$ within convexChart $C$ }
	{
		$R$ = computeRealizations($p$)\\

		\For{ each realization $r$ in $R$}
		{
			\If{!aPosterioriConstraintViolated($r$)}
			{
				\If{ isBoundaryPoint($r$) \&\& hasNewActiveConstraint($r$, $G$) }
				{
					$e$ = newActiveConstraint($r$, $G$);\\
					$G'$ := $G_{E \cup \{e\}}$ ;\\
					\If{ $G'$ is not already present in the current atlas}
					{
						childNode = new atlasNode($G'$)\\
						childNode.insertWitness($p$);\\
						sampleAtlasNode(childNode);\\
					} \Else{
						childNode = findNode($G'$);
					}
					node.setChildNode(childNode);
				} 
			}
		}
	}
		findAndSampleMissingAncestors(List of 0D regions);
\caption{High level pseudocode of the core algorithm}
\label{alg:sampleAtlasNode}
\end{algorithm}
This section details the core algorithm described briefly in Section
\ref{sec:coreAlgorithm}.  The exploration of the atlas is done by the recursive
\textbf{sampleAtlasNode} algorithm using one of the generated atlas root nodes
as input. Using depth first search, this algorithm samples the atlas node and
all its descendants.

\noindent\textbf{Base case of recursion:} If active constraint graph $G = (V, E)$ of
the node is minimally rigid i.e., the active constraint region is 0D, then
there is only 1 Cayley configuration (with finitely many Cartesian
configurations).  We have no more sampling to do, hence return.

\noindent\textbf{The recursion step:} If $G = (V, E)$ is not minimally rigid, the
core algorithm applies the \textbf{complete3Tree} algorithm  
to first check if $G$ is a partial 3-tree, and if so, find a set of non-edges $F$ that complete $G$ to a 3-tree.
$F$ is the set of Cayley parameters.
This leverages the convex parametrization theory~\cite{SiGa:2010} of Section
\ref{sec:convexification} and additionally ensures that a graph with edge set $E \cup F$
is minimally rigid, and the corresponding Cartesian configurations can be easily found. 

The \textbf{computeConvexChart} algorithm detects the tetrahedral bounds
on these Cayley parameters as described below and samples uniformly within this
region using a user specified step size.  

\begin{itemize} 

\item If there is only one Cayley parameter in a tetrahedron, the tentative
range of that parameter is computed by the intersection of tetrahedral
inequalities.

\item If there is more than one unfixed Cayley parameter in a tetrahedron,
then the tentative ranges of a parameters are computed in a specific sequence
\cite{ugandhar}. The tentative range of a parameter in the sequence is computed
through tetrahedral inequalities using fixed values for the parameters
appearing earlier in the sequence. Since the range of the parameter is affected
by the previously fixed parameters, more precise range computation of the
unfixed parameter is required for every iteration/assignment of fixed
parameters.  

\end{itemize}

\noindent \textbf{Cayley Sampling Efficiency and Decoupling:\\}
The actual range for each parameter for a given active constraint region is obtained by taking the intersection of
the tentative range and the range of Constraint \ctwo. The order
in which Cayley parameters are fixed have an effect on the efficiency of the
range computation. We pick parameters in the order that gives
the best efficiency\cite{ugandhar}. Once we choose the parameters $F$ and the sequence, the
explicit bounds can be computed in quadratic time in $|G|$.  Once explicit
bounds for each Cayley parameter have been found, we populate this region by
sampling it uniformly using a user specified step size. For roadmap generation,
coarse sampling is 
sufficient since the boundary being detected is only one dimension less.
Robustness against coarse sampling is also provided by the 
\textbf{findAndSampleMissingAncestors} algorithm, which ensures that all 
for any non-empty active constraint region, all ancestor regions are 
are added to the roadmap and sampled, even if they were initially missed 
due to coarse sampling. This effectively
decouples roadmap generation from sampling.

\noindent\textbf{Computing Cartesian Configurations:\\}
For each Cayley configuration, the sampleAtlasNode algorithm computes its
Cartesian configurations using the \textbf{computeRealization} algorithm.  The
\textbf{computeRealization} algorithm takes in a Cayley configuration and generates all
its 8 possible Cartesian configurations. There are 2 cases depending on
whether the active constraint graph is a partial 3-tree or not.  Cartesian
configurations for partial 3-trees is straightforward as described in Section
\ref{sec:realization}. 

\noindent\textbf{Boundary Detection:\\}
After finding the Cartesian configurations, we perform the
\textbf{aPosterioriConstraintViolated} check to discover a boundary region.
{\sl This is the crucial test that indicates that a new constraint has become
active.} The Cayley configuration, one of whose Cartesian configurations caused a child boundary region to
be found at a parent is called a \emph{witness} point, since it witnesses the
boundary, and is placed in the child boundary region clearly labeled as a
witness point coming from each parent region. 

The core algorithm relies on Cayley parameter grid sampling to find the child boundary
regions of each active constraint region. However, boundary detection is not
guaranteed by Cayley parameter grid sampling alone, since the sampling step
size may be too large to identify a close-by atom pair that causes a 
constraint to become active. That is, the constraint violation could occur between 2
feasible sample realizations or between a feasible and an infeasible
realization on the same flip in the sampling sequence. In the former case, the
missed boundary region is ``small,'' and will later be discovered by
the {\bf findAndSampleMissingAncestors}. In the latter case, the newly active
constraint has been flagged but exploration (by way of binary search) is
required to find the exact Cayley parameter values at which new constraints
became active. The binary search is on the Cayley parameter value, with
direction determined by whether the realization is feasible or not.

\noindent \textbf{Avoiding Repeat Sampling:\\}
In both cases, once a new active constraint $e$ is discovered, we add the new
constraint to $G$ and create an new active constraint graph $G'$ whose 
edge set is ${E \cup \{e\}}$. A boundary region could be detected via multiple parents.
However, since regions have unique labels, namely the active constraint graphs,
no region is sampled more than once. If $G'$ has already been sampled, we just
add the node for $G'$ into the atlas, as a child of $G$. Otherwise, we create
a new atlas node with $G'$, sample it using the recursive
\textbf{sampleAtlasNode} algorithm and then add it as a child of $G$.

\section{Symmetries in Assembly}
\label{sec:app:symmetries}
This section discusses symmetries used in the algorithm variant
described in Section \ref{sec:AlgorithmVariantn1}.
The concepts are developed in the paper \cite{sym8010005}.
An \emph{assembly configuration} is an ordered set $\mathcal{B} = (B_1, B_2,
\dots B_k)$, where each $B_i$, called a \emph{bunch}, is a set of $n$ identical
points. Two assembly configurations $\mathcal{B}$ and $\mathcal{B}'$ are
configurations of the same assembly system if $B_i$ is congruent to
$B'_{\sigma(i)}$ for some permutation $\sigma \in S_k$, for all $i$.  The set
of all assembly configurations of an assembly system is called an assembly
configurational region.

Two assembly configurations $\mathcal{B}$ and $\mathcal{B}'$ are
\emph{isomorphic} if there is a permutation $\sigma \in S_k$ such that for all
$i$, $B'_{\sigma(i)}$ is isomorphic to $B_i$.  Two assembly configurations
$\mathcal{B}$ and $\mathcal{B}'$ are \emph{strictly isomorphic}, if there is a
permutation $\sigma \in S_k$, such that for all $i$, $B'_{\sigma(i)}$ is
isomorphic to $B_i$. The weak automorphism group of $\mathcal{B}$ is the group
of all transformations that take $\mathcal{B}$ to a strictly isomorphic
$\mathcal{B}'$. It is clear that all assembly configurations in the same
assembly configurational region $\mathcal{A}$ have the same weak automorphism
group.

An active constraint graph $G(\mathcal{B})$ of an assembly configuration
$\mathcal{B} = (B_1, \dots, B_k)$ is a graph $(V, E)$ where the vertex set $V$
has one vertex for each $B_i$, and a vertex pair $\{x, y\} \in E$ if $x \ne y$
and $x$ and $y$ are at a preferred distance from each other. Two active
constraint graphs $G_1$ and $G_2$ are isomorphic if there is a $\sigma \in S_k$
such that $\{x, y\} \in E(G_1) \Leftrightarrow \{\sigma(x), \sigma(y)\} \in
E(G_2)$. The automorphism group of an active constraint graph $G$ is the group
of elements $\sigma \in S_k$ such that $\sigma(G) = G$. This is also called the
\emph{stabilizer group} of $G$. It is easy to see that the stabilizer group of
$\mathcal{B}$ is a subset of the stabilizer group of its active constraint
graph $G(\mathcal{B})$.  \figref{fig:v6e12} shows all the non-isomorphic active
constraint graphs with 12 edges of an assembly system of 6 bunches each
containing identical singleton spheres.

An active constraint region $R_G$ of the assembly configurational region contains 
all assembly  configurations $\mathcal{B}$ with active constraint graph 
$G(\mathcal{B}) = G$.

\begin{theorem}
\label{thm:symmetry}
\cite{sym8010005} For an active constraint graph $G = G(\mathcal{B})$ of an assembly configurational
region 
$\mathcal{A}$, it holds that:
$$stab_{Waut_\mathcal{A}} \mathcal{B}  \subseteq stab_{Waut_\mathcal{A}} G = stab_{Waut_\mathcal{A}} R_G$$
In addition, there exist active constraint graphs $G$ of assembly configurational region 
$\mathcal{A}$ where the above containment is strict, i.e.,
for every $\mathcal{B}$ such that $G=G(\mathcal{B})$,  
$$stab_{Waut_\mathcal{A}} \mathcal{B}  \subsetneq stab_{Waut_\mathcal{A}} G = 
stab_{Waut_\mathcal{A}} R_G$$
\end{theorem}

A \emph{fundamental domain} of a stratification $\mathcal{S}(\mathcal{A})$ is
the minimal sub-stratification $\overline{\mathcal{S}}(\mathcal{A})$ such that
$\bigcup_{\pi \in S_k} \pi(\overline{\mathcal{S}}(\mathcal{A}) = \mathcal{S}
(\mathcal{A})$, where each $\pi$ acts on $\overline{\mathcal{S}}(\mathcal{A})$
via its action on the active constraint region of
$\overline{\mathcal{S}}(\mathcal{A})$. 

\section{Algorithm Variant for $n=1$ and $2 < k \le 24$}
\label{sec:app:stickySphereAlgorithm}

\begin{algorithm}
\SetKwInOut{Input}{input}\SetKwInOut{Output}{output}
{\bf symmEASAL}\\
		\Input{$k$: number of $n=1$, identical \rmc s, \\$d_\mathcal{C}$: dimension of assembly landscape}
 \Output{Atlas}
		Generate all non-isomorphic graphs $G_{ij} = (V, E_{ij})$ of $k$ vertices and $|E_{ij}| = 3k -j$, $j= 6 \dots 6+d_\mathcal{C}$\\
		\For{each $G_{ij} = (V, E_{ij})$ with $j=6+d_\mathcal{C}$(root nodes)}{
				sampleAtlasNode($G_{ij}$) 
	}
		\For{$j=6+d_\mathcal{C}$ to $j=6$}{
				Remove all nodes with graph $G_{ij}$ not in the atlas and descendants 
			}
\caption{symmEASAL}
\label{alg:Sticky}
\end{algorithm}

\end{document}